\documentclass[11pt]{iopart}
\usepackage{graphicx,amssymb,amsfonts,units,bm,amsthm} 

\newtheorem{theorem}{Theorem}[section]
\newtheorem{lemma}[theorem]{Lemma}
\newtheorem{proposition}[theorem]{Proposition}
\newtheorem{corollary}[theorem]{Corollary}
\newenvironment{definition}[1][Definition.]{\begin{trivlist}
\item[\hskip \labelsep {\bfseries #1}]}{\end{trivlist}}

\begin{document}

\topical[Disordered Topological Insulators]{Disordered Topological Insulators: A Non-Commutative Geometry Perspective}

\author{Emil Prodan}

\address{Department of Physics, Yeshiva University, New York, NY 10016}

\begin{abstract}
The progress in the field of topological insulators is impetuous, being sustained by a suite of exciting results on three fronts: experiment, theory and numerical simulation. Very often, the theoretical characterizations of these materials involve advance and abstract techniques from pure mathematics, leading to complex predictions which nowadays are tested by direct experimental observations. Many of these predictions have been already confirmed. What makes these materials topological is the robustness of their key properties against smooth deformations and onset of disorder. There is quite an extensive literature discussing the properties of clean topological insulators, but the literature on disordered topological insulators is limited. This review deals with strongly disordered topological insulators and covers some recent applications of a well established analytic theory based on the methods of Non-Commutative Geometry (NCG) and developed for the Integer Quantum Hall-Effect. Our main goal is to exemplify how this theory can be used to define topological invariants in the presence of strong disorder,  other than the Chern number, and to discuss the physical properties protected by these invariants. Working with two explicit 2-dimensional models, one for a Chern insulator and one for a Quantum spin-Hall insulator, we first give an in-depth account of the key bulk properties of these topological insulators in the clean and disordered regimes. Extensive numerical simulations are employed here. A brisk but self-contained presentation of the non-commutative theory of the Chern number is given and a novel numerical technique to evaluate the non-commutative Chern number is presented. The non-commutative spin-Chern number is defined and the analytic theory together with the explicit calculation of the topological invariants in the presence of strong disorder are used to explain the key bulk properties seen in the numerical experiments presented in the first part of the review.
\end{abstract}


\date{\today}

\maketitle

\tableofcontents

\section{Introduction}

The field of topological insulators is progressing extremely fast on both theoretical and experimental fronts and in the past few years it attracted an unprecedented attention from the condensed matter community. This expedited but self-contained review is concerned with a less studied aspect of the field, namely the effect of strong disorder on topological materials. A material is called topological insulator if it behaves like an insulator when probed deep into the bulk and as a metal when probed near any edge or surface cut into the material \cite{Kane:2006xu}. This behavior is not triggered by any externally applied field and, instead, it is an intrinsic property of the material. For a topological insulator, the metallic character of the edges or surfaces is robust against smooth deformations of the material, as long as the insulating character is maintained in the bulk of the material. Examples of topological insulators are the Chern and Quantum spin-Hall (QSH) insulators which will be extensively discussed later.

The idea behind this review was to bring the attention to a set of analytic tools developed for the Integer Quantum Hall-Effect (IQHE) by Bellissard and his collaborators in the late 1980's, reviewed in the excellent paper from 1994 \cite{BELLISSARD:1994xj}. As we shall see, these analytic tools can be applied quite directly to many classes of topological insulators \cite{Prodan:2010cz}, therefore providing a natural theoretical framework to analytically treat the effect of strong disorder in topological materials (for alternative approaches we point the reader to Refs.~\cite{HastingsJMP2010gy,Hastings2010by}). Is the work by Bellissard et al relevant to the field of topological insulators? We think it is more than ever, at both conceptual and practical levels.

It is important at the conceptual level because the main claim in the field of topological insulators is the robustness of the topological properties against disorder. This is a huge claim, holding a lot of promises as most of the newly envisioned applications are based on it. But to date, we are still missing the hard evidence for it, counting experiment, numerics and theory all together. The transport experiments on HgTe/CdTe quantum wells \cite{Koenig:2007ko,RothSc2009cu}, the only 2D QSH insulator discovered to date, consistently showed a decrease of the conductance with the increase of separation between the ohmic contacts. The quantization of the conductance, as predicted by theory, was observed only for short contact distances of less than 1 micron (one should be aware that HgTe and CdTe materials are extremely difficult to work with so it is hard to pinpoint the cause of this behavior). For 3D samples, angle resolved photoemission spectroscopy (ARPES) measurements for disordered surfaces \cite{Xia2008fk,HsiehNature2009io,Xia2009xu} seem to indicate robust topological extended surface bands, but these conclusions need to be confirmed by transport measurements. The transport measurements in 3D topological samples have been notoriously difficult \cite{TaskinPRB2009xu,CheckelskyPRL2009fy,PengNatMat2010gy,AnalytisPRB2010hv} because of a metallic bulk. Recent characterizations also showed that band bending near the crystal's surface can trap conventional 2D electron states that coexist but also mix with the topological surface states \cite{BianchiNatComm2010fy}, making the experiments even harder. But most recently, high quality single crystals grown by molecular beam epitaxy have been reported \cite{ChengPRL2010gj,ChenPRL2010gy,ZhangNatPhys2010vy,Wang2010lf} (see also the important new development in Ref.~\cite{SatoPRL2010yv}). When properly doped, these crystals display insulating bulk \cite{Wang2010lf} and accurate transport measurements of the surface can be recorded. Unfortunately, the first such transport measurements indicate that the topological surface states are slightly localized (the localization is believed to be induced by inelastic scattering processes). Ref.~\cite{Wang2010lf} also reported ARPES measurements, which look very similar to the previously reported data \cite{Xia2008fk,HsiehNature2009io,Xia2009xu}, despite the fact that the surface states are localized. The character of the surface states has been also probed by STM measurements \cite{RoushanNature2009vu,Yazdani:2010cz}, which gave evidence that at short scales (the ``field of view" of the STM is less than 1 $\mu$m$^2$) the states are extended despite the presence of strong defects. Shubnikov-de Haas oscillation measurements, which allow one to map the Fermi surface (if any), gave an inconclusive picture so far, with few studies \cite{QuScience2010gy,RenPRB2010vu} reporting clear signal coming from a metallic surface and another study \cite{ButchPRB2010gl}, done with ultra high-quality crystals, reporting no significant contributions from the surface, implying an extremely low surface conductivity. The conclusion here is that the experimental measurements are converging to a point where the robustness against disorder can be rigorously tested but we don't have yet the experimental confirmation of this property.

The number of existing numerical simulations for disordered topological insulators is quite small compared with the number of simulations done in the 80's and 90's for the Anderson localization. Until recently, there was only one short numerical study \cite{Onoda:2007xo} on the robustness of the bulk extended states in QSH insulators. This study used the transfer matrix analysis and was completed for small quasi-one dimensional samples. Recently, the transfer matrix analysis was repeated for much larger systems in Ref.~\cite{Yamakage2010xr}, re-confirming the existence of robust extended bulk states  against disorder in QSH insulators. This study seems to contain the most accurate computations to date. The transfer matrix analysis was also adopted in \cite{Obuse:2007qo,Obuse:2008ff} for a representative network model in the symplectic class, with emphasis on the critical exponents at the mobility edges. The properties of the bulk states in a 2D QSH insulator were also probed in Ref.~\cite{Essin:2007ij} by computations of certain topological invariant in the presence of disorder (the systems considered in this study are extremely small). Same method was adopted in Ref.~\cite{GuoPRB2010fu} for a 3D QSH insulator with disorder (probably the first simulation in 3D; also very small). Ref.~\cite{Prodan2010ew} presented a level statistics analysis for Chern insulators and computations of the Chern number in the presence of disorder. The robustness of the edge states against disorder was studied in Refs.~\cite{LiPRL2009xi,GrothPRL2009xi,JiangPRB2009sf} by computing the Landauer conductance for a disordered ribbon connected to clean leads. All these three studies worked with $S_z$ conserving models and were limited to small systems (the length of the ribbons was about 200 lattice sites; this number was $10^8$ in the transfer matrix computations of Ref.~\cite{Yamakage2010xr}; we also want to mentioned that the theory of the edge states in $S_z$ conserving models with disorder is firmly established \cite{Prodan:2009lo}). But despite all these fine numerical simulations, we are still lacking systematic studies that combine more than one method and where the numerical convergence is rigorously analyzed. From our experience, we can attest that, no matter how elaborate these numerical experiments are, there will always be a margin of doubt about the localization-delocalization issues, until an analytic proof will be available.

The non-commutative methods have also a great practical value. As pointed out in Ref.~\cite{HastingsJMP2010gy}, the non-commutative formulas of the invariants can lead to extremely efficient numerical algorithms, allowing computations of the invariants in the presence of disorder for system sizes that are orders of magnitude larger that what was possible with more traditional methods \cite{Loring2010vy,Prodan2010ew,Hastings2010by}. Developing accurate and efficient methods for mapping the phase diagram of disordered topological insulators is extremely important for practical applications especially that, as pointed out in Refs.~\cite{LiPRL2009xi,GrothPRL2009xi,Yamakage2010xr}, the phase diagram of a topological insulator can be strongly deformed by the presence of disorder.  

There are additional reasons for writing this review. By its very nature, the field of topological insulators can lead to an unprecedented cross fertilization between condensed matter physics and various fields in pure mathematics with tremendous potential benefits for both fields. We have already seen applications from classic Topology \cite{Kane:2005np,Moore:2007ew,Hatsugai2010nj,Roy2010nj}, Chern-Simons Theory \cite{qi2008B}, Conformal Field Theory, K-Theory \cite{Kane:2005zw,kitaev:22}, Random Matrix Theory and nonlinear $\sigma$-models \cite{Schnyder:2008qy,Ryu:2010tq}. One hope is that we will see many more contributions of this sort from theoretical condensed matter physicists and from pure mathematicians. For this reason, we have tried to keep the presentation pedagogical and appealing to a wide audience, especially to the undergraduate and graduate students looking for good projects, to the theoretical condensed matter researchers who like to compute things explicitly, and to specialists in Non-Commutative Geometry looking for exciting applications of their field. The targeted audience is quite broad and choosing the style of the exposition was not easy. Our final choice might not satisfy all readers, but at least we want to let the readers know that a great deal of effort was spent on this issue. 

Our discussion is restricted only to the bulk properties of topological insulators in two dimensions. Although the current and broadly accepted definition of a topological insulator highlights only the robust metallic character of the edge or surface states, every known topological insulator seems to display extended bulk states that are robust against disorder. This is an extraordinary behavior, especially in two dimensional models. When an edge or a surface is cut in a sample of topological insulator, the emerging edge/surface states seem to be connected to these bulk states. In fact, the edge/surface states can be viewed as these extended bulk states terminating at boundary. For this reason, understanding the bulk and the edge/surface properties of the topological insulators is equally important.  

We will present several numerical experiments, involving straightforward applications of classic techniques such as level spacing statistics. We will follow the standard interpretation of the numerical outputs, which will show a clear difference between the behavior of a normal and a topological insulator in the presence of disorder. This together with a detailed introduction of two models of topological insulators will occupy the first part of the review. We also describe here how to define a robust spin-Chern invariant for $S_z$ non-conserving models (that is, systems that do not decouple into independent copies of Chern insulators). Several interesting questions will emerge, which will set the direction for the rest of the review.

The analytic part of the review presents a brisk account of the Non-Commutative Theory of the Chern number developed by Bellissard, van Elst and Schulz-Baldes \cite{BELLISSARD:1994xj}. We have reworked certain parts to give the exposition a more ``calculatoristic" flavor, so that condensed matter physicists who like to compute could easily follow the arguments. We have complemented the proofs with discussions and remarks, and tried to keep the reader informed at all times about where the calculation or the argument is heading and why do we need to go there. We summarize the arguments before each lengthy proof to provide more guidance. We decided to collect the important statements in Theorems, Lemmas and Propositions, something to the taste of the mathematicians but that could easily irritate other people. One reason for why we chose to do so was to alert the reader that these statements have a rigorous proof and that they can be applied with absolute confidence. Another reason was that, by doing so, we can state in one place the result and the conditions when the result is valid. The last part is especially important for our subject because our main goal is precisely to find the most general conditions that assures the quantization and invariance of the topological numbers.

The review includes a presentation of a numerical technique to evaluate the Chern \cite{Prodan2010ew} and spin-Chern numbers in the presence of disorder. This technique steams directly from NCG Theory and allows one to compute the invariants for finite systems without imposing the traditional twisted boundary conditions. The finite size formulas converge exponentially fast to the thermodynamic limit, given by the NCG Theory. The technique allows us to compute the Chern and spin-Chern numbers for large lattice systems and large number of disorder configurations (at least one order of magnitude over what is currently available in the published literature).

The review has a section devoted entirely to applications. The class of Chern insulators was chosen as the ``control case," because they closely resemble the Integer Quantum Hall-Effect, already extremely well understood. In this case, the NCG Theory gives a full account of all the effects seen in the numerical experiments on Chern insulators, presented at the beginning of the review. Calculations of the Chern number will allow us to witness explicitly its quantization when the Fermi level is located inside the localized part of the energy spectrum, and the failure of such quantization when the Fermi level is located inside the delocalized part of the energy spectrum. For Quantum spin-Hall insulators, we define the non-commutative spin-Chern number, following Ref.~\cite{Prodan:2009oh}, and discuss the conditions when its quantization occurs. Explicit calculations of the spin-Chern number indicate again quantization when the Fermi level is located in the localized part of the energy spectrum. 

\section{Topological insulators: A brief account} 

This will be a brief account, indeed. The reason we kept it short is because there are now several reviews surveying the evolution of the field and its current status, from both theoretical and experimental point of views \cite{Qi2010,ZHassanRevModPhys2010du,Qi2010hg,Hasan2010by}. Nevertheless, through this brief account we want to let the readers know about the impetuous advances that are happening right now in the field of topological insulators.

It is probably a good idea to start from the Integer Quantum Hall Effect, (IQHE) which is now extremely well understood. Discovered at the beginning of the 1980's \cite{Klitzing:1980vh}, IQHE revealed a truly spectacular manifestation of ordinary matter, displayed in the quantization of the Hall conductance and the emergence of dissipationless charge currents flowing around the edges of any finite IQHE sample. The intellectual activity spurred by this effect has led to some of the greatest leaps in condensed matter theory. Working with a clean periodic system and using Kubo's formula for the Hall conductance $\sigma_H$, Thouless, Kohmoto, Nightingale and den Nijs made the famous connection between $\sigma_H$ and a topological invariant now known as the TKNN invariant. Using general charge-pumping arguments, the Hall conductance was also linked to the classic Chern number (see Avron in Physics Today 2003). But it was already clear from the early works \cite{Laughlin1981fe,Prange1981uc,Halperin1982fs,Joynt1984hc} that impurity states are essential for explaining the Hall plateaus seen in the IQHE experiments. The quest for an analytic theory of IQHE that includes the disorder has led Bellissard, van Elst and Schulz-Baldes \cite{BELLISSARD:1994xj} to one of the most amazing applications of a new and exciting branch of mathematics called Non-Commutative Geometry \cite{Connes:1994wk}. This work gives an explicit optimal condition that assures the quantization and invariance of the bulk Hall-conductance in the presence of strong disorder. Further homotopy arguments for quantization and invariance of the bulk Hall-conductance were developed in Refs.~\cite{Aizenman1998bf} and \cite{Richter2001jg}.

The progress about the edge physics of IQHE satarted with the works by Hatsugai, who established in 1993 a fundamental result \cite{Hatsugai:1993cs,Hatsugai:1993jt} saying that the number of conducting edge channels, forming in an energy gap of the Landau Hamiltonian due to the presence of an edge, is equal to the total Chern number of the Landau levels below that gap. The technique developed by Hatsugai can deal only with clean systems, rational magnetic flux and  homogeneous edges with Dirichlet boundary conditions. It was only about 10 years later when, using the methods of Non-Commutative Geometry, Kellendonk, Richter and Schulz-Baldes established \cite{SchulzBaldes:2000p599,Kellendonk:2002p598,Kellendonk:2004p597} a new link between the bulk and edge theory, which ultimately allowed them to generalize Hatsugai's statement to cases with weak random potentials, irrational magnetic flux and general boundary conditions. The equality between bulk and edge Hall conductance was also demonstrated by  Elbau and Graf \cite{Elbau:2002qf}, soon after the publication of Ref.~\cite{Kellendonk:2002p598}, this time using more traditional methods.  Further progress was made in Ref.~\cite{Combes:2005qd}, which treated continuous magnetic Schr\"odinger operators and potentials that can assume quite general forms, in particular, they can include strong disorder. A similar result was established for discrete Schr\"odinger operators in Ref.~\cite{Elgart:2005rc}. We mentioned that some of these ideas were formalized in an abstract setting in Ref.~\cite{Prodan:2009od} and applications to the edge states problem in topological insulators were given in Refs.~\cite{Prodan:2009mi,Prodan:2009lo}.

The  IQHE can be observed only in the presence of an externally applied magnetic field. In 1988, Haldane presented a model of a condensed matter phase that exhibits IQHE without the need of a macroscopic magnetic field \cite{HALDANE:1988rh}. The systems that behave like the one described by Haldane are now called Chern insulators. The time-reversal symmetry in these systems is broken like in the IQHE, but it is broken by the presence of a net magnetic moment in each unit cell rather than by an external magnetic field, as it is the case for IQHE. As we shall see, the techniques developed for IQHE can be directly applied to Chern insulators, whose bulk and edge physics \cite{Prodan:2009lo} is very well understood now. The Chern insulators were never found experimentally, even thought there is not one known physical reason for this not to happen one day. The Haldane model truly describes the first topological insulator, but because of the lack of the experimental evidence, the field of topological insulators took off many years after the work of Haldane. 

The interest for a related phenomenon, the spin-Hall-Effect, was picking up in the mid 2000's. The effect was predicted decades ago \cite{Dyakonov1971sc,Dyakonov1972er}, and says that a bar made of a semiconductor with strong spin-orbit interaction will display spin-polarized edge states when an electric charge current is forced through it. The effect was finally observed in 2004-2005 \cite{Kato2004ht,Wunderlich2005hg}. The search for a quantized version of the spin-Hall-Effect started at the time when the results on the classical spin-Hall-Effect were making the news \cite{Zhang2001vc,Murakami:2003zf,Murakami:2004tv,Murakami:2004dx,Sinova:2004ad,Culcer:2004zi,Wunderlich:2005io,Sheng:2005mg,Sheng:2005lj}, but at that time nobody could imagine that there are samples displaying a quantized spin-Hall-Effect in the absence of externally applied fields. This changed after the discovery of graphene \cite{Novoselov:2004oc,Novoselov:2005lo,Novoselov:2005rk,Zhang:2005mu}, which inspired Kane and Mele to propose an explicit model \cite{Kane:2005np} displaying topological edge modes carrying a net spin current around the edges. The emergence of the spin-carrying edge states is triggered solely by the intrinsic spin-orbit interaction and the flow of the spin current is protected by the time reversal symmetry of the model. All materials that are not magnetically ordered display the time-reversal symmetry, and there is a large number of materials with strong spin-orbit interaction that are not magnetically ordered. Therefore, the chance of observing the Quantum spin-Hall-Effect in real materials is quite high. The materials exhibiting this effect are now called Quantum spin-Hall (QSH) insulators and their hallmark is a dissipation-less spin current flowing along the edges of  the samples, an effect due to the non-trivial topological properties of the bulk \cite{Kane:2005zw}. 

The original calculations suggested that the newly discovered graphene could be a QSH insulator. Unfortunately, the spin-orbit interaction is very weak in graphene and that makes the experimental detection of the effect very difficult. Nevertheless, the race for the discovery of the first QSH insulator was on. HgTe/CdTe quantum wells were predicted to display QSH-Effect in 2006 \cite{Bernevig:2006hl} and confirmed experimentally in the following year \cite{Koenig:2007ko}. The first QSH insulator was discovered and a new field emerged \cite{Kane:2006xu}, that of topological insulators defined as materials that are insulators in the bulk but metallic along any edge or surface that is cut into the material. Unfortunately, the HgTe quantum wells remain the only two dimensional QSH insulators discovered to date. In three dimensions, the list of confirmed QSH insulators is quite impressive \cite{Hsieh:2008vm,Kuga:2008hv,Shitade2008,Xia:2009cu,Hsieh:2009wq,Zhang:2009mm,Hor:2009zd,Chen:2009vs,Lin2010gh,Lin2010vb,Lin2010fg,Yan2010gl,Xiao2010kc,Xu2010ci,Xu2010jf,SatoPRL2010yv,KimPRB2010ru,Wang2010bu} and the experimental characterization of these materials is vigorously underway \cite{koenig2008,hsieh2009a,hsieh2009b,taskin2009,gomes2009,checkelsky2009,alpichshev2009,tzhang2009,hor2009,hor2009a,EtoPRB2010xy,TaskinPRB2010vi,Nishide2010nj,Jenkins2010te,Zhao2010hf,Wray2010ce} (see also the references cited in the Introduction). Additional classes of topological insulators are expected to emerge in the future \cite{Fu2010jy,Hughes2010gh,Turner2010cu}. 

\section{Introduction to Chern Insulators}

\subsection{Chern insulators in the clean limit}

\begin{figure}
 \center
 \includegraphics[width=8cm]{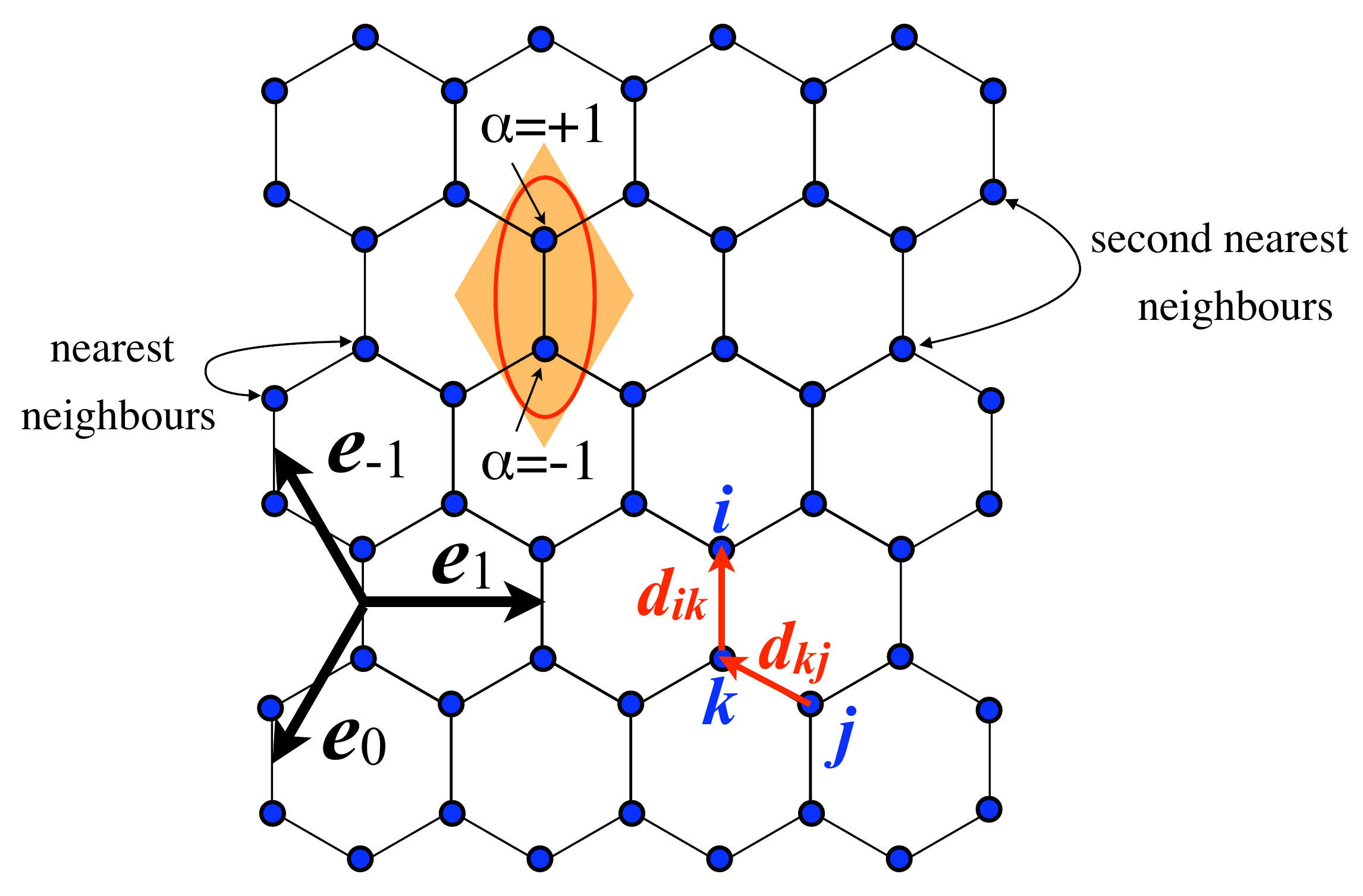}
 \caption{The honeycomb lattice, its unit cell and generating vectors, together with some notations used in the main text.}
 \label{Lattice1}
\end{figure}

A Chern insulator is a periodic band insulator with broken time reversal symmetry, with the distinct property of having a net charge current flowing around the edges of any finite sample. The time reversal symmetry is broken not by an externally applied magnetic field, but by some intrinsic property of the material, such as the occurrence of a net magnetic moment in each unit cell. In the following, we will use an explicit model to exemplify some of the most important features of these materials.     

The first model of a Chern insulator was introduced by Haldane in 1988 \cite{HALDANE:1988rh}, who worked with the honeycomb lattice shown in Fig.~\ref{Lattice1}. The honeycomb lattice can be viewed as a triangular lattice with two sites per unit cell (see the shaded region in Fig.~\ref{Lattice1}). The two sites of the unit cell will be labeled by $\alpha=\pm 1$ as in Fig.~\ref{Lattice1}. The triangular lattice is generated by the vectors ${\bm e}_{\pm 1}$. An additional vector ${\bm e}_0=-({\bm e}_{1}+{\bm e}_{-1})$ is shown in Fig.~\ref{Lattice1}, which will play a certain role later. The Haldane model involves spinless electrons and assumes only one quantum state (orbital) per site, denoted by $|{\bm n}\rangle$. The linear combinations of these states generate a Hilbert space ${\cal H}$, which is equipped with the scalar product $\langle {\bm n}|{\bm m} \rangle=\delta_{{\bm n}{\bm m}}$. The system is assumed half-filled, which means there is one electron per unit cell .

\begin{figure}
 \center
 \includegraphics[width=13cm]{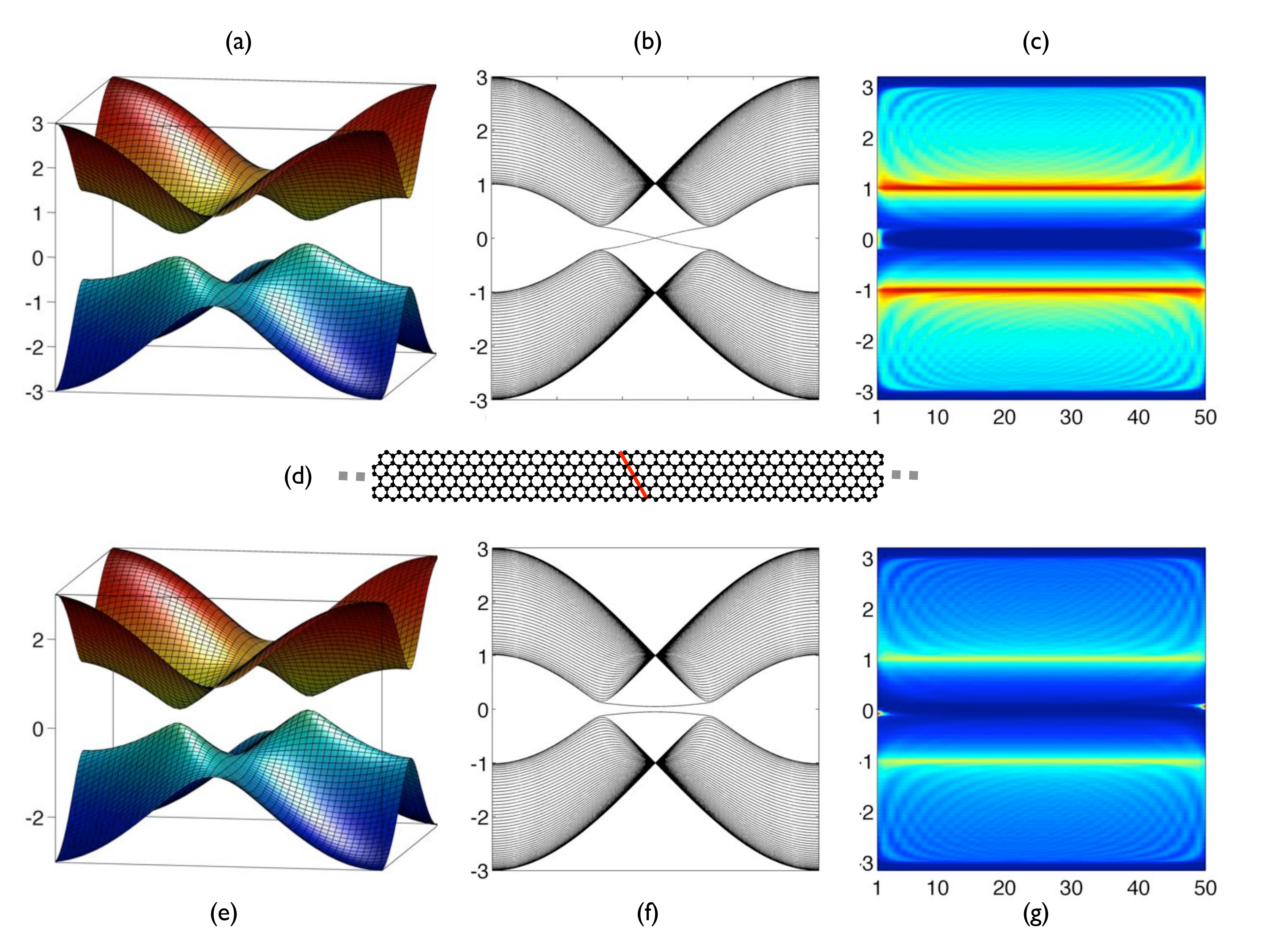}
 \caption{(a) The bulk spectrum of Haldane Hamiltonian Eq.~\ref{HaldaneModel} ($t=0$ and $\eta=0.1$) as function of $(k_1,k_2)$. (b) The energy spectrum of the same Hamiltonian when restricted on an infinitely long ribbon with open boundary conditions at the two edges. The spectrum is represented as function of $k$ parallel to the ribbon's edges. (c) The local density of states (see Eq.~\ref{ldos}) of the ribbon, plotted as an intensity map in the plane of energy (vertical axis) and unit cell number along the red line shown in panel (d) (horizontal axis). Blue/red colors corresponds to low/high values. (d) Illustration of the ribbon used in the calculations shown in panels (b, c) and (f, g). The ribbon was 50 unit cells wide. (e-g) Same as (a-c) but for $t=0.1$ and $\tau=0$.}
 \label{ChernSpectrum}
\end{figure}

In its simplest form, the Haldane's Hamiltonian reads:
\begin{equation}\label{HaldaneModel}
\begin{array}{c}
H_0^{\mbox{\tiny{Chern}}}=\sum\limits_{\langle {\bm n \bm m} \rangle} |{\bm n}\rangle \langle {\bm m}| + \sum\limits_{\langle \langle {\bm n \bm m} \rangle \rangle} \{ \zeta_{\bm n} |{\bm  n}\rangle \langle {\bm  m}|+\zeta_{\bm n}^* |{\bm  m}\rangle \langle {\bm  n}|\},
\end{array}
\end{equation}
where $\zeta_{\bm n}=\frac{1}{2}(t+i\eta)\alpha_{\bm n}$ with $\alpha_{\bm n}$ being the $\pm 1$ label attached to each site ${\bm n}$, depending on how ${\bm n}$ is positioned in the unit cell (see Fig.~\ref{Lattice1}), also known as the isospin in the condensed matter community. The single (double) angular parenthesis indicate that the sum over ${\bm n}$ runs over all the lattice sites while the sum over ${\bm m}$ is restricted to the first (second) near neighboring sites to ${\bm n}$. Notice that a second neighbor hopping always connect sites with same $\alpha$.  If we view the honeycomb lattice as a triangular lattice with two sites per unit cell, then the Hamiltonian takes the form: 
\begin{equation}\label{TriangChernModel}
\begin{array}{c}
H_0^{\mbox{\tiny{Chern}}}=\sum\limits_{{\bm n},\alpha,\gamma} |{\bm n},\alpha \rangle \langle {\bm n}+\alpha \gamma{\bm e}_\gamma,-\alpha|  \medskip \\
+\sum\limits_{{\bm n},\alpha,\gamma} \{\xi_\alpha |{\bm n},\alpha\rangle \langle {\bm n}+{\bm e}_\gamma,\alpha| + \xi_\alpha^* |{\bm n}+{\bm e}_\gamma,\alpha \rangle \langle {\bm n},\alpha| \},
\end{array}
\end{equation} 
where now ${\bm n}$ denotes the position of the unit cell in the triangular lattice and $\alpha=\pm 1$ is the isospin labeling the two sites of a unit cell. The variable $\gamma$ takes the values 0 and $\pm 1$, and ${\bm e}_\gamma$ are the vectors shown in Fig.~\ref{Lattice1}. We actually prefer this later form of the Hamiltonian, which will be used from now on. The Hamiltonian depends on the two parameters $(t,\eta)$. We will omit the label ``Chern" and use the simplified notation $H_0$ for the Hamiltonian of Eq.~\ref{HaldaneModel} throughout the current section.

In the absence of disorder, we can perform the Bloch decomposition using the isometric transformation $U$ from ${\cal H}$ into a continuum direct sum of ${\bm C}^2$ spaces:
\begin{equation}
\begin{array}{c}
U: {\cal H} \rightarrow \bigoplus_{{\bm k}\in {\cal T}} {\bm C}^2, \ \ U|{\bm n},\alpha\rangle = \frac{1}{2\pi} \bigoplus_{{\bm k}\in {\cal T}}  e^{-i {\bm k}\cdot{\bm n}} \xi_\alpha,
\end{array}
\end{equation}
where ${\cal T}$ is the Brillouin torus ${\cal T}$=$[0,2\pi]$$\times$$[0,2\pi]$ and: 
\begin{equation}
\begin{array}{c}
\xi_1=\left (
\begin{array}{c}
1 \\
0
\end{array}
\right ), \ \ 
\xi_{-1}=\left (
\begin{array}{c}
0 \\
1
\end{array}
\right ).
\end{array}
\end{equation}
 Under this transformation, $UH_0U^{-1}=\bigoplus_{{\bm k}\in {\cal T}}\hat{H}_{\bm k}$ with:
\begin{equation}\label{KHaldane}
\hat{H}_{\bm k}=\sum\limits_{\gamma}\left(
\begin{array}{cc}
 t \cos({\bm k} {\bm e}_\gamma)-\eta\sin({\bm k}{\bm e}_\gamma) &  e^{i\gamma{\bm k}\cdot{\bm e}_\gamma} \medskip\\
 e^{-i\gamma{\bm k}\cdot{\bm e}_\gamma} & - t \cos({\bm k}{\bm e}_\gamma)+\eta\sin({\bm k}{\bm e}_\gamma) 
\end{array}
\right )
\end{equation}
We denote by $\epsilon_{\bm k}^{1,2}$ the two eigenvalues of $\hat{H}_{\bm k}$. The plot of $\epsilon_{\bm k}^{1,2}$ as function of ${\bm k}$ will be referred to as the bulk band structure of the model.

\begin{figure}
 \center
 \includegraphics[width=4cm]{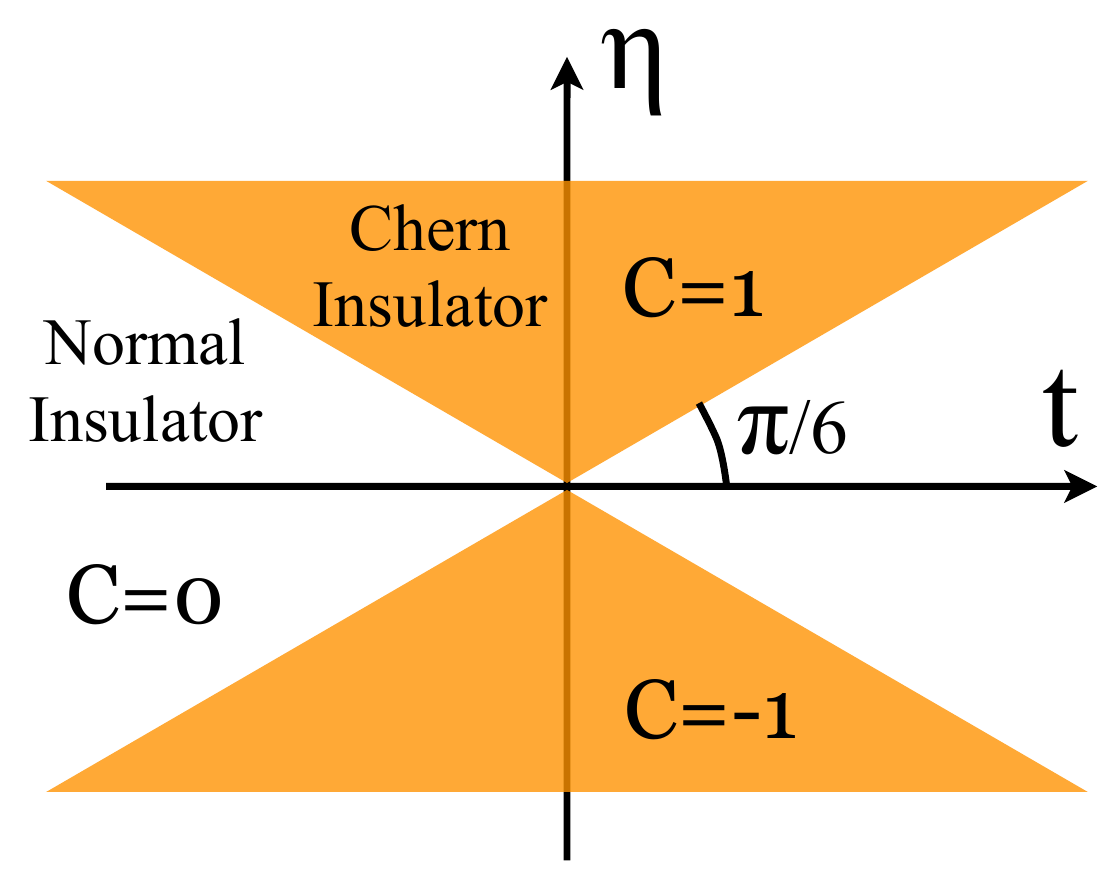}
 \caption{The region of the parameter space where the model of Eq.~\ref{HaldaneModel} is in the Chern insulating phase and displays the topological edge bands.}
 \label{ChernPhaseBoundary}
\end{figure}

We now imagine the following numerical experiment. We let the computer pick random points in the $(t,\eta)$ plane and then perform a computation of the energy spectrum for an infinite sample (the bulk spectrum), a computation of the energy spectrum for a ribbon shaped sample and a computation of the local density of states (LDOS) for the ribbon. The experiment will reveal that, with probability one, the system is an insulator because the occupied states are separated by a finite energy gap from the un-occupied states, as it is exemplified in panels (a) and (e) of Fig.~\ref{ChernSpectrum}. The bulk band spectrum will not reveal major differences between various regions of the $(t,\eta)$ parameter plane. However, the calculations for the ribbon geometry will bring major qualitative differences. For some values such as $t$=0.1 and $\eta$=0, the energy spectrum for the ribbon geometry displays an insulating energy gap, while for values like $t$=0 and $\eta$=0.1 it doesn't. Things become even more intriguing if we look at this spectrum as function of the momentum parallel to the direction of the ribbon. Examining panels (b) and (f) of Fig.~\ref{ChernSpectrum}, we see that, when $t$=0 and $\eta$=0.1, the spectrum displays two solitary energy bands crossing the bulk insulating gap. For $t$=0.1 and $\eta$=0, we can still see two solitary bands but they don't cross the bulk insulating gap. If we let the computer run for a while, picking random points in the $(t,\eta)$ plane, it will slowly reveal that this plane splits into regions were the model displays bands that cross the insulating gap like in Fig.~\ref{ChernSpectrum}(b) and region where the insulating gap remains open like in Fig.~\ref{ChernSpectrum}(f). These regions are shown in Fig.~\ref{ChernPhaseBoundary}.

It is instructive to also take a look at the maps of the local density of states (LDOS):
\begin{equation}\label{ldos}
\begin{array}{l}
\rho(\epsilon,{\bm n})=\frac{1}{\pi} \mbox{Im} \{ (H_0-\epsilon-i 0^+)^{-1}({\bm n},{\bm n}) \},
\end{array}
\end{equation}
which will reveal the spatial distribution of the quantum states. The $\rho(\epsilon,{\bm n})$ written above depends on 3 variables, the energy plus the two spatial coordinates, but for a homogeneous ribbon, like the one shown in Fig.~\ref{ChernSpectrum}(d), $\rho(\epsilon,{\bm n})$ is independent of the coordinate parallel to the edge. Hence $\rho$ is only a function of energy and one lattice coordinate, chosen to be along the red line of Fig.~\ref{ChernSpectrum}(d), in which case we can display $\rho$ using an intensity map. Such maps are shown in Figs.~\ref{ChernSpectrum}(c) and (g). Here, one can see that, if the spatial coordinate is away from the edges of the ribbon, there are clear regions of practically zero density of states, regions that are perfectly aligned with the bulk gaps seen in the band spectra of Figs.~\ref{ChernSpectrum}(a) and (e). When the spatial coordinate approaches the edges of the ribbon, the LDOS inside the bulk gap starts to pick up appreciable values in Fig.~\ref{ChernSpectrum}(c). This part of the LDOS can come only from the two bands crossing the bulk insulating gap in Fig.~\ref{ChernSpectrum}(b). In other words, the quantum states associated with these two bands are localized near the edges of the ribbon and, for this reason, they are called edge bands. Since the slope $d\epsilon_k/dk$ of a band gives the group velocity of an electron wave-packet generated from that band, we can label the edge bands as right and left moving. A more detailed analysis of the LDOS will reveal that the right/left moving bands are localized on the lower/upper edges of the ribbon, respectively (the correspondence will switch if we change the sign of $\eta$). Of course, there is a hybridization between two edge bands and a tiny energy gap is opened at the apparent band crossing, but this hybridization becomes exponentially small as the width of the ribbon is increased. For the ribbon considered in Fig.~\ref{ChernSpectrum}, this hybridization can be practically ignored. In fact, if we keep one edge at the origin and send the other edge to infinity, that is, we consider a semi-infinite sample, we will observe just one edge band crossing the insulating bulk gap. 

If $(t,\eta)$ is in the shaded region of Fig.~\ref{ChernPhaseBoundary}, the ribbon is in a metallic state, while if in the non-shaded region the ribbon remains in an insulating state. The edge bands seen in Fig.~\ref{ChernSpectrum}(b) are called chiral because they connect the valance and the conduction states. Due to this feature and provided the bulk insulating gap remains open, those bands will not disappear when the Hamiltonian is deformed by either changing the existing coupling constants or by turning on additional interaction terms. For this reason, we can say that the metallic state of the ribbon is topologically protected. In the trivial case, the bands can totally disappear when additional terms are turned on, and what typically happens is that the bands sink into the bulk spectrum. When that happens, there will be little or no trace of edge spectrum in the LDOS.

\begin{figure}
 \center
 \includegraphics[width=10cm]{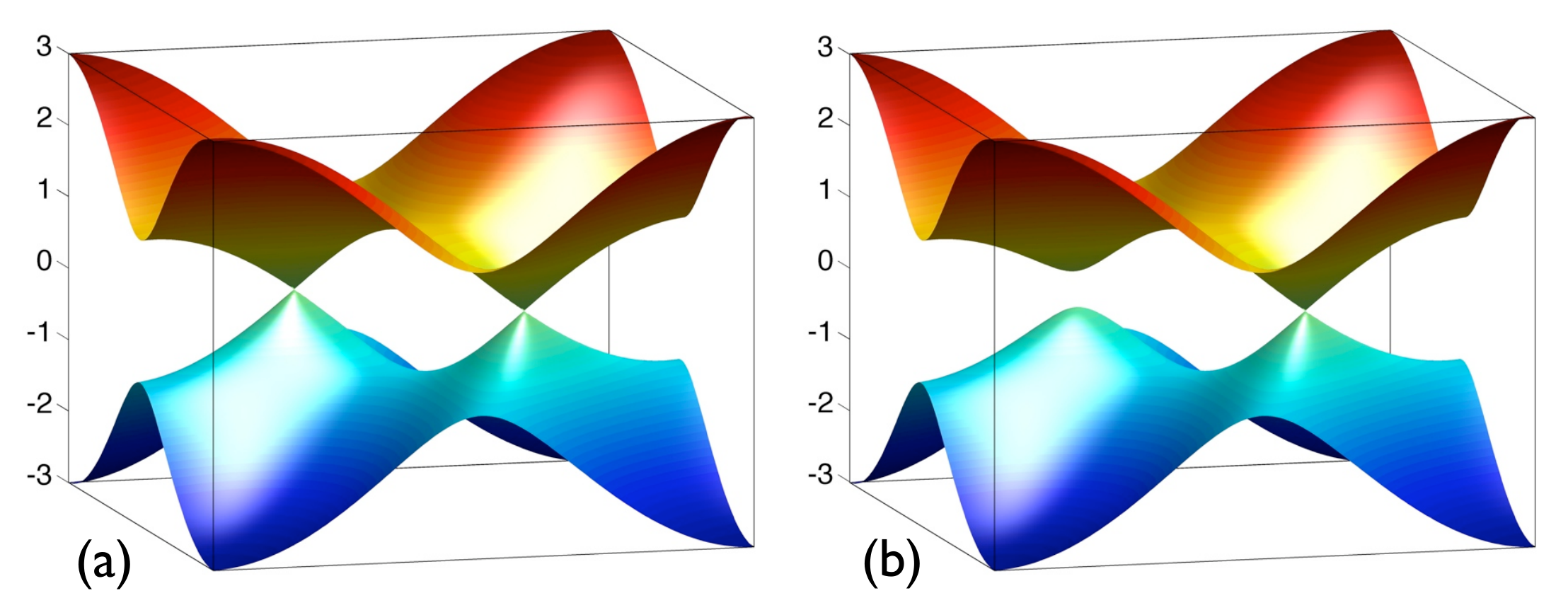}
 \caption{The bulk spectrum when $(t,\eta)$ is located at the boundary between the topological and normal insulating phases. (a) corresponds to $(t=0,\eta=0)$; (b) corresponds to $(t=\cos \frac{\pi}{6},\eta=\sin \frac{\pi}{6})$.}
 \label{ChernDiracPoint}
\end{figure}

As shown in Fig.~\ref{ChernDiracPoint}, if we pick any point $(t,\eta)$ at the boundary of the shaded region of  Fig.~\ref{ChernPhaseBoundary}, we will find that the bulk insulating gap is reduced to zero. We can also see some very distinct features emerging, namely, conic points where the bulk bands touch. These singular points are called Dirac points and they are actually at the origin of the topological properties of the model. When ($t$=0,$\eta$=0) there are two Dirac points, while for any other point of the phase boundary there is just one Dirac point. 

\subsection{The Chern number}

\begin{figure}
 \center
 \includegraphics[width=10cm]{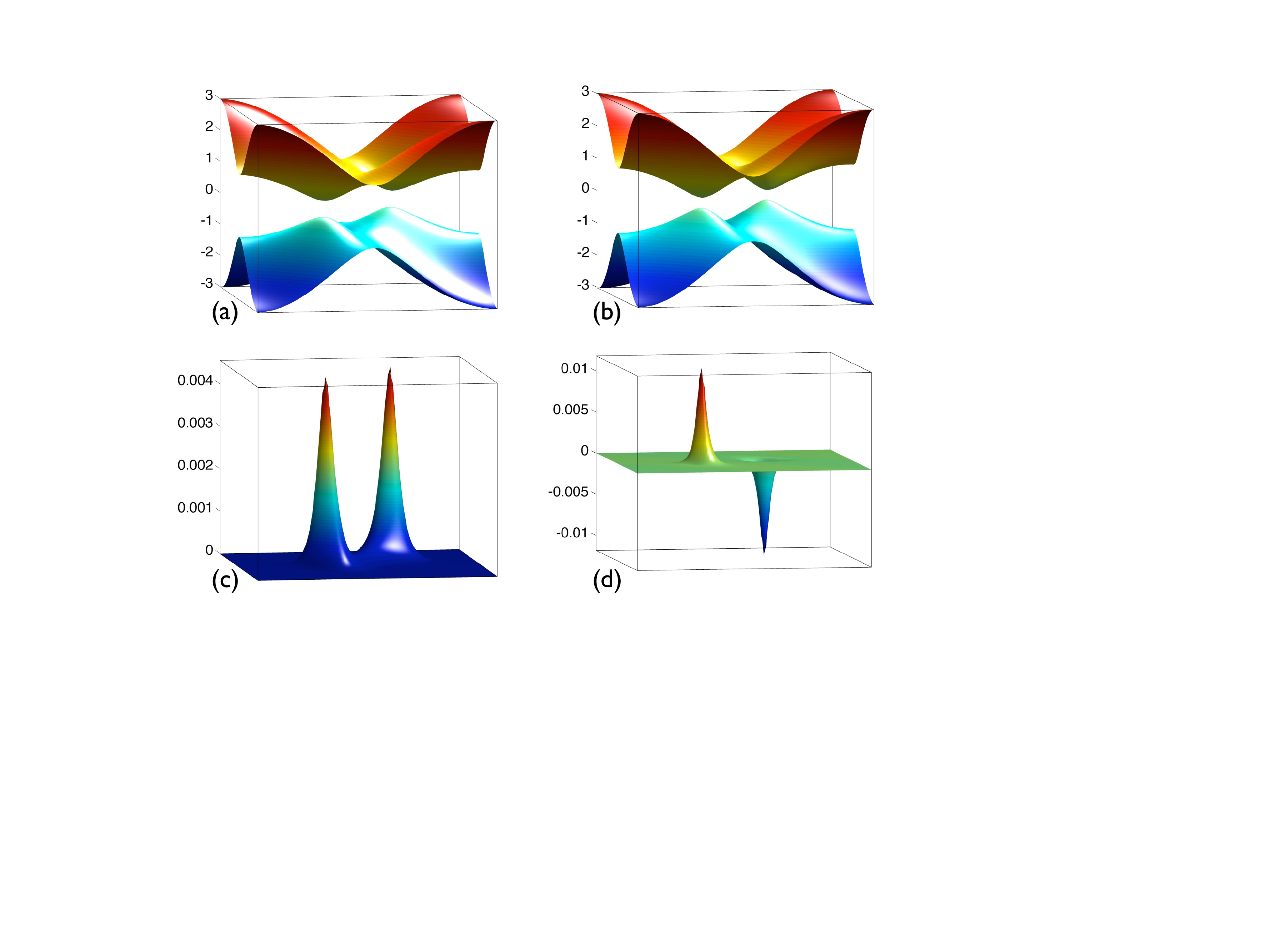}
 \caption{Plots of the adiabatic curvature as function of ${\bm k}$, when (a) $(t,\eta)$ is in the topological insulator region and (b) normal insulator region. The top plots show the bulk spectrum and they have been aligned with the plots of the curvature in order to show that the peaks seen in the curvature plots occur near those ${\bm k}$ points where the bulk bands tend to touch each other. The spikes become less prominent if the insulating gap is increased.}
 \label{CurvatureChernIns}
\end{figure}

We give here a brief and formal introduction of the Chern number. Let $P$ denote the projector onto the occupied spectrum:
\begin{equation}
\begin{array}{l}
P=\frac{1}{2\pi i}\oint_{\cal C} (z-H_0)^{-1}dz,
\end{array}
\end{equation}
where ${\cal C}$ is a contour in the complex plane surrounding the occupied energy spectrum. Under the Bloch transformation, $P$ decomposes in a direct sum of projectors: $UPU^{-1} = \oplus_{{\bm k}\in {\cal T}} \hat{P}_{\bm k}$, where $\hat{P}_{\bm k}$ is a finite matrix acting on ${\bm C}^2$.  It is analytic of ${\bm k}$, except when $(t,\eta)$ is on the phase boundary. The Chern number is given by the following formula:
\begin{equation}\label{Chern1}
\begin{array}{l}
C=\frac{1}{2\pi i}\int_{\cal T} \mbox{tr} \{\hat{P}_{\bm k}[\partial_{k_1}\hat{P}_{\bm k},\partial_{k_2}\hat{P}_{\bm k}]\}d^2{\bm k},
\end{array}
\end{equation}
where ``tr" means trace over the two dimensional ${\bm C}^2$ space. The integrand in Eq.~\ref{Chern1} is called the adiabatic curvature and the integral of Eq.~\ref{Chern1} can be shown to take only integer values, provided the family of projectors $P_{\bm k}$ are smooth of ${\bm k}$ over the entire Brillouin torus. A plot of the adiabatic curvature is shown in Fig.~\ref{CurvatureChernIns} for the topological phase $t$=0, $\eta$=0.1 (panel a), and for the trivial phase $t$=0.1, $\eta$=0 (panel b). The plot was generated by computing $\hat{P}_{\bm k}$ on a mesh-grid of 150$\times$150 points and by approximating the $k$-derivatives by the second-lowest order finite difference. Fig.~\ref{CurvatureChernIns} shows a distinct behavior of the curvature when the topological and trivial phases of the Haldane model are compared. In both cases, the curvature peaks near the split Dirac points, but in the topological phase the peaks have same signature, while in the normal phase the peaks have opposite signatures. Consequently, the curvature integrates to a non-zero value for the topological case, which is precisely 1, and to 0 for the normal case (plus/minus a small numerical error for both cases). A direct calculation will reveal that $C$ takes the value 0 inside the trivial phase and the $\pm 1$ values inside the topological phases as shown in Fig.~\ref{ChernPhaseBoundary}.

\subsection{Chern insulators with disorder}

In this section, we present several numerical experiments on the bulk of Chern insulators. They will reveal one of the flagship properties of these materials, manifested in the existence of spectral energy regions that contain delocalized states, even in the presence of strong disorder. Recall that we are dealing with a 2-dimensional model where, in general, the quantum states are localized in the presence of disorder \cite{Abrahams:1979et}. We will work with the following random Hamiltonian:
\begin{equation}\label{HaldaneModelDisorder}
\begin{array}{c}
H_\omega=H_0+W\sum_{{\bm n},\alpha} \omega_{\bm n} |{\bm n},\alpha\rangle \langle {\bm n},\alpha |,
\end{array}
\end{equation}
where $\omega_{\bm n}$ are randomly distributed amplitudes taking values in the interval $[-\frac{1}{2},\frac{1}{2}]$. We can think of the index $\omega$ in $H_\omega$ as the collection of all $\omega_{\bm n}$'s, which in turn can be regarded as a point in an infinite dimensional configuration space $\Omega$.

\begin{figure*}
 \center
 \includegraphics[width=15cm,angle=270]{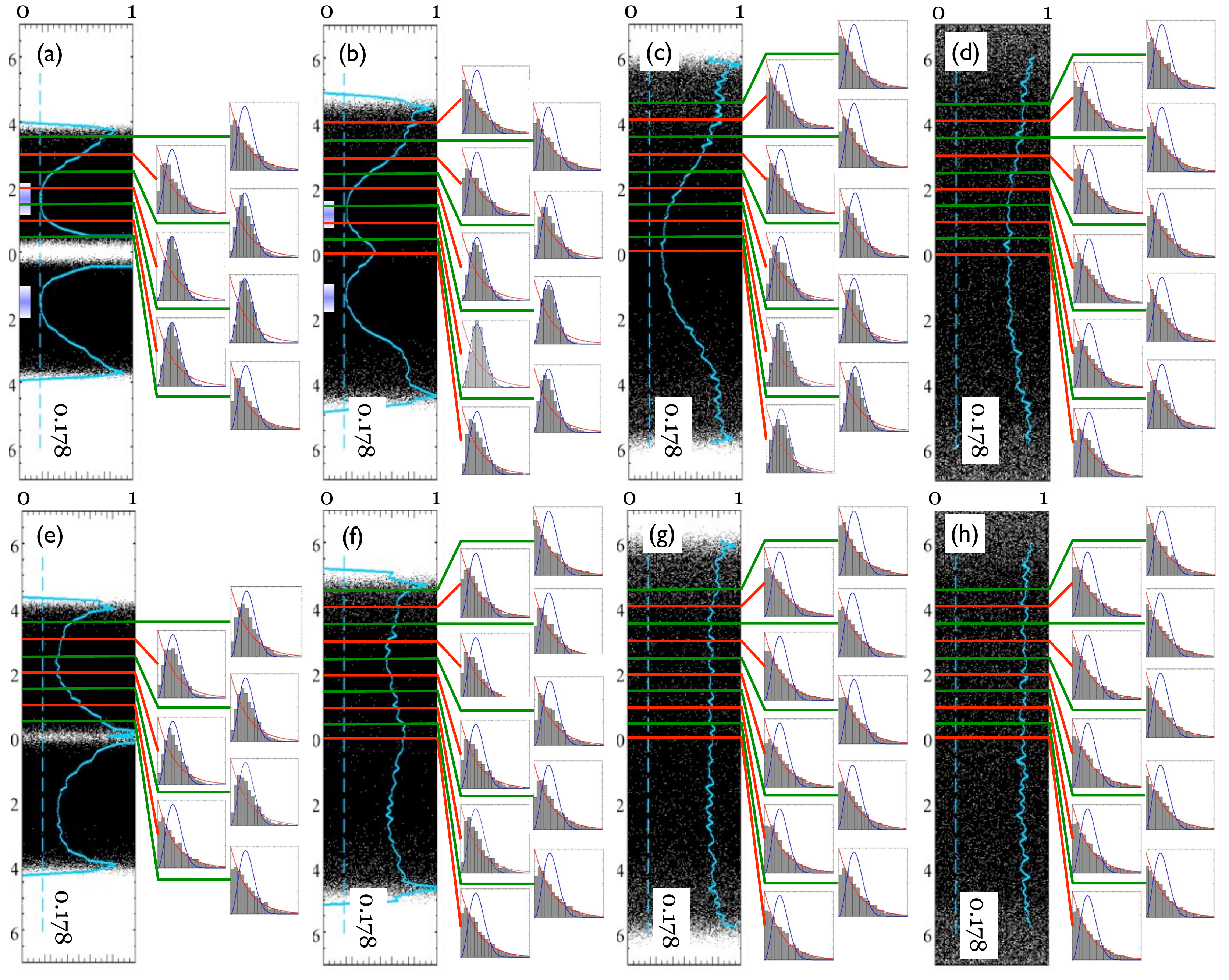}
 \caption{(Please rotate at 90 degrees.) Level statistics for the Chern insulator $\zeta$=$0.6i$ (upper panels) and for the normal insulator $\zeta$=$0.6$ (lower panels) at disorder strengths $W$=3, 5, 8 and 11. The main panels show the spectrum of $H_\omega$ for many disorder configurations. Level spacings were recorded from a small window around an energy $\epsilon$. Shown in light blue is the variance of these level spacings ensembles, when $\epsilon$ was brushed over the spectrum of $H_\omega$. The dotted line marks the value of 0.178 corresponding to the variance of $P_{\mbox{\tiny{GUE}}}$ ensemble. The small panels show histograms of the level spacings recorded at the marked energies. The histograms are compared with the $P_{\mbox{\tiny{GUE}}}$ (blue curve) and $P_{\mbox{\tiny{Poisson}}}$ (red curve) distributions.}
 \label{ChernLevelStatistics}
\end{figure*}

In the following experiments, we used a random number generator to build the $H_\omega$ of Eq.~\ref{HaldaneModelDisorder} on a lattice containing 30$\times$30 unit cells.  Periodic conditions were imposed at the boundaries of the lattice. We diagonalized $H_\omega$ and placed its eigenvalues $\{\epsilon_\omega^i\}_{i=1,2,\ldots}$ on a vertical axis, repeating the calculation $10^3$ times, every time updating the random potential. The result is a sequence of $10^3$ vertical sets containing the eigenvalues $\{\epsilon_\omega^i\}$ for each run, as illustrated in Fig.~\ref{ChernLevelStatistics} for different disorder amplitudes $W$. The level statistics was performed in the following way. We picked an arbitrary energy $\epsilon$ and, for each disorder configuration, we identified the unique $\epsilon_\omega^i$ and $\epsilon_\omega^{i+1}$ that satisfy: $\epsilon_\omega^i$$<$$\epsilon$$<$$\epsilon_\omega^{i+1}$. Then we computed the level spacings: $\Delta \epsilon$=$\epsilon_\omega^{i+j+1}$$-$$ \epsilon_\omega^{i+j}$, letting $j$ take consecutive values between $-k$ and $k$. We have experimented with $k$=1$\div$5 and the results were virtually the same. Fig.~\ref{ChernLevelStatistics} was generated with $k$=2, in which case, after repeating the procedure for all $10^3$ disorder configurations, we generated an ensemble of $5\times 10^3$ level spacings for each $\epsilon$, level spacings that were subsequently normalized by their average. Each diagram to the right of the energy spectrum in Fig.~\ref{ChernLevelStatistics} shows the distribution (histogram) of these $5\times 10^3$ level spacings. We picked several values for $\epsilon$ and we computed a histogram for each value. Imposed over the histograms are two continuous lines, one representing the Poisson distribution $P(s)=e^{-s}$ and another one representing the Wigner surmise for Gaussian Unitary Ensemble (GUE), $P_{\mbox{\tiny{GUE}}}=\frac{32}{3\pi}s^2 e^{-\frac{5}{3\pi}s^2}$. Imposed over the energy spectrum is the variance $\langle s^2\rangle$ -$\langle s \rangle^2$ of the $5\times 10^3$ level spacings recorded at a large (continuous) number of energies. We marked the theoretical value of 0.178 for the GUE variance by a dashed line in Fig.~\ref{ChernLevelStatistics}. The level statistics was performed for several disorder amplitudes: $W$=3, 5, 8, 11, for both the topological phase, $t$=0, $\eta$=0.6, and for the normal phase $t$=0.6, $\eta$=0.

\begin{figure}
 \center
 \includegraphics[width=5cm]{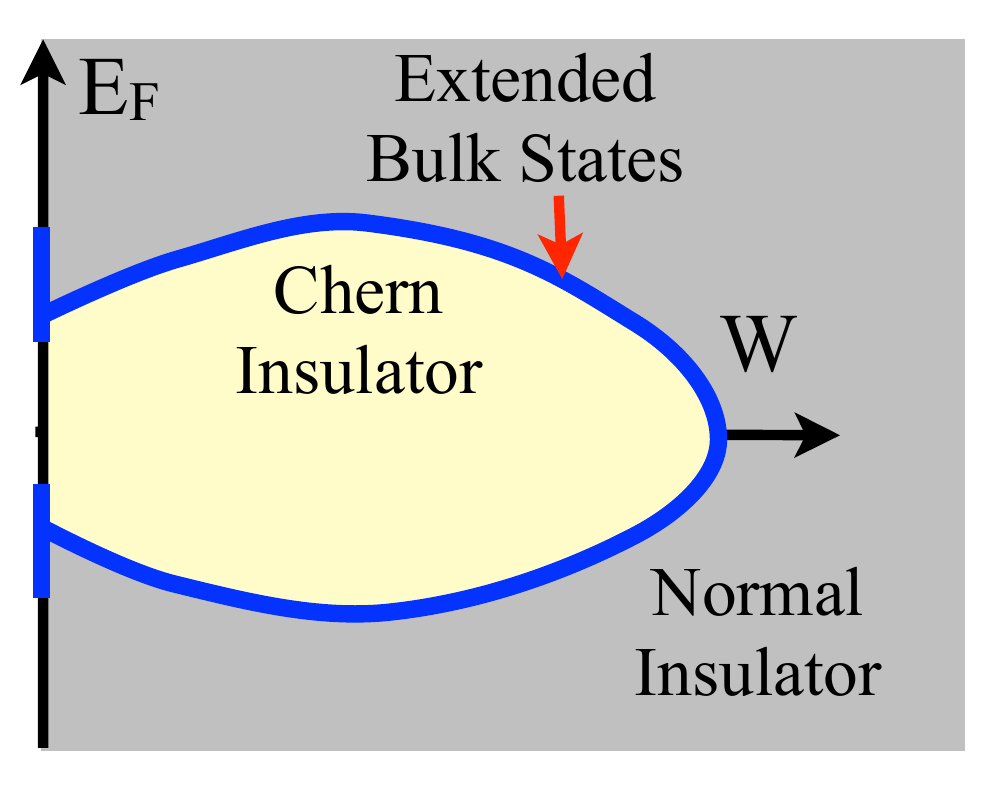}
 \caption{The phase diagram of a Chern insulator as inferred from the numerical calculations of Fig.~\ref{ChernLevelStatistics}.}
 \label{ChernPhaseDiagram}
\end{figure}

Let us focus on the topological case first, shown in panels (a)-(d). Inspecting the histograms and the variance in Fig.~\ref{ChernLevelStatistics}, one can see energy regions where the level distribution is Poisson, thus the states in these regions are very likely to be localized. But one can also see sharp energy regions where the histograms overlap quite well with the $P_{\mbox{\tiny{GUE}}}$ distribution and where the variance converges to the 0.178 value. These regions are very likely to contain delocalized states \cite{Evangelou1996yc,Cuevas1998xc}, which is quite remarkable since the disorder amplitude in all these panels is larger than the bandwidth of the clean energy bands seen in Fig.~\ref{ChernSpectrum}. One can also observe in Fig.~\ref{ChernLevelStatistics} that, as the disorder amplitude is increased, the spectral regions supporting the extended states do not abruptly vanish and instead they drift towards each other until they meet and only then they disappear. This is the so called levitation and pair annihilation phenomenon, which is a general characteristic feature of the extended states carrying a non-zero topological number. This will be discussed in depth later. Based on our current observations, the phase diagram in the $(E_{\mbox{\tiny{F}}},W)$ plane of a Chern insulator with $(t,\eta)$ fixed in the topological region should look like in Fig.~\ref{ChernPhaseDiagram}. If we examine the normal insulator in Fig.\ref{ChernLevelStatistics}, panels (e)-(h), we see that the spectral regions containing delocalized states are completely absent. There is no levitation and annihilation in this case, and instead the extended states become localized the moment we turn the disorder on. 

While all that has been said so far about the Chern insulators is just an introduction, we already reached the core of our investigation: To establish that the topological property of the Chern insulators (to carry a non-zero Chern number) has highly unusual  physical consequences, manifested in the existence of extended bulk states that resist localization even in the presence of strong disorder. One of our goals will be to demonstrate that the phase diagram in the $(E_{\mbox{\tiny{F}}},W)$ plane can be derived analytically, using the methods of NCG.

\section{Introduction to Quantum spin-Hall (QSH) insulators}

\subsection{QSH insulators in the clean limit} 

\begin{figure}
 \center
 \includegraphics[width=13cm]{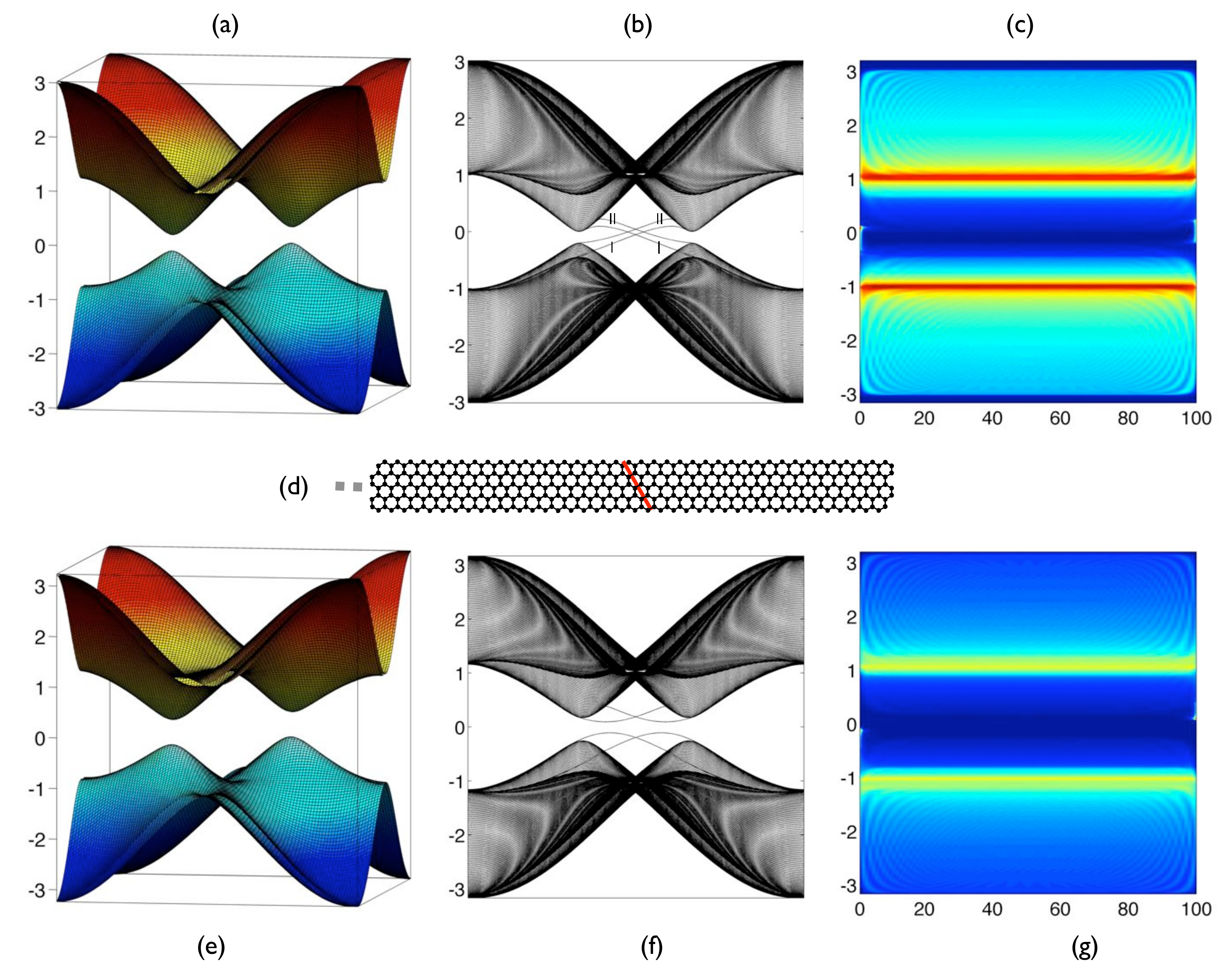}
 \caption{(a) The bulk spectrum of the Kane-Mele model of Eq.~\ref{KaneMele} ($t=0.1$ and $\eta=0.1$ and $\lambda=0.15$) as function of $(k_1,k_2)$. (b) The energy spectrum of the same Hamiltonian when restricted to an infinitely long ribbon with open boundary conditions at the two edges. The spectrum is represented as function of $k$ parallel to the ribbon's edges. (c) The local density of states (see Eq.~\ref{ldos}) of the ribbon, plotted as an intensity map in the plane of energy (vertical axis) and number of unit cells (horizontal axis) along the red line shown in panel (d). Blue/red colors corresponds to low/high values. (d) Illustration of the ribbon used in the calculations shown in panels (b-c) and (f-g). The ribbon was 100 unit cells wide. (e-g) Same as (a-c) but for ($t=0.4$ and $\eta=0.1$ and $\lambda=0.15$).}
 \label{QSHSpectrum}
\end{figure}

The first model of a QSH insulator was introduced by Kane and Mele \cite{Kane:2005np,Kane:2005zw}, who worked on the same honeycomb lattice of Fig.~\ref{Lattice1}, but considered also the spin degree of freedom. The Kane-Mele Hamiltonian reads:
\begin{equation}\label{KaneMele}
\begin{array}{c}
H_0^{\mbox{\tiny{QSH}}}=\sum\limits_{\langle {\bm n \bm m} \rangle,\sigma} |{\bm n},\sigma\rangle \langle {\bm m},\sigma| \medskip \\
+\sum\limits_{\langle \langle {\bm n \bm m} \rangle \rangle,\sigma}  \alpha_{\bm n}(t/2+i\eta [\hat{ {\bm S}} \cdot \underline{{\bf d}_{\bm k \bm m}\times{\bf d}_{\bm n \bm k} }]_{\sigma,\sigma} )|{\bm  n},\sigma\rangle \langle {\bm  m},\sigma| \medskip \\
+i\lambda\sum\limits_{\langle {\bm n \bm m} \rangle,\sigma \sigma'}  [ {\bf e}_z\cdot (\hat{{\bm S}}\times \underline{{\bf d}_{\bm n \bm m}})]_{\sigma,\sigma'} |{\bm n},\sigma\rangle \langle {\bm m},\sigma'|.
\end{array}
\end{equation}
Here, $\hat{{\bm S}}={\bm e}_x\hat{S}_x+{\bm e}_y\hat{S}_y+{\bm e}_z\hat{S}_z$ is the spin operator ($\hat{S}_x|{\bm n},\sigma\rangle = \frac{1}{2}|{\bm n},-\sigma \rangle$, $\hat{S}_y|{\bm n},\sigma\rangle = \frac{i}{2}\sigma|{\bm n},-\sigma \rangle$ and $\hat{S}_z|{\bm n},\sigma\rangle = \frac{1}{2}|{\bm n},\sigma \rangle$) and $\sigma$ and $\sigma'$ the electron spin degrees of freedom, taking the values $\pm 1$. The Hamiltonian acts on the Hilbert space ${\cal H}$ spanned by the orthonormal basis $|{\bm n},\sigma\rangle$, where ${\bm n}$ is a site of the honeycomb lattice. The simple and double angular brackets below the sums in Eq.~\ref{KaneMele} have same meaning as before. Inside the second sum, ${\bm k}$ represents the unique common nearest-neighbor of ${\bm m}$ and ${\bm n}$ and ${\bf d}_{\bm n \bm k}$, ${\bf d}_{\bm k \bm m}$ are the displacements shown in Fig.~\ref{Lattice1}. The underline on the vectors in Eq.~\ref{KaneMele} means normalization to unity. The model was built specifically for graphene and the three terms in Eq.~\ref{KaneMele} represent the nearest neighbor hopping, the second nearest neighbor hopping where the intrinsic spin-orbit coupling occurs, and the Rashba potential induced by the substrate supporting the graphene sheet or by an externally applied electric field. Half-filling will be assumed, that is, two electrons per unit cell, until it is specified otherwise. If we view the honeycomb lattice as a triangular lattice with two sites per unit cell, then the Hamiltonian takes the form: 
\begin{equation}\label{TriangQSHModel}
\begin{array}{c}
H_0^{\mbox{\tiny{QSH}}}=\sum\limits_{{\bm n},\alpha,\gamma,\sigma} |{\bm n},\alpha,\sigma \rangle \langle {\bm n}+\alpha \gamma{\bm e}_\gamma,-\alpha,\sigma|  \medskip \\
+\sum\limits_{{\bm n},\alpha,\gamma,\sigma} \{\xi_{\alpha,\sigma} |{\bm n},\alpha,\sigma\rangle \langle {\bm n}+{\bm e}_\gamma,\alpha,\sigma| + \xi_{\alpha,\sigma}^* |{\bm n}+{\bm e}_\gamma,\alpha,\sigma \rangle \langle {\bm n},\alpha,\sigma| \} \medskip \\
+ \frac{1}{2}\lambda \sum\limits_{{\bm n},\alpha,\gamma,\sigma} \alpha \sigma e^{i\frac{2\pi}{3}\gamma\sigma} |{\bm n},\alpha,\sigma\rangle \langle {\bm n}+\alpha\gamma{\bm e}_\gamma,-\alpha,-\sigma|,
\end{array}
\end{equation} 
where $\xi_{\alpha,\sigma}=\frac{1}{2}(t+i\sigma \eta)\alpha$ and the rest of the notation was already explained. As before, we prefer to work with this form of the Hamiltonian, which we will actually do from now on.

In the absence of disorder, we can perform the Bloch decomposition, given by the isometry $U$ from the Hilbert space ${\cal H}$ into a continuum direct sum of ${\bm C}^4$ spaces:
\begin{equation}
\begin{array}{c}
U: {\cal H} \rightarrow \bigoplus_{{\bm k}\in {\cal T}} {\bm C}^4, \ \ U|{\bm n},\alpha,\sigma\rangle = \frac{1}{2\pi} \bigoplus_{{\bm k}\in {\cal T}}  e^{-i {\bm k}\cdot{\bm n}} \xi_{\alpha,\sigma}
\end{array}
\end{equation}
where 
\begin{equation}
\xi_{1,1}$=$\left (
\begin{array}{c}
1 \\ 0 \\0 \\ 0
\end{array} \right ), \ 
\xi_{-1,1}$=$\left (
\begin{array}{c}
0 \\ 1 \\0 \\ 0
\end{array} \right ), \
\xi_{1,-1}$=$\left (
\begin{array}{c}
0 \\ 0 \\1 \\ 0
\end{array} \right ), \ 
\xi_{-1,-1}$=$\left (
\begin{array}{c}
0 \\ 0 \\0 \\ 1
\end{array} \right ).
\end{equation}
We have:
\begin{equation}\label{QSHBloch}
UH_0 U^{-1}=\bigoplus_{{\bm k}\in {\cal T}} \hat{H}_{\bm k}',\ \mbox{with} \ \hat{H}_{\bm k}'=\sum\limits_{\gamma}\left(
\begin{array}{cc}
\hat{H}_{\bm k} &  \hat{M}_{\bm k} \medskip\\
\hat{M}^\dagger_{\bm k} & \hat{H}_{\bm k}|_{\eta\rightarrow -\eta} 
\end{array}
\right ),
\end{equation}
where $\hat{H}_{\bm k}$ was given in Eq.~\ref{KHaldane} and
\begin{equation}
\begin{array}{c}
\hat{M}_{\bm k}=\frac{\lambda}{2} \sum_{\gamma} \left ( 
\begin{array}{cc}
0 &  e^{i\frac{2\pi \gamma}{3}+i\gamma{\bm k} \cdot {\bm e}_\gamma} \\
- e^{i\frac{2\pi \gamma}{3}-i{\bm k} \cdot {\bm e}_\gamma} & 0 
\end{array} \right ).
\end{array}
\end{equation}
As before, $\gamma$ takes the values 0, and $\pm 1$. We denote by $\epsilon_{\bm k}^{\overline{1,4}}$ the four eigenvalues of the Hamiltonian $\hat{H}_{\bm k}'$. The bulk band spectrum contains 4 bands, out of which 2 are occupied and 2 un-occupied (assuming a half-filled system).

The parameter space of the model is 3 dimensional $(t,\eta,\lambda)$. We will let again the computer choose random points in this parameter space and then instruct it to repeat the numerical experiments already discussed for the Chern insulators. Such experiment will reveal that, with probability one, the bulk system is an insulator (see panels (a) and (e) of Fig.~\ref{QSHSpectrum}). Again, by looking only at the bulk band spectrum, we will not be able to distinguish any major qualitative difference between different parts of the parameter space $(t,\eta,\lambda)$, but the calculation for the ribbon reveals again major qualitative differences. For some values such as $(t=0.1,\eta=0.1,\lambda=0.15)$, the energy spectrum for the ribbon geometry displays 4 distinct energy bands that cross the bulk insulating gap (see Fig.~\ref{QSHSpectrum}(b)). For other values such as $(t=0.4,\eta=0.1,\lambda=0.15)$, the spectrum  still displays 4 distinct energy bands but they don't cross the bulk insulating gap (see Fig.~\ref{QSHSpectrum}(f)). If we let the computer for a while to pick random points in the $(t,\eta,\lambda)$ space, it will slowly reveal a distinct region were the model displays bands that cross the insulating gap like in Fig.~\ref{QSHSpectrum}(b) when restricted to the ribbon, and another region where the ribbon has an insulating gap like in Fig.~\ref{QSHSpectrum}(f).

The plot of the local density of states shown in Fig.~\ref{QSHSpectrum}(f) reveals that the two solitary bands marked by ($|$) are localized on the top edge and the bands marked by ($\|$) are localized on the bottom edge of the ribbon. Since the ribbon was 100 units wide, there is practically no hybridization between the bands localized at different edges. The fact that each edge supports two bands steams from the time reversal symmetry of the model and the half-integer value of the spin. The time reversal operation is implemented by the anti-unitary operator:
\begin{equation}
T=e^{-i\pi \hat{S}_y}K, \ \ ([H_0,T]=0),
\end{equation}
where $\hat{S}_y$ is the $y$ component of the spin and $K$ is the complex conjugation. The fact that $T^2=-1$ has a distinct consequence in that if $\psi$ is an eigenvector of a time reversal symmetric Hamiltonian, than $T\psi$ is also an eigenvector that is orthogonal to $\psi$ because:
\begin{equation}
\langle \psi, T\psi \rangle=\overline{\langle T\psi,T^2 \psi \rangle}=-\langle \psi, T\psi \rangle.
\end{equation}
The conclusion is that the spectrum of a time reversal symmetric spin $\frac{1}{2}$ Hamiltonian is always doubly degenerate, a phenomenon known as Krames' degeneracy. For this reason, even when considering a semi-infinite sample with one edge, one will necessarily observe pairs of right and left moving bands. The Bloch edge Hamiltonian inherits the time-reversal symmetry at $k$=0 and $k$=$\pi$ points. At this $k$ points, the spectrum of the Bloch edge Hamiltonian is necessarily doubly degenerate, which means the edge band crossings occurring at $k$=0 and $k$=$\pi$ cannot be split by any deformation that preserves the time-reversal symmetry. Edge band crossings occurring at any other $k$ points can and are in general split by such deformations. Now, if the number of pairs of chiral edge bands is odd, like in the Kane-Mele model, then a simple exercise will show that one cannot open a gap in the edge band spectrum by performing all the allowed splittings of the edge band crossings. The situation is different if the number of pairs of chiral edge bands is even, in which case a gap can be opened, and generically will open under deformations that preserve the time-reversal symmetry. This leads to the celebrated ${\bm Z}_2$ topological classification of the time-reversal invariant insulators introduced by Kane and Mele \cite{Kane:2005zw}. For our simple model, the conclusion is that the ribbon is in a protected metallic state.

Returning to our specific model, if $\lambda$ is set to zero in Eq.~\ref{KaneMele}, the spin up and spin down sectors are left invariant by the Hamiltonian, which is reduced to a direct sum of two copies of Haldane Hamiltonian with $\zeta=\frac{1}{2}(t+i\eta)\alpha$ for $\sigma$=1 and $\zeta=\frac{1}{2}(t-i\eta)\alpha$ for $\sigma$=$-$1. Concentrating for a moment on the bottom edge of the ribbon in Fig.~\ref{QSHSpectrum} and recalling our discussion from the previous section, one can see that the right/left moving edge bands belong to the $\sigma$=$\pm$1 sectors. Therefore, the edge bands generate a spin flow, because one band carries a $\sigma$=1 spin in one direction and the other carries a $\sigma$=-1 spin but in the opposite direction. When $\lambda$ is turned on, the spin sectors are no longer invariant under the action of the Hamiltonian and, as a consequence, the edge bands will acquire a finite opposite spin component, but still the picture remains practically the same.

\subsection{The spin-Chern number for $S_z$ non-conserving models}

\begin{figure*}
 \center
 \includegraphics[width=8cm,angle=0]{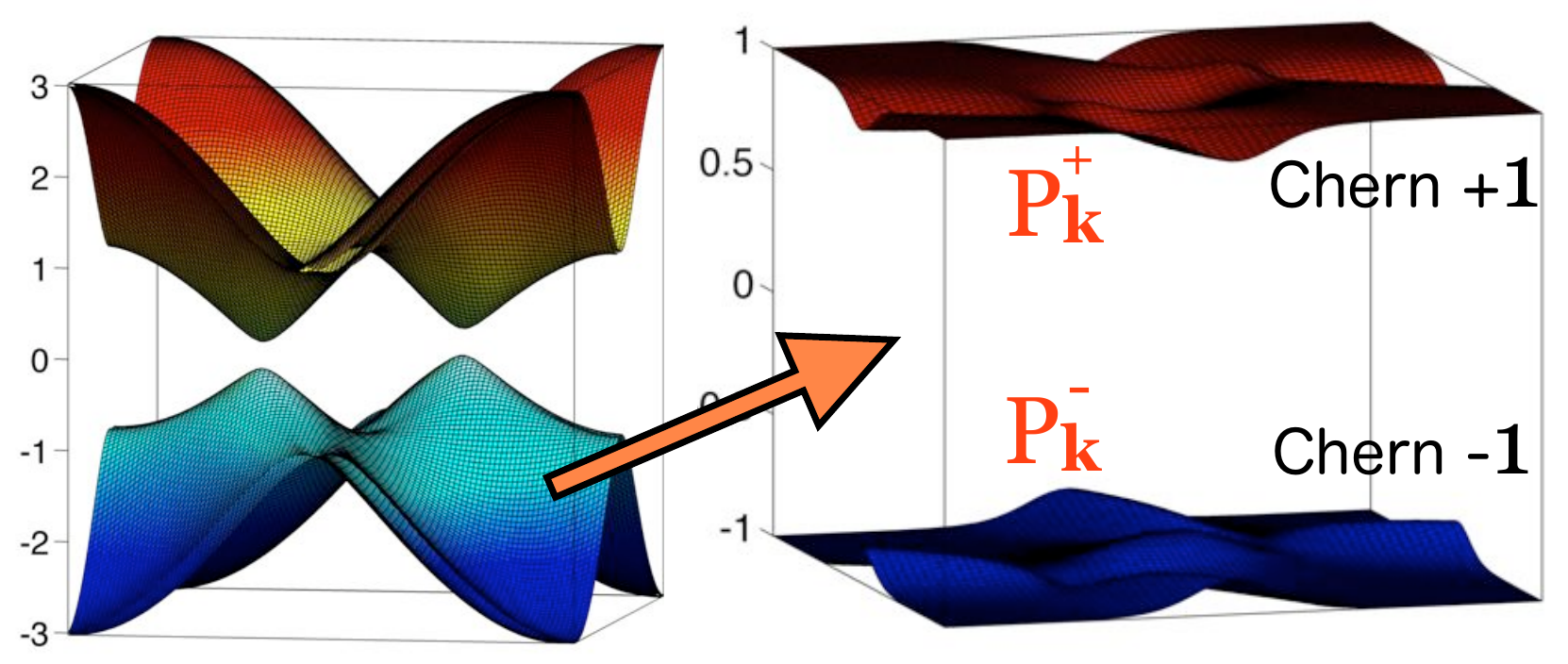}
 \caption{The left panel shows the spectrum of $H_0^{\mbox{\tiny{QSH}}}$. Its bands below the gap are highly entangled. The right panel shows the spectrum of $P\hat{\sigma}_z P$. One can see that the occupied states have been disentangled into two widely separated bands with non-zero Chern numbers.}
 \label{DisEntangle}
\end{figure*}

Time reversal invariant insulators have trivial Chern number. According to Refs.~\cite{Panati:2007sy,Brouder:2007p233}, these systems are trivial from the general homotopy point of view. However, if one insists on preserving the time reversal symmetry, these insulators still display topological properties, as we've already seen. There is quite a variety of approaches when it comes to the classification of time reversal symmetric insulators \cite{Volovik:1989eb,Moore:2007ew,Kane:2005zw,Prodan:2009oh,Fu:2006ka,Murakami:2006bw,Sugimoto:2006zr,Fu:2007ti,Fu:2007vs,Teo:2008zm,Sheng:2006nq,Sheng:2006na,Murakami:2007zd,Essin:2007ij,Fukui:2007dz,Fukui:2008mi,Qi:2008cg,Moore:2008ok,Subramaniam:2008tt,Obuse2008,Roy:2009cc,Roy:2009am,Gurarie2010by,RachelPRB2010vy,Ryu2010cu}. Here we discuss the topological properties of the QSH insulators using the spin-Chern number \cite{Sheng:2006na}. While many of the proposed invariants can be, formally, extended to the strong disordered case, and this is done usually by employing finite samples and twisted boundary conditions, it is not clear at all that the quantization and invariance of these extensions survive in the thermodynamic limit when the system becomes gapless due to the strong disorder, like in Fig.~\ref{ChernLevelStatistics}(b) for example. Note that most of these extensions tacitly assume that a small spectral gap remain open. In reality, or better said, in the numerical simulations with strong disorder, the typical cases are those in which a multitude of eigenvalues cross the Fermi level, from below and above when the boundary conditions are twisted (hence one cannot avoid Fermi level crossings by moving the Fermi level). In this case, the projector onto the occupied states is no longer smooth when twisting the boundary conditions, and ad-hoc numerical solutions must be implemented to deal with this fact. This is a serious problem, because, even when the states are localized near the Fermi level, the number of states crossing the Fermi level can be very large and one needs a criterion that specifies when this can create a problem (in some sense this is what NCG gives us). Another weak point of these extensions is that their quantization can be probed only numerically and the algorithms are not efficient because one needs to repeat the diagonalization of the Hamiltonian of a whole sample for a large number of boundary conditions. The spin-Chern number is the only invariant to date that can be defined in the presence of strong disorder and directly in the thermodynamic limit and this is why we focus here exclusively on it. 

Originally \cite{Sheng:2006na}, the spin-Chern invariant was defined for a large but finite, squarely shaped sample. Twisted boundary conditions were imposed at the boundaries:
\begin{equation}
\psi(L,y)=e^{ik_xL}\psi(0,y), \ \ \psi(x,L)=e^{2iS_z k_y L}\psi(x,0),
\end{equation}
 and the projector $\tilde{P}_{k_x,k_y}$ onto the occupied states was calculated for each $(k_x,k_y)$ on the Brillouin torus. The spin-Chern number was then computed via Eq.~\ref{Chern1}. The same original work has shown, through impressive numerical calculations, that the spin-Chern number remains quantized and invariant when disorder is added, even after the insulating gap was completely filled with localized spectrum.
 
 \begin{figure*}
 \center
 \includegraphics[width=9cm,angle=0]{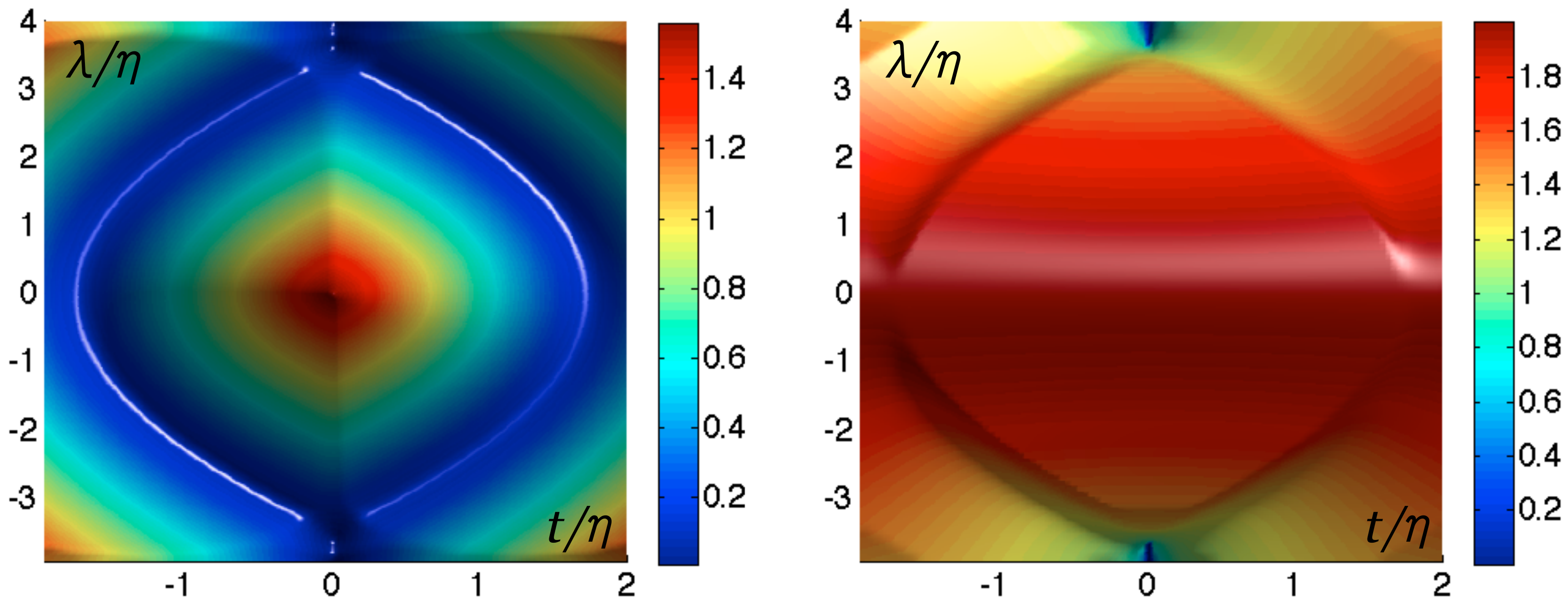}
 \caption{The left panel shows the energy gap of $H_{\mbox{\tiny{QSH}}}$ for various values of $t$ and $\lambda$ and $\eta$ fixed at $0.3$. The right panel shows the gap of $P\hat{\sigma}_z P$. In the left panel, one can see a fine white line where the gap of $H_{\mbox{\tiny{QSH}}}$ is closed. This line delimitates the QSH phase. The gap of $P\hat{\sigma}_z P$ is positive inside the QSH phase.}
 \label{Gaps}
\end{figure*}
 
The present discussion follows an alternative definition of the spin-Chern number \cite{Prodan:2009oh}, which is more convenient for analytic calculations. The idea is to split the occupied space into two or more sectors with non-trivial Chern numbers. For the Kane-Mele model, one can use the spectral properties of $P\hat{\sigma}_zP$ to achieve just that, where $\hat{\sigma}_z|{\bm n},\sigma\rangle = \sigma |{\bm n},\sigma\rangle$. Indeed, given the Bloch decomposition Eq.~\ref{QSHBloch}, one can easily compute the Bloch representation $\hat{P}_{\bm k}$ of the projector onto the occupied states and then form the matrix $\hat{P}_{\bm k}\hat{\sigma}_z\hat{P}_{\bm k}$. In Fig.~\ref{DisEntangle} we chose $(t=0,\eta=0.6,\lambda=0.3)$ and plotted the energy spectrum of $\hat{H}_{\bm k}'$ and the spectrum of $\hat{P}_{\bm k}\hat{\sigma}_z\hat{P}_{\bm k}$. Looking at the energy spectrum, one can see the two energy bands below the insulating gap being highly entangled. For this reason, no topological invariant can be associated with the individual bands. In contradistinction, the bands in the spectrum of $\hat{P}_{\bm k}\hat{\sigma}_z\hat{P}_{\bm k}$ are separated by a seizable gap.  As long as this gap and the gap in the energy spectrum remain open, a Chern number can be associated to each individual bands of $\hat{P}_{\bm k}\hat{\sigma}_z\hat{P}_{\bm k}$. If we denote by $\hat{P}_{\bm k}^{\pm}$ the projector onto the upper/lower eigenvalue of $P_{\bm k}\hat{\sigma}_zP_{\bm k}$, we can compute the corresponding Chern numbers via 
\begin{equation}\label{SChern}
C_{\pm}=\frac{1}{2\pi i}\int_{\cal T} \mbox{tr} \{\hat{P}_{\bm k}^\pm[\partial_{k_1}\hat{P}_{\bm k}^\pm,\partial_{k_2}\hat{P}_{\bm k}^\pm]\}d^2{\bm k}.
\end{equation}
Since the total Chern number $C_-+C_+$ is zero, $C_+-C_-$ is an even number and we can define the spin-Chern number as the integer:
\begin{equation}\label{SpinChern}
\begin{array}{c}
C_s=\frac{1}{2}(C_+-C_-).
\end{array}
\end{equation}

In Fig.~\ref{Gaps} we present the insulating energy gap of $H_0^{\mbox{\tiny{QSH}}}$ and the spectral gap of $P\hat{\sigma}_zP$ for $\eta$ fixed at 0.3 and $t$ and $\lambda$ varied over a wide range. One can see in panel (a) the insulating energy gap closing along a certain line in the $(t,\lambda)$ plane, line that delimitates the QSH phase. As one can see, the spectral gap of $P\hat{\sigma}_zP$ remains open for all $\lambda$ and $t$ values inside the QSH phase. The picture remains for any other value of $\eta$, showing that bands of $P\hat{\sigma}_zP$ are always separated by a finite gap and consequently the spin-Chern number is well defined. The spin-Chern number takes the values $C_s=\pm 1$ (depending on the sign of $\eta$) in the QSH region of the phase diagram, and $C_s=0$ in the trivial region of the phase diagram.

As a concluding remark, we mention Ref.~\cite{LiPRB2010uv} where analytic calculations of the spin-Chern number were carried out for a $S_z$ non-conserving model of a QSH insulator. These analytic calculations are interesting because they show that the Pfaffian function needed in the computation of the ${\bm Z}_2$ invariant \cite{Kane:2005zw} and the integrand in Eq.~\ref{SChern} are closely related.

\begin{figure*}
 \center
 \includegraphics[width=15cm,angle=270]{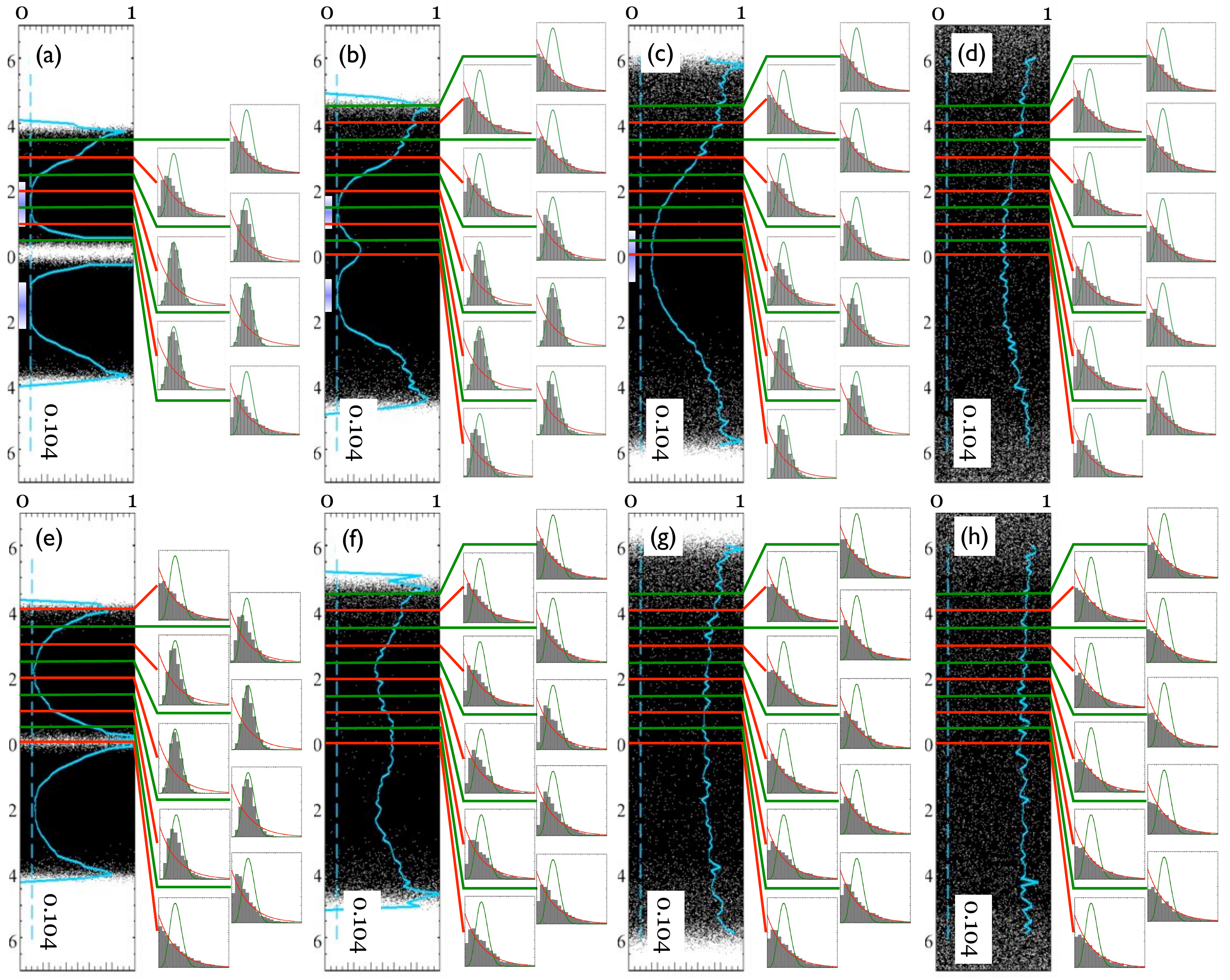}
 \caption{(Please rotate at 90 degrees.) Level statistics for the QSH insulator $t$=0, $\eta$=0.6 and $\lambda$=0.3 (upper panels) and for the normal insulator $t$=0.6, $\eta$=0 and $\lambda$=0.3 (lower panels) at disorder strengths $W$=3, 5, 8 and 11. The main panels show the spectrum of $H_\omega$ for many disorder configurations. Level spacings were recorded from a small window around an energy $\epsilon$. Shown in light blue is the variance of these level spacings ensembles, when $\epsilon$ was brushed over the spectrum of $H_\omega$. The dotted line marks the value of 0.104 corresponding to the variance of $P_{\mbox{\tiny{GSE}}}$ ensemble. The small panels show histograms of the level spacings recorded at the marked energies. The histograms are compared with the $P_{\mbox{\tiny{GSE}}}$ (green curve) and $P_{\mbox{\tiny{Poisson}}}$ (red curve) distributions.}
 \label{QSHLevelStatistics}
\end{figure*}

\subsection{QSH insulators with disorder}

We consider here the Kane-Mele model with diagonal disorder:
\begin{equation}
\begin{array}{c}
H^{\mbox{\tiny{QSH}}}_\omega=H_0+W\sum_{n,\alpha,\sigma} \omega_{\bm n} |{\bm n},\alpha,\sigma\rangle \langle {\bm n},\alpha,\sigma|.
\end{array}
\end{equation}
We have repeated the numerical experiments presented in the previous section and the results are show in Fig.~\ref{QSHLevelStatistics}. The computed histograms are now compared with Wigner surmise for the symplectic case: $P_{\mbox{\tiny{GSE}}}=\frac{2^{18}}{3^6 \pi^3}s^4e^{-\frac{64}{9\pi}s^2}$. The variance of this distribution is 0.104. For the topological case, shown in panels (a-d), the numerical experiment reveals again the existence of energy regions where the histograms of the level spacings overlap quite well with $P_{\mbox{\tiny{GSE}}}$ and the numerically computed variance takes values extremely close to 0.104. These indicate again the existence of delocalized states \cite{Evangelou1996yc}, which persist even for large disorder amplitudes. The levitation and pair annihilation  is still visible in the upper panels of Fig.~\ref{QSHLevelStatistics}. The regions of extended states are absent in the trivial case shown in panels (e-h). The levitation and annihilation is also absent in the trivial case.

\begin{figure}
 \center
 \includegraphics[width=5cm]{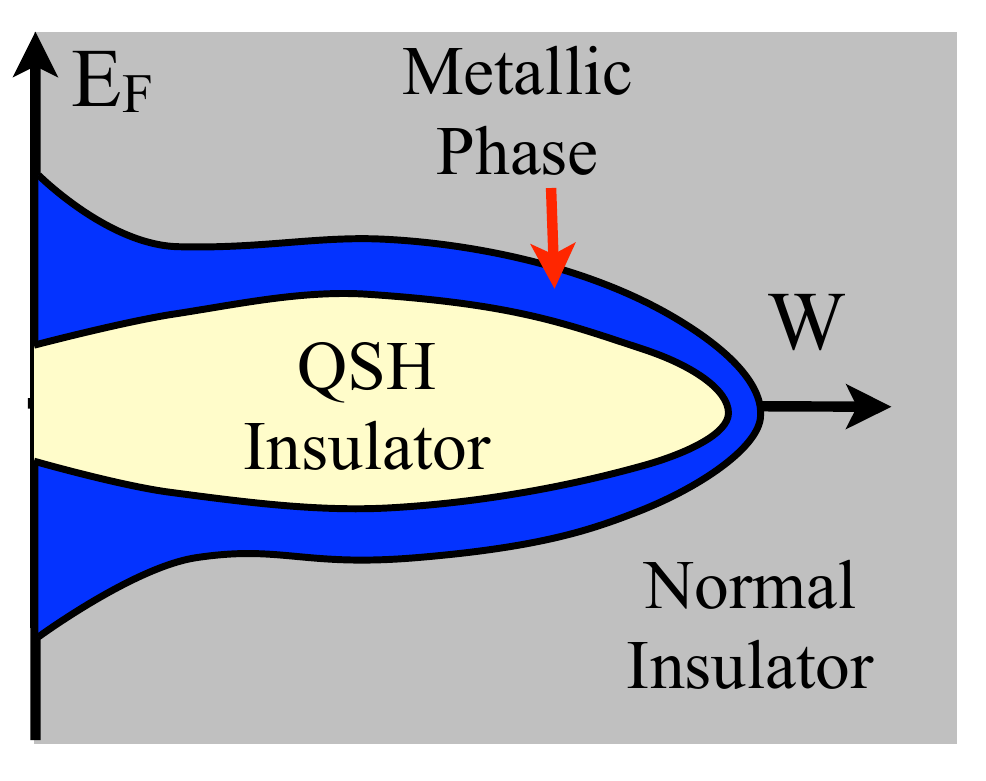}
 \caption{The phase diagram of a QSH insulator as inferred from the numerical calculations of Fig.~\ref{QSHLevelStatistics}.}
 \label{QSHPhaseDiagram}
\end{figure}

There is one distinct difference between the Chern and QSH insulators, regarding their bulk properties. The spectral regions containing the extended states are reduced to a single point for the Chern insulators, while they remain of finite width for QSH insulators. Hence, our numerical experiments imply the phase diagram shown in Fig.~\ref{QSHPhaseDiagram} for the Kane-Mele model when $E_F$ and $W$ are varied.

We would like to end this section with a discussion of the relation between the bulk and edge topological properties of QSH insulators. It is now well established that the QSH insulators with odd spin-Chern number do present robust edge modes, while those with even spin-Chern number do not. This gives the connection between the spin-Chern number and the ${\bm Z}_2$ classification of the QSH insulators, which is based on the edge physics. For example, numerical experiments indicate that the Kane-Mele model displays extended edge states over the entire QSH phase drawn in Fig.~\ref{QSHPhaseDiagram}. But the story does not end here. As we will argue in the following, the spin-Chern number protects a set of extended bulk states against disorder, like the ones revealed in Fig.~\ref{QSHLevelStatistics}, regardless of its parity. We have now gathered enough numerical evidence to announce here with confidence that explicit models with $C_s$=2 \cite{Shulman2010cy} or with $C_s$=1 but broken time-reversal symmetry do posses robust extended bulk states in the presence of strong disorder, even though the edge spectrum displays a mobility gap.

\section{The Chern invarint for disordered systems: The Non-Commutative Geometry approach of Bellissard, van Elst and Schulz-Baldes}

The definition of the spin-Chern number given in the previous section is based on the Chern number. Consequently, the Non-Commutative Theory of the Chern number is relevant for both the Chern and Quantum spin-Hall insulators and will be presented in depth in this section. 

We start by setting some basic notations. The symbol $\|A\|$ will denote the operator norm: $\|A\|:=\sup \sqrt{\langle A\Psi| A\Psi \rangle}$, where the supremum is taken over all vectors of norm one in the underlying Hilbert space. We will often make reference to the space of bounded operators, ${\cal B}({\cal H})$, which is the linear space of all $A$'s for which the operator norm is finite. When we use the wording ``continuous deformation" of an operator we mean variations of that operator that are continuous with respect to the operator norm. Several additional classes of operators and norms will be introduced later. 

We restrict the discussion to a Hilbert space ${\cal H}$ spanned by orthonormal vectors of the form: $|{\bm n},\alpha\rangle$, where ${\bm n}\in {\bm Z}^2$ is a site of a 2D lattice and $\alpha=1,\ldots,K$ labels the orbitals associated with a particular site. We denote the projector onto these elementary states by $\pi_{{\bm n},\alpha}$ and the projector onto the quantum states at a site ${\bm n}$ by $\pi_{\bm n}$:
\begin{equation}
\begin{array}{c}
\pi_{{\bm n},\alpha}=|{\bm n},\alpha \rangle \langle {\bm n},\alpha|, \ \ \pi_{\bm n}=\sum_{\alpha} |{\bm n},\alpha\rangle \langle {\bm n},\alpha|.
\end{array}
\end{equation}
The theory will be developed for generic orthogonal projectors $\Pi$ that act on this Hilbert space ($\Pi^*=\Pi$ and $\Pi^2=\Pi$).

\subsection{The Fredholm class and the Index of a Fredholm operator}

The Index is one of the main tools of modern topology and will be heavily invoked in the following, so we need to begin with the basics of the Index. Given a bounded linear operator $F$, we define its null space as the linear space of its zero modes:
\begin{equation}
\mbox{Null}(F)=\{\phi\in {\cal H}|F\phi = 0\}.
\end{equation}
The Index of the operator $F$ is defined as the difference between the number of its zero modes and the number of the zero modes of its conjugate:
\begin{equation}\label{Index}
\mbox{Ind}\{F\}=\mbox{dim} \ \mbox{Null}(F)-\mbox{dim} \ \mbox{Null}(F^*) .
\end{equation}
To have a meaning, at least one of the above null spaces must be of finite dimensionality. In fact, if we want the Index to be of any use, we must require that both null spaces have finite dimensions. We must also rule out the existence of the so called generalized wave-functions which obey $F\phi=0$, as it happens when 0 is in the continuum spectrum of $F$. This can be done by restricting ourselves to operators $F$ for which the range $F{\cal H}$ is a closed space. So we alreay singled out a very special class of operators, called the Fredholm class, defined by the following properites \cite{Gilkey1994tr}:

\begin{definition}
The Fredholm class contains all bounded operators $F$ with the property that $\mbox{dim} \ \mbox{Null}(F)<\infty$, $\mbox{dim} \ \mbox{Null}(F^*)<\infty$, and $F{\cal H}$ and $F^*{\cal H}$ are closed spaces.
\end{definition}

\noindent The Index, as defined by Eq.~\ref{Index}, takes finite integer values when evaluated on operators $F$ from the Fredholm class and it has no meaning when evaluated on operators outside of the Fredholm class. For this reason, we always need to make sure that the operators belong to the Fredholm class before we evaluate their Index. This is probably the appropriate moment to introduce another class of operators, which contains all operators for which the trace operation makes sense. This class is called the trace class and is defined as:

\begin{definition} The trace class ${\cal S}^1$ contains all compact operators $A$ for which $\sum_i \mu_i < \infty$, where $\{\mu_i\}$ are the singular values of $A$, that is, the eigenvalues of $\sqrt{AA^*}$. The functional:
\begin{equation}
\begin{array}{l}
\|A\|_{{\cal S}^1}= \sum_i \mu_i  
\end{array}
\end{equation}
defines a norm on ${\cal S}^1$, which becomes a Banach spaces. The trace class is a double ideal among the bounded operators, that is, $BA$ and $AB$ are both in the trace class if $A$ is in the trace class and $B$ is any bounded operator (not necessarily in the trace class).
\end{definition}

\noindent The trace operation is finite when evaluated on operators from the trace class, and it can be computed as $\mbox{Tr} A =\sum_i \langle \phi_i|A|\phi_i \rangle$, with $\{\phi_i\}$ being any orthonormal basis in ${\cal H}$. The sum is independent of the chosen basis if $A$ is in the trace class. For $A$ outside the trace class, the sum $\sum_i \langle \phi_i|A|\phi_i \rangle$ can diverge, be oscillatory or change with the change of basis. For this reason, whenever we plan to compute a trace of an operator, we will make sure first that the operator is in the trace class, even if this sometimes can be more difficult than computing the trace itself. 

We now turn to the question of how to evaluate the Index. If the action of $F$ is known explicitly and is simple enough, the Index can be evaluated by using its very definition given in Eq.~\ref{Index}. But this is not the case in a large number of situations, in which case we must rely on more computationally friendly  methods. One such method was derived by Fedosov and will be used here.

\begin{proposition} [Fedosov's formula] If, for some finite positive integer $n$, the operators $(I-FF^*)^n$ and $(I-F^*F)^n$ are in the trace class, then $F$ is in the Fredholm class and its Index can be computed as:
\begin{equation}\label{Fedosov}
Ind\{F\}=\mbox{Tr} \ (I-FF^*)^n - \mbox{Tr} \ (I-F^*F)^n.
\end{equation}
\end{proposition}

\noindent It is a fact that, if $(I-FF^*)^n$ and $(I-F^*F)^n$ are in trace class for some $n$, then they are in the trace class for any other integer larger than $n$. The computation will lead to the same Index, independent of which allowed $n$ we choose to work with. In practice, however, one tries to work with the smallest possible value for this $n$.

When the right part of Eq.~\ref{Fedosov} is evaluated, it often leads to explicit formulas that involve geometric data. In general, a successful and useful Index calculation, which is conditioned by the choice of the operator $F$ and the ability to compute the right hand side of Fedosov's formula Eq.~\ref{Fedosov} (or whatever formula we choose to work with), leads to statements of the form:
\begin{equation}\label{equality1}
\begin{array}{c}
\mbox{Ind}\{F\} = \int \{...\}
\end{array}
\end{equation}
where the integral involves objects with explicit geometric meaning, such as the curvature in the case of the Chern number. The equality written in Eq.~\ref{equality1} establishes that this integral is an integer, something that might be extremely difficult to see by just looking at the integral itself. In modern Topology, the left hand side of Eq.~\ref{equality1} is called the analytic Index, while the right hand side is called the geometric or topologic Index. The analytic Index provides the quantization (because from its very definition, the Index is an integer) and, as we shall see in a moment, also the topological invariance, while the geometric Index provides the geometric interpretation and an explicit way to compute the actual value of the Index. This is precisely the philosophy that will guide us throughout the present paper. The topological invariance follows from the following remarkable property of the Index \cite{Gilkey1994tr,Murphy1994hf}.

\begin{proposition}\label{IndProp} Suppose that the operator $F(\lambda)$ changes continuously with $\lambda$ and for all $\lambda$'s the operator $F(\lambda)$ stays in the Fredholm class. In this case, $\mbox{Ind}\{F(\lambda)\}$ is well defined and takes the same integer value for all $\lambda$'s. 
\end{proposition}

\noindent In other words, the Index of an operator remains unchanged under continuous deformations that keep the operator inside the Fredholm class. This principle not only gives a very general way to prove the topological invariance of the geometric Index, but also allows one to figure out the precise conditions that assures this invariance. For this one has to find out how far can $F$ be deformed and still remain inside the Fredholm class. This will be exemplified on explicit models, shortly. We end this section by listing some additional properties of the Index \cite{Murphy1994hf}.

\begin{proposition} Let $T$, $S$ be any Fredholm operators, $C$ any compact operator and $U$ any unitary operator. Then:\smallskip

i) $\mbox{Ind}\{TS\}=\mbox{Ind}\{T\}+\mbox{Ind}\{S\}$.\smallskip

ii) $\mbox{Ind}\{T+C\}=\mbox{Ind}\{T\}$.\smallskip

iii) $\mbox{Ind}\{UTU^{-1}\}=\mbox{Ind}\{T\}$
\end{proposition}

\subsection{The Chern number as an analytic Index: The translational invariant case.}

Assume the existence of a translational invariant projector $\Pi$ acting on the Hilbert space ${\cal H}$, with exponentially decaying matrix elements:
\begin{equation}\label{ExpProp}
|\langle {\bm n},\alpha|\Pi|{\bm m},\beta\rangle | < \mbox{ct}. e^{-\gamma |{\bm n}-{\bm m}|},
\end{equation}
with $\gamma>0$ fixed. Consider the Bloch transformation $U$ from ${\cal H}$ into a continuum direct sum of ${\bm C}^K$ complex spaces:
\begin{equation}
\begin{array}{c}
U: {\cal H} \rightarrow \bigoplus_{{\bm k}\in {\cal T}} {\bm C}^K, \ \ U|{\bm n},\alpha\rangle = \frac{1}{2\pi} \bigoplus_{{\bm k}\in{\cal T}} e^{-i {\bm k}\cdot{\bm n}} \xi_\alpha,
\end{array}
\end{equation}
where $\xi_\alpha$ are the single column matrices with entry 1 in row $\alpha$ and zero everywhere else. The transformation takes the projector $\Pi$ into a direct sum of projectors on ${\bm C}^K$: $U\Pi U^{-1} = \bigoplus_{{\bm k}\in {\cal T}} \hat{\Pi}_{\bm k}$. The projectors $\hat{\Pi}_{\bm k}$ are finite matrices acting on ${\bm C}^K$ space and they are analytic of ${\bm k}$, a property that follows from Eq.~\ref{ExpProp}. The following formula for the Chern number is familiar to most of the condensed matter theorists and we already mention it when we discussed the Chern insulators:
\begin{equation}\label{ChernX}
C=\frac{1}{2\pi i}\int_{\cal T} \mbox{tr} \{\hat{\Pi}_{\bm k}[\partial_{k_1}\hat{\Pi}_{\bm k},\partial_{k_2}\hat{\Pi}_{\bm k}]\}d^2{\bm k}.
\end{equation}
This formula packs geometric and physical content, since the integrand is precisely the adiabatic curvature, derived from the adiabatic connection, both having precise and deep physical meanings (see Avron in Physics Today 2003). The following result links $C$ with an analytic Index. 
\begin{theorem}
Let $u:{\cal H}\rightarrow {\cal H}$ be the following unitary transformation:
\begin{equation}
\begin{array}{l}
u|{\bm n},\alpha \rangle =\frac{n+n_0}{|n+n_0|}|{\bm n},\alpha \rangle \ \ ({\bm n}=(n_1,n_2), \ n\equiv n_1+in_2, \ n_0 \equiv \frac{1}{2}+\frac{i}{2}).
\end{array}
\end{equation}
Then, if Eq.~\ref{ExpProp} holds true, the operator $(\Pi u\Pi)\oplus \Pi^\bot$ is in the Fredholm class and:
\begin{equation}\label{Chern2}
Ind\{(\Pi u\Pi)\oplus \Pi^\bot\}=\frac{1}{2\pi i}\int_{\cal T} \mbox{tr} \{\hat{\Pi}_{\bm k}[\partial_{k_1}\hat{\Pi}_{\bm k},\partial_{k_2}\hat{\Pi}_{\bm k}]\}d^2{\bm k}.
\end{equation}
Therefore, the Chern number is quantized and invariant to continuous deformations of the projector $\Pi$, as long as Eq.~\ref{ExpProp} holds true for some finite $\gamma$.
\end{theorem}

\noindent The unitary transformation $u$ describes the effect of an infinitely thin quantum flux threaded through the lattice at the spatial point located at $n_0$. Recalling our previous discussion, it is quite clear that the Chern number appears in Eq.~\ref{Chern1} as the geometric Index corresponding to the analytic Index of $(\Pi u\Pi)\oplus \Pi^\bot$ ($\Pi^\bot=1-\Pi$ is the orthogonal complement of $\Pi$). With this relation established, the topological properties of the Chern number follow solely from the extremely general properties of the analytic Index. If $\Pi$ is the projector onto the occupied states of an insulator, then the existence of the energy gap assures the exponential localization of this projector and we can safely conclude that the Chern number of the occupied states is quantized and invariant under continuous deformations of the insulator that keep the energy gap open. If the energy gap closes and opens again, then $(\Pi u\Pi)\oplus \Pi^\bot$ exits and re-enters the Fredholm class, in which case the Chern number can assume a different value. This concludes our short exercise showcasing the strength of the general strategy that we next apply to the disordered case. The above statement will then appear as a particular case of the disordered case, which will be treated in full detail.

\subsection{Differential and integral non-commutative calculus for the disordered case}

A disordered configuration will be labeled by the random variable $\omega$, which takes values in a configuration space $\Omega$. We assume that the disordered configurations cannot be macroscopically discerned, therefore any laboratory measurement involves averaging over the random variable $\omega$. The averaging is done with respect to a model dependent probability measure $dP(\omega)$ on $\Omega$. For example, for the white noise considered in the previous sections, $dP(\omega)=\prod_{\bm n}d\omega_{\bm n}$ (this formal expression can be given a rigorous meaning). We also consider homogeneous disorder, that is, various translations of a sample generate allowed disordered configurations of the original sample. Mathematically, this means that each lattice translation by a vector ${\bm n}$ induces a map  $t_{\bm n}$ of $\Omega$ onto itself and, for consistency, these maps must satisfy:
\begin{equation}
t_{\bm n}\circ t_{\bm m}=t_{{\bm n}+{\bm m}}.
\end{equation}
Moreover, the probability measure must be invariant with respect to these maps, that is, if $Q$ is a subset of $\Omega$ and $Q'=t_{\bm n}Q$, then $\int_QdP(\omega)$ = $\int_{Q'} dP(\omega)$. Formally, we can write this as $dP(\omega)=dP(t_{\bm n}\omega)$. We will also require that any subset $Q$ of $\Omega$ left invariant by the maps $t_{\bm n}$ (${\bm n}\in {\bm Z}^2$) have zero measure: $\int_Q dP(\omega)=0$. In other words, we assume that the measure $dP(\omega)$ is ergodic with respect to the flow $t$.

We now introduce the notion of a covariant family of bounded operators. A collection of bounded operators $\{A_\omega\}_{\omega \in \Omega}$ is called a covariant family if:
\begin{equation}
T_{\bm n}A_{\omega}T_{\bm n}^{-1}=A_{t_{\bm n}\omega}, \ \mbox{for all} \ \omega \in \Omega \ \mbox{and} \ {\bm n}\in {\bm Z}^2.
\end{equation}
Here $T_{\bm n}$ represents the implementation in the Hilbert space ${\cal H}$ of the lattice translation by ${\bm n}$: $T_{\bm n}|{\bm m},\alpha\rangle = |{\bm m}+{\bm n},\alpha\rangle$. The covariant families form an algebra, since:
\begin{equation}
\{A_\omega+B_\omega\}_\omega \ \mbox{and} \ \{A_\omega B_\omega\}_\omega
\end{equation}
are also covariant families if $\{A_\omega\}_{\omega \in \Omega}$ and $\{B_\omega\}_{\omega \in \Omega}$ are covariant families. To make sense of infinite sums and infinite products within this algebra, one defines the following norm:
\begin{equation}
\|\{A_\omega\}_\omega\|=\sup_{\omega\in\Omega} \|A_\omega\|.
\end{equation}
The infinite sums and products can be seen as limits, like for example:
\begin{equation}
\sum_n \{A_\omega^n\}_\omega=\{S_\omega\}_\omega \ \mbox{if} \ \lim_{N\rightarrow \infty}\|\{S_\omega\}_\omega - \sum_{n<N} \{A_\omega^n\}\|\rightarrow 0.
\end{equation}
The definition of the bounded covariant families, the algebraic operations and the norm defined above leads to a well defined algebra ${\cal A}$, which replaces the algebra of bounded translationally invariant operators.

We are now in position to describe the transition from commutative to non-commutative calculus. In the translational invariant case, the Bloch decomposition allows us to view the operators as matrices, whose entries are functions defined over the Brillouin torus. The existence of this classical manifold allows one to define two operations over the space of translational invariant operators: a) the differentiation with respect to ${\bm k}$ and b) the integration with respect to ${\bm k}$. Let us first describe how this operations extend to the disordered case and then comment on the notion of non-commutative Brillouin torus.

The first observation is that, for the Bloch decomposition $UAU^{-1}=\bigoplus_{{\bm k}\in {\cal T}} \hat{A}(k)$, we have
\begin{equation}
\begin{array}{c}
U^{-1}[\bigoplus_{{\bm k}\in {\cal T}} \partial_{k_j} \hat{A}(k)] U=-i [x_j,A], \ j=1,2.
\end{array}
\end{equation}
The second observation is that:
\begin{equation}
\begin{array}{c}
\frac{1}{(2\pi)^2}\int \mbox{Tr}\{\hat{A}({\bm k})\}d^2{\bm k}=\lim_{\mbox{\tiny{Area}}\rightarrow \infty} \frac{1}{\mbox{\tiny{Area}}}\mbox{Tr}_{\mbox{\tiny{Area}}} \{A\},
\end{array}
\end{equation}
where $\mbox{Tr}_{\mbox{\tiny{Area}}}$ is taken only over the sites included in a finite area. These two observations provides the clue on how to proceed. The partial derivatives of a covariant family $\{A_\omega\}_\omega$ are the covariant families $\{\partial_j A_\omega\}_\omega$ ($j=1,2$) defined by:
\begin{equation}\label{derivative}
\partial_j A_\omega \equiv-i[x_j,A_\omega], \ \mbox{for all} \ \omega\in\Omega 
\end{equation}
Given that with probability one,
\begin{equation}
\begin{array}{c}
\lim_{\mbox{\tiny{Area}}\rightarrow \infty} \frac{1}{\mbox{\tiny{Area}}}\mbox{Tr}_{\mbox{\tiny{Area}}} \{A\}=\int dP(\omega) \ \mbox{Tr}\{\pi_0 A_\omega \pi_0\},
\end{array}
\end{equation}
the integration over the Brillouin torus translates into the non-commutative version:
\begin{equation}\label{integral}
\begin{array}{c}
{\cal T}(\{A_\omega\}) \equiv \int dP(\omega) \ \mbox{Tr}\{\pi_0 A_\omega \pi_0\},
\end{array}
\end{equation}
where $\pi_0=\sum_\alpha |0,\alpha\rangle \langle 0,\alpha|$ is the projector onto the quantum states at the origin. To ease the notation, we will use $\mbox{tr}_0\{A\}$ to denote $\mbox{Tr}\{\pi_0 A \pi_0\}$.

The algebra ${\cal A}$ together with the differentiation and integration defined above can be regarded as a non-commutative Brillouin torus \cite{BELLISSARD:1994xj}. The terminology becomes clear if one remembers that the classic manifolds are in one to one correspondence with the commutative algebra of smooth functions defined over them, that is, one can reconstruct a manifold if he is given this algebra. The Brillouin torus can be recovered from the algebra of translationally invariant operators, whose Bloch decomposition leads to ordinary functions defined over the classic Brillouin torus. The algebra of translationally invariant operators was enlarged here to that of covariant families of bounded operators. No classic manifold can be recovered from this algebra, yet we can define a differential calculus for it. 

\subsection{The Chern number as a analytic Index: The disordered case.}

Given the non-commutative rules of calculus, we can easily spell out the non-commutative version of the Chern number:
\begin{equation}
C=2\pi i \ {\cal T}\{\Pi_\omega [\partial_1 \Pi_\omega,\partial_2 \Pi_\omega]\}.
\end{equation}
The main question now is if the non-commutative Chern number remains quantized and invariant to deformations, and under what exact conditions is this happening? 

\begin{theorem}\label{NCGChern}
[Bellissard et al, 1994 \cite{BELLISSARD:1994xj}] Consider a covariant family of projectors $\{\Pi_\omega\}_{\omega\in \Omega}$. If
\begin{equation}\label{Condition1}
{\cal T}\{| [{\bm x},\Pi_\omega]|^2\}<\infty,
\end{equation}
then $(\Pi_\omega u \Pi_\omega)\oplus \Pi_\omega^\bot$ is, with probability one, in the Fredholm class. Moreover, with probability one, its Index is independent of $\omega$ and:
\begin{equation}\label{Chern2}
Ind\{(\Pi_\omega u \Pi_\omega)\oplus \Pi_\omega^\bot\} = 2\pi i \ {\cal T}\{\Pi_\omega [\partial_1 \Pi_\omega,\partial_2 \Pi_\omega]\}.
\end{equation}
\end{theorem}

\noindent The above result gives the quantization of the Chern number in the presence of disorder but does not provide yet the invariance, which will be discussed later in the paper. We want to point out several things. First, one should notice that the condition written in Eq.~\ref{Condition1}  is extremely simple and transparent. It can be reformulated as:
\begin{equation}
\Lambda^2=\int dP(\omega) \ \mbox{Tr}\{\pi_0 \Pi_\omega {\bm x}^2 \Pi_\omega \pi_0\} <\infty
\end{equation}
and now one can see that $\Lambda$ can be thought as a localization length associated to $\Pi_\omega$. Second, the condition written in Eq.~\ref{Condition1} is optimal, that is, Eq.~\ref{Condition1} provides the most general condition that assures the quantization of the non-commutative Chern number. The remaining of the section is devoted to proving the Theorem \ref{NCGChern}.

\begin{lemma}\label{Step1}
Let $\Pi$ be an orthogonal projection such that $(\Pi-u\Pi u^*)^{2n-1}$ is in the trace class for some integer $n$. Then $(\Pi u \Pi)\oplus \Pi^\bot$ is in the Fredholm class and
\begin{equation}\label{formula1}
Ind\{(\Pi u \Pi)\oplus \Pi^\bot\}=-Tr \ (\Pi-u \Pi u^*)^{2n-1}.
\end{equation}
\end{lemma}

\proof We will invoke Fedosov's formula. If we take $F=(\Pi u \Pi)\oplus \Pi^\bot$, then $I-FF^* = \Pi-\Pi u \Pi u^*\Pi$, which can be put into the form
\begin{equation}
\Pi-\Pi u \Pi u^*\Pi=\Pi (\Pi-u \Pi u^*)^2.
\end{equation}
Since $\Pi$ commutes with $(\Pi-u\Pi u^*)^2$, we have
\begin{equation}
(\Pi-\Pi u \Pi u^*\Pi)^n=\Pi (\Pi - u \Pi u^*)^{2n}.
\end{equation}
Similarly, $I-F^*F = \Pi - \Pi u^* \Pi u \Pi$ and:
\begin{equation}
(\Pi - \Pi u^*\Pi u \Pi)^n=\Pi (\Pi-u^*\Pi u)^{2n}=\Pi u^*(\Pi - u \Pi u^*)^{2n}u.
\end{equation}
Therefore $(I-FF^*)^n$ and $(I-F^*F)^n$ are in the trace class and we can use Fedosov's method to compute the index:
\begin{equation}\label{partial1}
\mbox{Ind}\{(\Pi u \Pi)\oplus \Pi^\bot\}=\mbox{Tr}(\Pi-\Pi u \Pi u^*\Pi)^n - \mbox{Tr}(\Pi-\Pi u ^* \Pi u \Pi)^n.
\end{equation}
We will use simple algebraic manipulations to bring the above expression to the desired form. For this, let us define:
\begin{equation}
\hat{U}=\left (
\begin{array}{cc}
0 & u \\
u^* & 0
\end{array}
\right ), \ \hat{G}=\left (
\begin{array}{cc}
1 & 0 \\
0 & -1
\end{array}
\right ), \ \mbox{and} \ \hat{\Pi}=\left (
\begin{array}{cc}
\Pi & 0 \\
0 & \Pi
\end{array}
\right ).
\end{equation}
A direct computation will show that:
\begin{equation}
\left (
\begin{array}{cc}
\Pi u \Pi u^*\Pi & 0 \\
0 & \Pi u^*\Pi u \Pi
\end{array}
\right )=\hat{\Pi}\hat{U}\hat{\Pi}\hat{U}\hat{\Pi}
\end{equation}
and consequently
\begin{equation}
\left (
\begin{array}{cc}
(\Pi-\Pi u \Pi u^*\Pi)^n & 0 \\
0 & (\Pi-\Pi u^*\Pi u \Pi)^n
\end{array}
\right )=(\hat{\Pi}-\hat{\Pi}\hat{U}\hat{\Pi}\hat{U}\hat{\Pi})^n.
\end{equation}
Inspecting Eq.~\ref{partial1}, we can see that the Index can be computed as $\mbox{Tr}\{\hat{G}(\hat{\Pi}-\hat{\Pi}\hat{U}\hat{\Pi}\hat{U}\hat{\Pi})^n\}$. Using the following two identities:
\begin{equation}
\hat{\Pi}-\hat{\Pi}\hat{U}\hat{\Pi}\hat{U}\hat{\Pi}=-[\hat{U},\hat{\Pi}]^2\hat{\Pi} \ \ \mbox{and} \ 
 \ [\hat{U},\hat{\Pi} ]^2 \hat{\Pi}=\hat{\Pi}[\hat{U},\hat{\Pi}]^2,
\end{equation}
which can be verified through direct computation, we can write:
\begin{equation}
\mbox{Ind}\{(\Pi u \Pi)\oplus \Pi^\bot\}=-\mbox{Tr}\{\hat{G}[\hat{U},\hat{\Pi}]^{2n-2}\hat{\Pi}(\hat{\Pi}-\hat{U}\hat{\Pi}\hat{U})\hat{\Pi}\}.
\end{equation}
Because $\hat{\Pi}$ commutes with $[\hat{U},\hat{\Pi}]^2$ and $\hat{G}$, we can erase one of the two flanking $\hat{\Pi}$'s. It is convenient to do that in a symmetric way (also note that $\hat{U}^2=1$):
\begin{equation}
\begin{array}{l}
\mbox{Ind}\{(\Pi u \Pi)\oplus \Pi^\bot\} \medskip \\
=-\frac{1}{2}\mbox{Tr}\{\hat{G}[\hat{U},\hat{\Pi}]^{2n-2}((\hat{\Pi}-\hat{U}\hat{\Pi}\hat{U})\hat{\Pi}+\hat{\Pi}(\hat{\Pi}-\hat{U}\hat{\Pi}\hat{U}))\} \medskip \\
=-\frac{1}{2}\mbox{Tr}\{\hat{G}[\hat{U},\hat{\Pi}]^{2n-2}(\hat{U}[\hat{U},\hat{\Pi}]\hat{\Pi}+\hat{\Pi}[\hat{\Pi},\hat{U}]\hat{U})\}.
\end{array}
\end{equation}
We can use the identities $[\hat{U},\hat{\Pi}]\hat{U}=-\hat{U}[\hat{U},\hat{\Pi}]$ and $\hat{G}\hat{U}=-\hat{U}\hat{G}$ to bring one $\hat{U}$ all the way to the front and then use the cyclic property of the trace to move it all the way to the back. After these maneuvers we arrive at:
\begin{equation}
\begin{array}{l}
\mbox{Ind}\{(\Pi u \Pi)\oplus \Pi^\bot\}
=\frac{1}{2}\mbox{Tr}\{\hat{G}[\hat{U},\hat{\Pi}]^{2n-2}([\hat{U},\hat{\Pi}]\hat{\Pi}+\hat{\Pi}[\hat{U},\hat{\Pi}])\hat{U}\}.
\end{array}
\end{equation}
Now, if we recognize that $[\hat{U},\hat{\Pi}]\hat{\Pi}+\hat{\Pi}[\hat{U},\hat{\Pi}]$ is nothing else but $[\hat{U},\hat{\Pi}]$, we can write the compact expression:
\begin{equation}
\mbox{Ind}\{(\Pi u \Pi)\oplus \Pi^\bot\}=\frac{1}{2}\mbox{Tr}\{\hat{G}[\hat{U},\hat{\Pi}]^{2n-1}\hat{U}\}.
\end{equation}
This last expression can be cast in the form
\begin{equation}
\begin{array}{l}
\mbox{Ind}\{(\Pi u \Pi)\oplus \Pi ^\bot\}=-\frac{1}{2}\mbox{Tr}\{\hat{G}(\hat{\Pi}-\hat{U}\hat{\Pi}\hat{U}]^{2n-1}\} \medskip \\
=-\frac{1}{2}\mbox{Tr}\left (
\begin{array}{cc} 
(\Pi-u \Pi u^*)^{2n-1} & 0 \\
0 & -(\Pi-u^* \Pi u)^{2n-1}
\end{array}
\right ) \medskip \\
=-\frac{1}{2}\mbox{Tr}\{(\Pi-u\Pi u^*)^{2n-1}+u^*(\Pi-u \Pi u^*)^{2n-1}u\} \medskip \\
=-\mbox{Tr}\{(\Pi-u \Pi u^*)^{2n-1} \}.\qed
\end{array}
\end{equation}

\medskip The above Lemma gives a very convenient way to compute the Index. We can apply this formula to our projectors $\{\Pi_\omega\}$ by virtue of the following fact.

\begin{lemma}\label{Step2} The condition ${\cal T}\{| [{\bm x},\Pi_\omega]|^2\}$$<$$\infty$ assures that $(\Pi_\omega-u\Pi_\omega u^*)^3$ is, with probability one, in the trace class.
\end{lemma}

\noindent We will give the proof of this Lemma at the end, since it is the most technical part of the discussion. We now can complete the main Theorem. We will apply the formula of Eq.~\ref{formula1} with $n=2$. The invariance of the Index with $\omega$ is crucial for the computation. It can be demonstrated in the following way. Since the translations act ergodically on $\Omega$, we need to show that the Indexes of $(\Pi_\omega u \Pi_\omega)\oplus \Pi_\omega$ and $(\Pi_{t_{\bm n}\omega} u \Pi_{t_{\bm n}\omega})\oplus \Pi_{t_{\bm n}\omega}$ are the same for all ${\bm n}\in {\bm Z}^2$. But due to the covariance of $\{\Pi_\omega\}_\omega$ and the invariance of the Index relative to the unitary transformations, we have:
\begin{equation}
\mbox{Ind}\{(\Pi_{t_{\bm n}\omega} u \Pi_{t_{\bm n}\omega})\oplus \Pi_{t_{\bm n}\omega}\}=\mbox{Ind}\{(\Pi_\omega T_{\bm n}^* u T_{\bm n} \Pi_\omega)\oplus \Pi_\omega\},
\end{equation}
hence we need to show that 
\begin{equation}\label{p1}
\mbox{Ind}\{(\Pi_\omega  u  \Pi_\omega)\oplus \Pi_\omega\}=\mbox{Ind}\{(\Pi_\omega T_{\bm n}^* u T_{\bm n} \Pi_\omega)\oplus \Pi_\omega\}.
\end{equation}
The difference between the two operators appearing above is $\Pi_\omega  (u-T_{\bm n}^* u T_{\bm n}) \Pi_\omega$ and 
\begin{equation}
\begin{array}{l}
u-T_{\bm n}^* u T_{\bm n}=\sum_{\bm m}  \left( \frac{m+n_0}{|m+n_0|}-\frac{n+m+n_0}{|n+m+n_0|}\right ) \pi_{\bm m}.
\end{array}
\end{equation}
But $ \frac{m+n_0}{|m+n_0|}-\frac{n+m+n_0}{|n+m+n_0|}$ decays as $\frac{1}{|m|}$ for large $|{\bm m}|$, which means $u-T_{\bm n}^* u T_{\bm n}$ is a compact operator, and therefore the difference between the two operators appearing in Eq~\ref{p1} is a compact operator and consequently the two Indexes are indeed the same (cf. the general properties of the Index).

Now, according to Lemma~\ref{Step1}, we have:
\begin{equation}
\mbox{Ind}\{(\Pi_\omega u \Pi_\omega)\oplus \Pi_\omega^\bot\}=- \mbox{Tr}\ \{\Pi_\omega-u\Pi_\omega u^*\}^3.
\end{equation}
Because the Index is independent of $\omega$, we can average the right part without breaking the equality. If we do that, we have:
\begin{equation}\label{p3}
\begin{array}{l}
 - \int dP(\omega) \ \sum\limits_{{\bm n}} \mbox{Tr}\{ \pi_{\bm n}(\Pi_\omega-u\Pi_\omega u^*)^3\pi_{\bm n}\} \medskip \\
= - \sum\limits_{{\bm n}} \int dP(\omega) \   \mbox{Tr} \{T_{\bm n}^* \pi_{\bm n}(\Pi_\omega-u\Pi_\omega u^*)^3\pi_{\bm n}T_{\bm n} \} \medskip \\
= - \sum\limits_{{\bm n}} \int dP(\omega) \   \mbox{Tr} \{ \pi_{\bm 0}(\Pi_{t_{-{\bm n}}\omega}-T_{\bm n}^*uT_{\bm n}\Pi_{t_{-{\bm n}}\omega} T_{\bm n}^* u^*T_{\bm n})^3\pi_{\bm 0} \} \medskip \\
= - \sum\limits_{{\bm n}} \int dP(\omega) \   \mbox{tr}_0 \{ (\Pi_{\omega}-T_{\bm n}^*uT_{\bm n}\Pi_{\omega} T_{\bm n}^* u^*T_{\bm n})^3 \},
\end{array}
\end{equation}
where in the last line we made a change of variable $t_{-{\bm n}}\omega\rightarrow \omega$ and used the invariance of $dP(\omega)$. One should note that we were able to interchange the integral and the sum precisely because $(\Pi_\omega-u\Pi_\omega u^*)^3$ is in the trace class. Let us use the shorthand $u_{\bm n}$ for $T_{\bm n}^*uT_{\bm n}$ and also point out that:
\begin{equation}\label{Pi}
\begin{array}{l}
u_{\bm n}\pi_{\bm m}=\pi_{\bm m}u_{\bm n}=\frac{n+m+n_0}{|n+m+n_0|}\pi_m.
\end{array}
\end{equation}
To ease the notation, we will write $n$ for $n$+$n_0$ in the following. At this point, we try to compute:
\begin{equation}\label{partial23}
\begin{array}{lll}
\sum\limits_{{\bm n}}  \mbox{tr}_0 \{ (\Pi_{\omega}-u_{\bm n}\Pi_{\omega} u_{\bm n}^*)^3 \} =\sum\limits_{\bm {n,k,m}} \mbox{tr}_0 \{ (\Pi_{\omega}-u_{\bm n}\Pi_{\omega} u_{\bm n}^*) \medskip \\
 \ \ \ \ \ \times \ \pi_{\bm k} (\Pi_{\omega}-u_{\bm n}\Pi_{\omega} u_{\bm n}^*)\pi_{\bm m}(\Pi_{\omega}-u_{\bm n}\Pi_{\omega} u_{\bm n} \} \medskip \\
=\sum\limits_{\bm {n,k,m}} A({\bm n},{\bm k},{\bm m}) \  \mbox{tr}_0 \{\Pi_\omega \pi_{\bm k} \Pi_{\omega}\pi_{\bm m} \Pi_{\omega} )\},
\end{array}
\end{equation}
with:
\begin{equation}
\begin{array}{l}
A({\bm n},{\bm k},{\bm m}) =\left (1-\frac{n\overline{(n+k)}}{|n(n+k)|} \right ) \left(1- \frac{(n+k)\overline{(n+m)}}{|(n+k)(n+m)|}\right ) \left(1-\frac{(n+m)\overline{n}}{|(n+m)n|}\right ).
\end{array}
\end{equation}
We will use the following remarkable identity:
\begin{equation}\label{Identity1}
\sum\limits_{\bm {n}}A({\bm n},{\bm k},{\bm m}) =2\pi i (k_1 m_2-m_1 k_2),
\end{equation}
to continue from Eq.~\ref{partial23}:
\begin{equation}
\begin{array}{l}
\ldots = 2\pi i   \sum\limits_{{\bm {k,m}}}(k_1 m_2-m_1k_2) \mbox{tr}_0 \{\Pi_\omega \pi_{\bm k} \Pi_\omega \pi_{\bm m} \Pi_\omega\} \medskip \\
=2\pi i \sum\limits_{{\bm {k,m}}}(k_1 m_2-m_1k_2)\mbox{tr}_0 \{\Pi_\omega[\pi_{\bm k},\Pi_\omega][\pi_{\bm m},\Pi_\omega]\} \medskip \\
=- 2\pi i  \ \mbox{tr}_0 \{\Pi_\omega\left [i[x_1,\Pi_\omega],i[x_2,\Pi_\omega]\right ]\}.
\end{array}
\end{equation}
One can also show that as long as ${\cal T}\{| [{\bm x},\Pi_\omega]|^2\}$$<$$\infty$, we can interchange the summation and integration in the last line of Eq.~\ref{p3}. In this case, we can use the above computation to finally conclude:
\begin{equation}
\mbox{Ind}\{(\Pi_\omega u \Pi_\omega)\oplus \Pi_\omega^\bot\}=2\pi i  \int dP(\omega)\ \mbox{tr}_0 \{\Pi_\omega\left [-i[x_1,\Pi_\omega],-i[x_2,\Pi_\omega]\right ]\}.
\end{equation}

\begin{figure*}
 \center
 \includegraphics[width=8cm,angle=0]{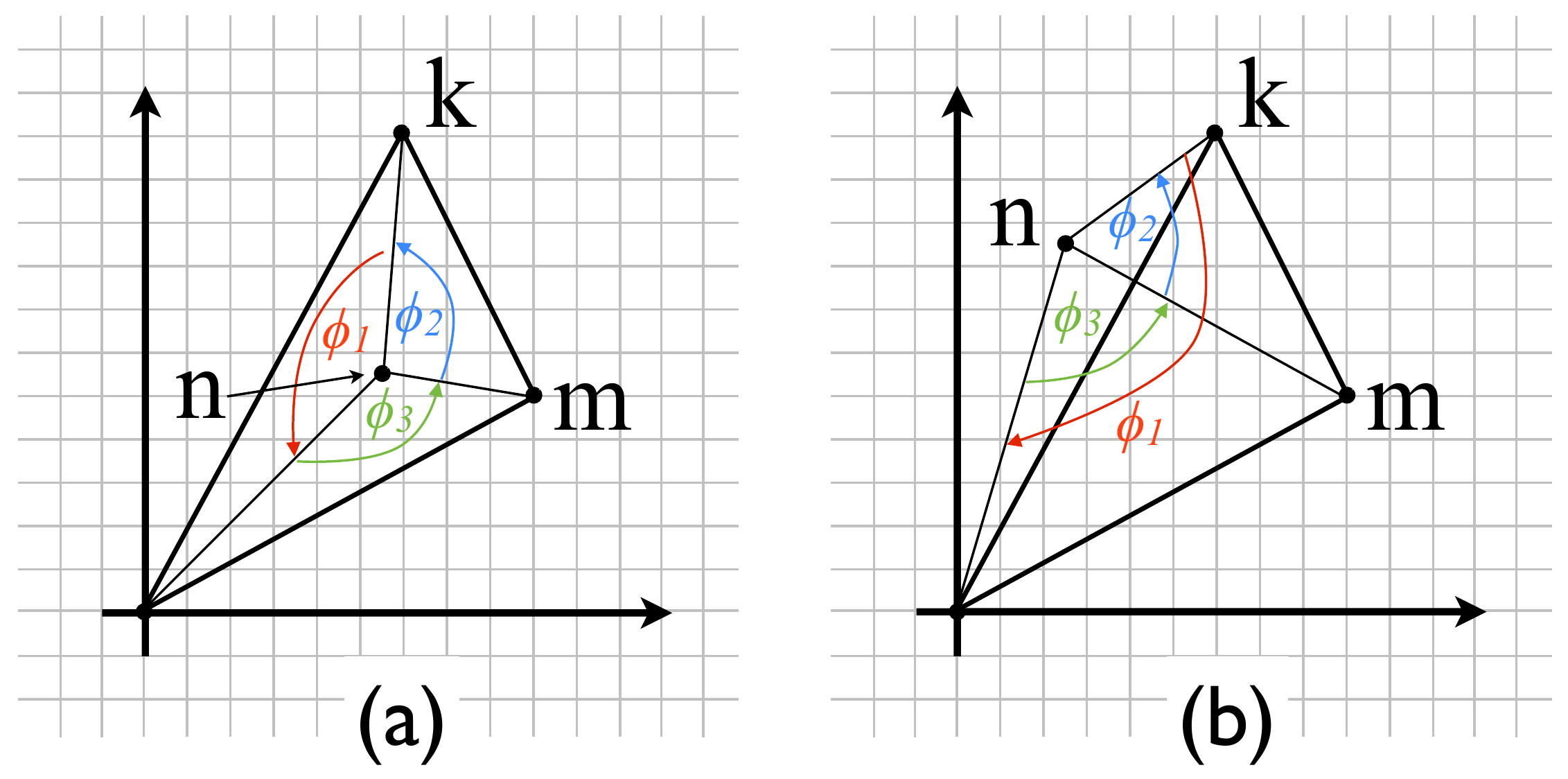}
 \caption{Diagrams used in the proof of identity of Eq.~\ref{Identity1}. The arrow marking an angle tells when the angle is positive or negative: a counterclockwise/clockwise arrow means positive/negative values.}
 \label{GeomInt}
\end{figure*}

It remains to prove the identity written in Eq.~\ref{Identity1} and for this we follow Ref.~\cite{Aizenman1998bf}. Let us consider $A(-{\bm n},{\bm k},{\bm m})$ because this quantity has a geometrical interpretation. To see that, we place $n$, $k$ and $m$ in the complex plane at the vertices of our lattice and shift ${\bm n}$ by $(\frac{1}{2},\frac{1}{2})$ to account for $n_0$, as in Fig.~\ref{GeomInt}. We can immediately see that:
\begin{equation}
\begin{array}{c}
\frac{-n}{|n|}\frac{\overline{k-n}}{|k-n|}=e^{i\phi_1}, \ \frac{k-n}{|k-n|}\frac{\overline{m-n}}{|m-n|}=e^{i\phi_2}, \ \frac{m-n}{|m-n|}\frac{\overline{-n}}{|n|}=e^{i\phi_3},
\end{array}
\end{equation}
where $\phi_i$'s are the angles defined in Fig.~\ref{GeomInt}. A direct calculation will show:
\begin{equation}
\begin{array}{c}
A(-{\bm n},{\bm k},{\bm m})=-2i( \sin \phi_1 + \sin \phi_2 + \sin \phi_3).
\end{array}
\end{equation}
We are going to write the summation over ${\bm n}$ in Eq.~\ref{Identity1} in the following way:
\begin{equation}\label{inter}
\begin{array}{c}
\frac{-1}{2i}\sum_{\bm n} A(-{\bm n},{\bm k},{\bm m}) = \sum_{\bm n} (\phi_1+\phi_2 +\phi_3)\medskip \\
-\sum_{\bm n} (\phi_1-\sin \phi_1)+\sum_{\bm n} (\phi_2 - \sin \phi_2 )- 2\sum_{\bm n} (\phi_3-\sin \phi_3).
\end{array}
\end{equation}
Such decomposition is possible because each of the last three sums are absolutely convergent, due to the fact that the $\phi$'s behave as $1/|n|$ for large $|n|$, and consequently $\phi - \sin \phi \sim 1/|n|^3$. In contradistinction, the first sum cannot be broken in partial sums because the resulting sums will not be convergent. Now, since $\phi_1-\sin \phi_1$ is antisymmetric with respect to the reflection of ${\bm n}$ relative to the midpoint between the origin and ${\bm k}$, and similarly for $\phi_2$ and $\phi_3$, each sum in the second row of Eq.~\ref{inter} is identically zero. Furthermore, as exemplified in Fig.~\ref{GeomInt}, we have the simple fact:
\begin{equation}
\phi_1+\phi_2+\phi_3=\left \{
\begin{array}{l}
2 \pi \ \mbox {if} \ {\bm n} \ \mbox{is inside the triangle} \\
\pi \ \mbox{if} \ {\bm n} \ \mbox{is on an edge} \\
0 \ \mbox {if} \ {\bm n} \ \mbox{is outside the triangle}
\end{array}
\right .
\end{equation}
At the end, we can conclude that
\begin{equation}
\frac{-1}{4\pi i}\sum_{\bm n} A(-{\bm n},{\bm k},{\bm m}) = \mbox{\# of grid points inside the triangle},
\end{equation}
with the points on the edges counted only as $\frac{1}{2}$. This number is precisely equal to the area of the triangle, that is, $\frac{1}{2}|k_1 m_2-m_1 k_2|$ (same as $-\frac{1}{2}(k_1 m_2-m_1 k_2)$). 
This concludes the proof of Theorem~\ref{NCGChern}.\qed

\subsection{Mac\`aev spaces and the Dixmier trace}

We are dealing here with the question of weather the operator $(\Pi_\omega-u\Pi_\omega u^*)^3$ is in the trace class. This question can be answered in a quite elegant way with the help of the structures and operation announced in the title of this section. To begin, it is instructive to see how $(\Pi_\omega-u\Pi_\omega u^*)^2$ fails to be in the trace class, by applying the following simple criterion: Given an operator $A$, $A^2$ is in the trace class if and only if:
\begin{equation}\label{Criterion}
\sum_{{\bm n},{\bm m},\alpha,\beta} |\langle {\bm m},\beta|A|{\bm n}, \alpha\rangle|^2 < \infty.
\end{equation}
No such simple criterion exists for probing if $A^3$ is in ${\cal S}^1$. So let's look at the above sum when $A$ is $\Pi_\omega-u\Pi_\omega u^*$:
\begin{equation}\label{partial45}
\begin{array}{l}
\sum\limits_{{\bm n},{\bm m},\alpha,\beta} |\langle {\bm m},\beta|\Pi_\omega-u\Pi_\omega u^*|{\bm n}, \alpha\rangle|^2 \medskip \\
=\sum\limits_{{\bm n},{\bm s},\alpha,\beta} \left|1-\frac{(n+s)\bar{n}}{|(n+s)n|)}\right |^2|\langle {{\bm n}+{\bm s}},\beta|\Pi_\omega|{\bm n}, \alpha\rangle|^2.
\end{array}
\end{equation}
Then, even if we assume an exponential decay of the projector: $|\langle {\bm n+s},\beta|\Pi_\omega|{\bm n}, \alpha\rangle|<ct. e^{-\gamma |s|}$, which will make the summation over ${\bm s}$ convergent, the sum still diverges because the front term decays only as $|{\bm n}|^{-2}$ and that is not enough to make the summation over ${\bm n}$ convergent. A second look at what we just said, reveals that the sum of Eq.~\ref{partial45} diverges only logarithmically, suggesting that the sum of the singular values $\sum_i \mu_i^2$ of $(\Pi_\omega-u\Pi_\omega u^*)^2$ has a weak logarithmic divergence.

The above discussion revealed the importance of a new class of operators, namely those for which $\sum_i \mu_i^2$ diverges logarithmically, or more generally, for which $\sum_i \mu_i^p$ diverges logarithmically. These classes are now fairly well understood, especially after the work of Mac\`aev. 

\begin{definition} For $p\in [1,\infty)$, the Mac\`aev space ${\cal M}^{p}$ contain all compact operators for which:
\begin{equation}
\limsup\limits_{N\rightarrow \infty} \ \frac{1}{\ln N}\sum_{i=1}^N \mu_i^p <\infty.
\end{equation}
\end{definition}
 
\noindent We collect the most important properties of these spaces in the following statements.

 \begin{proposition} \label{Macaev}
 i) ${\cal M}^{p}$ are two-sided ideals in ${\cal B}({\cal H})$, that is, $AB$ and $BA$ belongs to ${\cal M}^{p}$ if $A\in {\cal M}^{p}$ and $B$ is a bounded operator.\smallskip
 
\noindent ii) The functional
\begin{equation}\label{Norm}
\|A\|_{{\cal M}^p}=
\sup\limits_{N>1} \ \frac{1}{\ln N}\sum_{i=1}^N \mu_n^p
\end{equation}
defines a norm on ${\cal M}^{p}$. With this norm, the Mac\`aev spaces become Banach spaces.\smallskip
 
\noindent iii) If $A$ belongs to ${\cal M}^2$, then $A^{3}$ belongs to the trace class.\smallskip
\end{proposition}

\noindent The first statement is useful when one needs to establish that a product of operators is in ${\cal M}^p$ and the second statement is important when dealing with infinite sums and products of operators. The third statement is particularly important for our discussion because it tells us that if one can establish that $\Pi_\omega -u\Pi_\omega u^*$ is in ${\cal M}^2$, then he can automatically conclude that $(\Pi_\omega -u\Pi_\omega u^*)^3$ is in the trace class. This is in fact the strategy that we follow. The classical trace cannot be extended from ${\cal S}^1$ to ${\cal M}^1$ and the construction of a new trace over the ${\cal M}^1$ is paramount for our discussion. As we shall see, this trace will provide a computationally manageable criterion to test if  $\Pi_\omega -u\Pi_\omega u^*$ is in ${\cal M}^2$. 

We begin the discussion of this new trace and we follow the notes from Ref.~\cite{CONNES:1995rv} (see also the exposition in \cite{Connes:1994wk}). Let us consider first only the positive compact operators, i.e. those with the property that $\langle \phi |A|\phi \rangle \ge 0$ for all $\phi\in {\cal H}$.  If $A$ is such an operator and $\{\mu_i\}$ denote its singular values, ordered in decreasing order, we introduce the notation $\bar {\mu}_A(N)=\sum_{i<N} \mu_i$. The task is to use $\bar{\mu}_A(N)$'s and construct a linear functional on ${\cal M}^1$, which retains the main properties of the classical trace. To do so, one extrapolates $\bar{\mu}_A(N)$ to the real numbers by setting $\bar{\mu}_A(N+\lambda)$ equal to $(1-\lambda)\bar{\mu}_A(N)+\lambda \bar{\mu}_A(N+1)$ for $\lambda \in [0,1]$, and then defines:
\begin{equation}
\tau_A(\lambda) = \frac{1}{\log \lambda} \int_3^\lambda \frac{\bar{\mu}_A(u)}{\log u} \frac{du}{u}.
\end{equation}
Let us fix $\lambda$ for a moment and view $\tau_A(\lambda)$ as a functional that assigns positive numbers to the operators $A\in {\cal M}^1$. An important result \cite{CONNES:1995rv} says that
\begin{equation}\label{p5}
\tau_{A+B}(\lambda) - \tau_A(\lambda) - \tau_B(\lambda) =  O(\log(\log \lambda)/\log \lambda),
\end{equation}
for any $A, B \in {\cal M}^1$. According to this result, the functional appears to become linear when $\lambda$ is taken to infinity, which is exactly what is desired, but there is catch. While the whole left hand side of Eq.~\ref{p5} converges to zero as $\lambda \rightarrow \infty$, due to nice cancelations, the individual terms may not have a well defined limit when $\lambda \rightarrow \infty$. While it is true that, for $A\in {\cal M}^1$, $\tau_A(\lambda)$ is bounded for all values of $\lambda$, $\tau_A(\lambda)$ can oscillate like cosine or sine functions. To solve this shortcoming, one introduces an equivalent relation on the linear space $C_b[3,\infty)$ of continuous bounded functions, where the function $\lambda \rightarrow \tau_A(\lambda)$ belongs to, by saying that $f \sim g$ if the difference $f-g$ vanishes as $\lambda \rightarrow \infty$. If $\chi$ is any strictly positive linear functional on the quotient space $C_b[3,\infty)/\sim$, then
\begin{equation}
\mbox{Tr}_{\mbox{\tiny{Dix}}}\{A\} = \chi(\widehat{\tau_A}),
\end{equation}
defines a trace on the positive operators in ${\cal M}^1$, where $\widehat{\tau_A}$ represents the equivalence class of the function $\tau_A(\lambda)$ in $C_b[3,\infty)/\sim$. This newly introduced trace can be extended by linearity over the entire  ${\cal M}^1$, and this gives the Dixmier trace \cite{Dixmier1966jc}.

\begin{proposition} The functional 
\begin{equation}
{\cal M}^1 \ni A \rightarrow \mbox{Tr}_{\mbox{\tiny{Dix}}}\{A\}\in {\bm C}
\end{equation}
is: i) continuous on ${\cal M}^1$ when ${\cal M}^1$ is equipped with the norm of Eq.~\ref{Norm}; \smallskip 

\noindent ii) linear: $\mbox{Tr}_{\mbox{\tiny{Dix}}}\{A+B\}=\mbox{Tr}_{\mbox{\tiny{Dix}}}\{A\}+\mbox{Tr}_{\mbox{\tiny{Dix}}}\{B\}$;\smallskip

\noindent iii) cyclic: $\mbox{Tr}_{\mbox{\tiny{Dix}}}\{AB\}=\mbox{Tr}_{\mbox{\tiny{Dix}}}\{BA\}$, for $A\in {\cal M}^1$ and $B \in {\cal B}({\cal H})$;\smallskip

\noindent iv) unitarily invariant: $\mbox{Tr}_{\mbox{\tiny{Dix}}}\{UAU^{-1}\}=\mbox{Tr}_{\mbox{\tiny{Dix}}}\{A\}$, for $U$ unitary and $A\in {\cal M}^1$.\smallskip

\noindent v) positive: if $A\in {\cal M}^1$ is a positive operator, then $\mbox{Tr}_{\mbox{\tiny{Dix}}}\{A\}>0$. 
\end{proposition}

\noindent We can see that the Dixmier trace has all the characteristic properties of a trace. Computing the actual Dixmier trace may not be always easy, especially because of the involvement of the linear functional $\chi$. However, there are several simple rules which can simplify the computation of the Dixmier trace. For example, if $\lim_{N\rightarrow \infty} \bar{\mu}_A(N)/\log N$ exists for a positive operator $A$, then: 
\begin{equation}\label{partial59}
\mbox{Tr}_{\mbox{\tiny{Dix}}}\{A\} = \lim_{N\rightarrow \infty}\bar{\mu}_A(N)/\log N.\smallskip
\end{equation}
Since any operator can be written as a linear combination of four positive operators, the above rule can be extended by linearity beyond the class of positive operators. Another useful and important observation is that $\mbox{Tr}_{\mbox{\tiny{Dix}}}\{A\} = 0$ if $A$ is in the trace class. Also, for two positive operators $A$ and $B$ such that $A-B$ is positive, we have $\mbox{Tr}_{\mbox{\tiny{Dix}}}\{A\} < \mbox{Tr}_{\mbox{\tiny{Dix}}}\{B\}$. If $\mbox{Tr}_{\mbox{\tiny{Dix}}}\{B\}$ can be explicitly computed, maybe by formula given in Eq.~\ref{partial59}, then this relation gives an explicit upper bound on $\mbox{Tr}_{\mbox{\tiny{Dix}}}\{A\}$, which in many instances is as good as computing the trace itself. 

\subsection{Proof of Lemma~\ref{Step2}}

Let us lay down the main steps of our proof, which deviates from the original version of Ref.~\cite{BELLISSARD:1994xj}. Working with a truncated version $\Pi^R_\omega$ of $\Pi_\omega$, to be defined shortly, it is first shown that $\Pi^R_\omega-u\Pi^R_\omega u^*$ are in ${\cal M}^2$, equivalent to say that $(\Pi^R_\omega-u\Pi^R_\omega u^*)^2$ are in ${\cal M}^1$.  In the next step we establish that $\mbox{Tr}_{\mbox{\tiny{Dix}}}\{(\Pi^R_\omega-u\Pi^R_\omega u^*)^2\}$ is, with probability one, independent of $\omega$. This is an important observation because then we can average over $\omega$ and $\int dP(\omega)\mbox{Tr}_{\mbox{\tiny{Dix}}} \{(\Pi^R_\omega-u\Pi^R_\omega u^*)^2\}$ can be computed explicitly, the result being precisely $\frac{\pi}{2}{\cal T}(\{\Pi_\omega |{\bm x}|^2 \Pi_\omega \})$, in the limit $R\rightarrow \infty$. Thus, $\Pi_\omega-u\Pi_\omega u^*$ is in ${\cal M}^2$, and from the statement {\cal iii}) of Proposition~\ref{Macaev} we can conclude that $(\Pi_\omega-u\Pi_\omega u^*)^3$ is in the trace class. As a remark, notice that the main identity, found in Ref.~\cite{BELLISSARD:1994xj} and detailed below, can be written in the equivalent form:
\begin{equation}
\begin{array}{c}
\mbox{Tr}_{\mbox{\tiny{Dix}}}\{|\frac{1}{i}[u,\Pi_\omega]|^2\}=\frac{\pi}{2} {\cal T} \{ |\frac{1}{i}[{\bm x},\Pi_\omega]|^2 \},
\end{array}
\end{equation}
which can be seen as the translation to the present context of a central identity in Non-Commutative Geometry \cite{CONNES:1988sb}. Let us now formulate all these into a concise statement.

\begin{proposition}
Consider the truncated projector $\Pi_\omega^R$ obtained by setting to zero all matrix elements $\langle {\bm n}+{\bm s},\beta|\Pi_\omega|{\bm n},\alpha\rangle$ whenever $|{\bm s}|$ exceeds $R$. Then $\Pi_\omega^R - u\Pi_\omega^R u^*$ are in ${\cal M}^2$ and, with probability one, the Dixmier trace $\mbox{Tr}_{\mbox{\tiny{Dix}}}\{(\Pi^R_\omega-u\Pi^R_\omega u^*)^2\}$ is independent of $\omega$. Moreover,
\begin{equation}
\begin{array}{c}
\mbox{Tr}_{\mbox{\tiny{Dix}}}\{(\Pi^R_\omega-u\Pi^R_\omega u^*)^2\}=\frac{\pi}{2} \int dP(\omega)\ \mbox{tr}_0\{ \Pi_\omega |{\bm x}_R|^2 \Pi_\omega \},
\end{array}
\end{equation}
where ${\bm x}_R=\sum_{|{\bm s}|\leq R} {\bm s} \pi_{\bm s}$ is the truncated position operator. The equality remains valid in the limit $R\rightarrow \infty$, provided ${\cal T}(\{|[\Pi_\omega ,{\bm x}]|^2 \}) < \infty$.
\end{proposition}

\proof One has:
\begin{equation}
\begin{array}{l}
\Pi_\omega^R - u\Pi_\omega^R u^*=\sum_{|{\bm s}|\leq R} \sum_{\bm n} \pi_{{\bm n}+{\bm s}}(\Pi_\omega - u\Pi_\omega u^*)\pi_{\bm n} \equiv \sum_{|{\bm s}|\leq R} Q_{\bm s}.
\end{array}
\end{equation}
Because of the particular form of the $Q_s$ operators, one can easily compute their absolute values $|Q_{\bm s}|=\sqrt{Q_{\bm s}^* Q_{\bm s}}$: 
\begin{equation}
\begin{array}{l}
|Q_{\bm s}|=\sum_{\bm n} \left | \frac{n+s}{|n+s|}-\frac{n}{|n|} \right | \pi_{\bm n} \Pi_\omega\pi_{{\bm n}+{\bm s}} \Pi_\omega \pi_{\bm n},
\end{array}
\end{equation}
which are seen to be smaller than  
\begin{equation}
\begin{array}{l}
 \sum_{\bm n} \left |\frac{n+s}{|n+s|}-\frac{n}{|n|} \right |  \pi_{\bm n}.
 \end{array}
\end{equation}
Consequently:
\begin{equation}
\begin{array}{l}
\| Q_{\bm s}\|_{{\cal M}^2} \leq \left \|\sum_{\bm n} \left | \frac{n+s}{|n+s|}-\frac{n}{|n|} \right |  \pi_{\bm n} \right \|_{{\cal M}^2} < \infty,
\end{array}
\end{equation}
showing that each $Q_{\bm s}$ is in ${\cal M}^2$; therefore $\Pi_\omega^R-u\Pi_\omega^R u^*$, which is just a finite sum of $Q_{\bm s}$ operators, is also in ${\cal M}^2$.

We now start the computation of $\mbox{Tr}_{\mbox{\tiny{Dix}}}\{(\Pi^R_\omega-u\Pi^R_\omega u^*)^2\}$. One has: 
\begin{equation}
\begin{array}{l}
(\Pi^R_\omega-u\Pi^R_\omega u^*)^2=\sum_{|{\bm s}|\leq 2R}\sum_{\bm n} \pi_{{\bm n}+{\bm s}}(\Pi_\omega^R - u\Pi_\omega^R u^*)^2 \pi_{\bm n}.
\end{array}
\end{equation}
The Dixmier trace of the terms with ${\bm s}\neq 0$ is zero. This can be established by considering the unitary transformation (here we assume $s_1\neq 0$):
\begin{equation}
\begin{array}{l}
{\cal U}|{\bm n},\alpha \rangle = (-1)^{\left [ \frac{n_1}{s_1} \right ]} |{\bm n},\alpha \rangle,
\end{array}
\end{equation} 
which has the following useful properties:
\begin{equation}
\begin{array}{l}
{\cal U}\pi_{{\bm n}+{\bm s}} = (-1)^{\left [ \frac{n_1}{s_1} \right ]+1}\pi_{{\bm n}+{\bm s}} \ \ \mbox{and} \ \ \pi_{{\bm n}} {\cal U}^{-1} = (-1)^{\left [ \frac{n_1}{s_1} \right ]}\pi_{{\bm n}}.
\end{array}
\end{equation}
From these properties it follows that:
\begin{equation}
\begin{array}{l}
{\cal U} \left ( \sum_{\bm n} \pi_{{\bm n}+{\bm s}}(\Pi_\omega^R - u\Pi_\omega^R u^*)^2 \pi_{\bm n}\right ) {\cal U}^{-1} \medskip \\
= -\sum_{\bm n} \pi_{{\bm n}+{\bm s}}(\Pi_\omega^R - u\Pi_\omega^R u^*)^2 \pi_{\bm n}  
\end{array}
\end{equation}
which implies:
\begin{equation}
\begin{array}{l}
\mbox{Tr}_{\mbox{\tiny{Dix}}} \left \{ {\cal U} \left (\sum_{\bm n} \pi_{{\bm n}+{\bm s}}(\Pi_\omega^R - u\Pi_\omega^R u^*)^2 \pi_{\bm n}\right ) {\cal U}^{-1} \right \} \medskip \\
= -\mbox{Tr}_{\mbox{\tiny{Dix}}}\left \{ \sum_{\bm n} \pi_{{\bm n}+{\bm s}}(\Pi_\omega^R - u\Pi_\omega^R u^*)^2 \pi_{\bm n}  \right \},
\end{array}
\end{equation}
and this can happen only if the above traces are identically zero. At this step it was established that the Dixmier traces of $(\Pi^R_\omega-u\Pi^R_\omega u^*)^2$ and of $\sum_{\bm n} \pi_{{\bm n}}(\Pi_\omega^R - u\Pi_\omega^R u^*)^2 \pi_{\bm n}$ coincide. One can go one step further and show that they also coincide with the Dixmier trace of $\sum_{{\bm n},\alpha} \pi_{{\bm n},\alpha}(\Pi_\omega^R - u\Pi_\omega^R u^*)^2 \pi_{{\bm n},\alpha}$. Indeed, 
by using the unitary transformation (here we assume $\gamma \neq 0$):
\begin{equation}
\begin{array}{l}
{\cal W}|{\bm n},\alpha \rangle = (-1)^{\left [ \frac{\alpha}{\gamma} \right ]} |{\bm n},\alpha \rangle,
\end{array}
\end{equation} 
which has the following useful properties:
\begin{equation}
\begin{array}{l}
{\cal W}\pi_{{\bm n},\alpha+\gamma} = (-1)^{\left [ \frac{\alpha}{\gamma} \right ]+1}\pi_{{\bm n},\alpha+\gamma} \ \mbox{and} \ \pi_{{\bm n},\alpha} {\cal W}^{-1} = (-1)^{\left [ \frac{\alpha}{\gamma} \right ]}\pi_{{\bm n},\alpha},
\end{array}
\end{equation}
one can establish that:
\begin{equation}
\begin{array}{l}
\mbox{Tr}_{\mbox{\tiny{Dix}}} \left \{ {\cal W} \left (\sum_{\bm n} \pi_{{\bm n},\alpha + \gamma}(\Pi_\omega^R - u\Pi_\omega^R u^*)^2 \pi_{{\bm n},\alpha}\right ) {\cal W}^{-1} \right \} \medskip \\
= -\mbox{Tr}_{\mbox{\tiny{Dix}}}\left \{ \sum_{\bm n} \pi_{{\bm n},\alpha+\gamma}(\Pi_\omega^R - u\Pi_\omega^R u^*)^2 \pi_{{\bm n},\alpha} \right \},
\end{array}
\end{equation}
and this is again possible only if both traces are identically zero.

We are now in the position to show that the Dixmier trace of $(\Pi^R_\omega-u\Pi^R_\omega u^*)^2$ is independent of $\omega$. Since the flow $t$ induced by the translations acts ergodically on $\Omega$, it is enough to show that the Dixmier trace remains unchanged when $\omega$ is replaced by $t_{\bm r} \omega$, with ${\bm r}$ some arbitrary element of ${\bm Z}^2$. 
The invariance of the Dixmier trace to unitary transformations leads to:
\begin{equation}
\begin{array}{l}
\mbox{Tr}_{\mbox{\tiny{Dix}}}\{(\Pi^R_{t_{\bm r}\omega}-u\Pi^R_{t_{\bm r}\omega} u^*)^2\} - \mbox{Tr}_{\mbox{\tiny{Dix}}}\{(\Pi^R_\omega-u\Pi^R_\omega u^*)^2\}  \medskip \\
= \mbox{Tr}_{\mbox{\tiny{Dix}}}\{T_{\bm r}(\Pi^R_{\omega}-u_{\bm r}\Pi^R_{\omega} u_{\bm r}^*)^2 T_{\bm r}^*\} - \mbox{Tr}_{\mbox{\tiny{Dix}}}\{(\Pi^R_\omega-u\Pi^R_\omega u^*)^2\}  \medskip \\
=  \mbox{Tr}_{\mbox{\tiny{Dix}}}\{ \sum_{{\bm n},\alpha} \pi_{{\bm n},\alpha}[(\Pi_{\omega}^R - u_{\bm r}\Pi_{\omega}^R u_{\bm r}^*)^2 -(\Pi_\omega^R - u\Pi_\omega^R u^*)^2] \pi_{{\bm n},\alpha} \}. 
\end{array}
\end{equation}
The operator appearing inside the last Dixmier traces can be written in the following form:
\begin{equation}
\begin{array}{l}
\sum\limits_{{\bm s}\leq R,\beta} \sum\limits_{{\bm n},\alpha}|\langle {\bm n},\alpha|\Pi_{\omega}^R |{\bm n}+{\bm s},\beta \rangle |^2  \medskip \\
\times  \left ( \left |\frac{n+r+s}{|n+r+s|}-\frac{n+r}{|n+r|}\right |^2 - \left |\frac{n+s}{|n+s|}-\frac{n}{|n|}\right |^2 \right )   \pi_{{\bm n},\alpha} .
\end{array}
\end{equation}
The operators inside the first sum are in the trace class, since their singular values are less than:
\begin{equation}
\begin{array}{l}
\left | \left |\frac{n+r+s}{|n+r+s|}-\frac{n+r}{|n+r|}\right |^2 - \left |\frac{n+s}{|n+s|}-\frac{n}{|n|}\right |^2 \right |,
\end{array}
\end{equation}
which behave as $|n|^{-3}$ as $|n|\rightarrow \infty$, hence the singular values are summable. The Dixmier trace of trace class operators is zero and this concluded our argument.

The last steps of the computation go as follows:
\begin{equation}
\begin{array}{l}
\mbox{Tr}_{\mbox{\tiny{Dix}}}  \{ (\Pi^R_\omega-u\Pi^R_\omega u^*)^2\} \medskip \\
=\mbox{Tr}_{\mbox{\tiny{Dix}}} \left\{ \sum\limits_{{\bm n},\alpha} \langle {\bm n},\alpha |(\Pi_\omega^R-u\Pi_\omega^R u^*)^2|{\bm n},\alpha \rangle \pi_{{\bm n},\alpha} \right \}.
\end{array}
\end{equation}
Note that the last Dixmier trace can be computed via Eq.~\ref{partial59} and that the operators inside this last Dixmier trace, corresponding to different $\omega$'s, commute with each other.  Hence, we can average over $\omega$ and interchange the average with the Dixmier trace, to continue
\begin{equation}
\begin{array}{l}
\ldots =\int dP(\omega) \ \mbox{Tr}_{\mbox{\tiny{Dix}}} \left\{ \sum\limits_{{\bm n},\alpha} \langle {\bm n},\alpha |(\Pi_\omega^R-u\Pi_\omega^R u^*)^2|{\bm n},\alpha \rangle \pi_{{\bm n},\alpha} \right \} \medskip \\
= \mbox{Tr}_{\mbox{\tiny{Dix}}} \left\{ \sum\limits_{{\bm n},\alpha}  \int dP(\omega) \langle {\bm n},\alpha |(\Pi_\omega^R-u\Pi_\omega^R u^*)^2|{\bm n},\alpha \rangle \pi_{{\bm n},\alpha} \right \} \medskip \\
=\mbox{Tr}_{\mbox{\tiny{Dix}}} \left\{\sum\limits_{{\bm n},\alpha} \int dP(\omega) \langle {\bm 0},\alpha |T_{\bm n}^* (\Pi_\omega^R-u\Pi_\omega^R u^*)^2T_{\bm n}|{\bm 0},\alpha \rangle \pi_{{\bm n},\alpha} \right \} \medskip \\
=\mbox{Tr}_{\mbox{\tiny{Dix}}} \left\{ \sum\limits_{{\bm n},\alpha} \int dP(\omega) \langle {\bm 0},\alpha |(\Pi_{t_{-{\bm n}}\omega}^R-u_{\bm n}\Pi_{t_{-{\bm n}}\omega}^R u_{\bm n}^*)^2|{\bm 0},\alpha \rangle \pi_{{\bm n},\alpha} \right \} \medskip \\
\end{array}
\end{equation}
At this point we use the invariance of the measure $dP(\omega)$ to the flow $t$ to continue as:
\begin{equation}
\begin{array}{l}
\ldots=\mbox{Tr}_{\mbox{\tiny{Dix}}} \left\{\sum\limits_{{\bm n},\alpha} \int dP(\omega) \langle {\bm 0},\alpha |(\Pi_{\omega}^R-u_{\bm n}\Pi_{\omega}^R u_{\bm n}^*)^2|{\bm 0},\alpha \rangle \pi_{{\bm n},\alpha} \right \} \medskip \\
=\mbox{Tr}_{\mbox{\tiny{Dix}}} \left\{ \sum\limits_{{\bm n},\alpha} \int dP(\omega) \sum\limits_{|{\bm s}|\leq R} \langle {\bm 0},\alpha |(\Pi_{\omega}\pi_{\bm s}\Pi_{\omega}|{\bm 0},\alpha \rangle \left |\frac{n+s}{|n+s|}-  \frac{n}{|n|}\right |^2 \pi_{{\bm n},\alpha} \right \} \medskip \\
=\int dP(\omega) \sum\limits_{|{\bm s}|\leq R,\alpha} \langle {\bm 0},\alpha |(\Pi_{\omega}\pi_{\bm s}\Pi_{\omega}|{\bm 0},\alpha \rangle \mbox{Tr}_{\mbox{\tiny{Dix}}} \left\{ \sum\limits_{{\bm n}} \left |\frac{n+s}{|n+s|}-  \frac{n}{|n|}\right |^2 \pi_{{\bm n},\alpha} \right \} .
\end{array}
\end{equation}
The remaining Dixmier trace can be explicitly evaluated and the result is $\frac{\pi}{2} |{\bm s}|^2$. At this point we arrived at:
\begin{equation}
\begin{array}{c}
\mbox{Tr}_{\mbox{\tiny{Dix}}}  \{ (\Pi^R_\omega-u\Pi^R_\omega u^*)^2\} 
=\frac{\pi}{2} \int dP(\omega) \sum\limits_{|{\bm s}|\leq R} \langle {\bm 0},\alpha |(\Pi_{\omega} |{\bm s}|^2\pi_{\bm s}\Pi_{\omega}|{\bm 0},\alpha \rangle,
\end{array}
\end{equation}
which is precisely the statement of the Proposition.\qed

The limit $R\rightarrow \infty$ can be safely taken if $\int dP(\omega)\ \mbox{tr}_0\{ \Pi_\omega |{\bm x}|^2 \Pi_\omega \}<\infty$, in which case the result shows that  $\Pi_\omega-u\Pi_\omega u^*$ is in ${\cal M}^2$ class and consequently $ (\Pi_\omega-u\Pi_\omega u^*)^3$ is in the trace class.

\subsection{Invariance of the Chern number}\label{Invariance}

So far, the theory established the optimal conditions that assures the quantization of the non-commutative Chern number corresponding to a generic covariant family of projectors. We call these conditions optimal because they cannot be improved any further. The next question is under what conditions does the Chern number remain constant under various deformations of the projectors? The invariance of the Chern number with the shift of the Fermi level was already discussed in in the original work by Bellissard et al \cite{BELLISSARD:1994xj}, and was reconsidered in Ref.~\cite{Aizenman1998bf}. The invariance of the Chern number under more general deformations was considered in Ref.~\cite{Richter2001jg}. These works based their analysis on estimates done directly on the Chern number formula. Here we will base our analysis on the special properties of the Index. There are two complications. Even for smooth deformations of the Hamiltonian, the projector onto the spectrum below $E_F$ does not change smoothly, because eigenvalues will inherently cross the Fermi level. Secondly, some of the operators $\Pi_\omega u \Pi_\omega$ may exit and re-enter the Fredholm class during the deformation, because a finite localization length $\Lambda$ does not imply that all $\Pi_\omega u \Pi_\omega$ are in the Fredholm class (only almost all). 

\begin{theorem}\label{InvarianceNCG} Consider a deformation $\{\Pi_\omega(\tau)\}_{\omega\in\Omega}$ of a covariant family of projectors, such that:
\begin{enumerate}

\item ${\cal T}(\{| [{\bm x},\Pi_\omega(\tau)]|^2\}) \leq \bar{\Lambda}^2<\infty$, for all $\tau\in [0,\bar{\tau}]$.

\item There exists a covariant family of self-adjoint operators $\{Q_\omega(\tau)\}_{\omega\in \Omega}$ that are, with probability one, in the Fredholm class, and such that 
\begin{equation}\label{co1}
\| Q_\omega(\tau)-Q_\omega(\tau')\| \ \mbox{and} \ \ \| (Q_\omega\Pi_\omega)(\tau)-(Q_\omega\Pi_\omega)(\tau')\|<ct.|\tau-\tau'|,
\end{equation}
with the constant independent of $\omega$ or $\tau$.
\end{enumerate}
In these conditions, the Chern number of $\{\Pi_\omega(\tau)\}_{\omega\in\Omega}$ is well defined and assumes the same integer value for all $\tau\in[0,\bar{\tau}]$.
\end{theorem}\smallskip

\noindent {\it Proof.} Condition i) assures that $[\Pi_\omega(\tau) u \Pi_\omega(\tau)]\oplus \Pi_\omega(\tau)^\bot$ is in the Fredholm class and that $\mbox{Ind}\{[\Pi_\omega(\tau) u \Pi_\omega(\tau)]\oplus \Pi_\omega(\tau)^\bot\}$ is the same for all $\omega\in \Omega$, except a possible set of measure zero. Given the general properties of the Fredholm operators, same can be said for $Q_\omega(\tau)([\Pi_\omega(\tau) u \Pi_\omega(\tau)]\oplus \Pi_\omega(\tau)^\bot) Q_\omega(\tau)$ and, since the Index of a self-adjoint operator is zero, we have:
\begin{equation}\label{co1}
\begin{array}{c}
\mbox{Ind}\{[\Pi_\omega(\tau) u \Pi_\omega(\tau)]\oplus \Pi_\omega(\tau)^\bot\}\medskip \\
=\mbox{Ind}\{Q_\omega(\tau)([\Pi_\omega(\tau) u \Pi_\omega(\tau)]\oplus \Pi_\omega(\tau)^\bot )Q_\omega(\tau)\}.
\end{array}
\end{equation}
From the general theory of the Index \cite{Gilkey1994tr}, we know that each set $O_n$, containing all Fredholm operators with $\mbox{Ind}\{F\}=n$, is an open set in the space of bounded linear operators. The open sets $O_n$ are disjoint and one cannot continuously move from one $O_n$ to another without exiting the Fredholm class. In the following, we will use the wording ``almost all" to indicate that a statement holds true for all $\omega$'s, except for a set of measure zero. Now, because of the invariance of the Chern number with $\omega$, a set $O_n$ will either contain almost all operators $Q_\omega(\tau)[\Pi_\omega(\tau) u \Pi_\omega(\tau)]\oplus \Pi_\omega(\tau)^\bot Q_\omega(\tau)$, or almost none of them. There is no alternative. This means that if the value of the Chern number changes, almost all operators will have to simultaneously migrate from one $O_n$ to another. This is prohibited because all these operators change continuously with $\tau$.\qed

The complication with the operators $Q_\omega(\tau)$ is unavoidable because, in the concrete examples, the projectors $\Pi_\omega(\tau)$ will not change continuously with $\tau$, since eigenvalues will unavoidably cross the Fermi level during the deformation. Although this section dealt with a general family of projectors $\{\Pi_\omega\}_{\omega \in \Omega}$, we will like to switch now to the case when $\Pi_\omega$ is the spectral projector onto the spectrum below $E_F$: $\Pi_\omega$$\rightarrow$$P_\omega=\chi(H_\omega$$-$$E_F)$, with $\chi(-x)$ being the Heaviside function and $\{H_\omega\}_{\omega\in \Omega}$ a covariant family of Hamiltonians. In this case, the operators $Q_\omega$ can be constructed via a standard procedure.

\begin{corollary}\label{Coro} Let $\{H_\omega\}_{\omega \in \Omega}$ be a covariant family such that
\begin{equation}\label{ExpDec}
\begin{array}{c}
\langle {\bm n},\alpha|H_\omega|{\bm m},\beta\rangle =0 \ \mbox{if} \ |{\bm n}- {\bm m}|>R,
\end{array}
\end{equation}
with $R$ arbitrarily large but fixed. Consider a deformation $\{H_\omega(\tau)\}_{\omega \in \Omega}$ such that the above relation holds true for $\tau$ in some interval $[0,\bar{\tau}]$ and such that:
\begin{equation}\label{Aizenman2}
\begin{array}{c}
|\langle {\bm n},\alpha|H_\omega(\tau)-H_\omega(\tau')|{\bm m},\beta\rangle | \leq ct. |\tau-\tau'|,
\end{array}
\end{equation}
for all $\tau$ and $\tau'\in[0,\bar{\tau}]$. The constant above is assumed independent of $\omega$, $\tau$ or $\tau'$. If
\begin{equation}\label{LocLength}
{\cal T}(\{ [{\bm x},\chi(H_\omega(\tau) - E_F)]^2\}) \leq \bar{\Lambda}^2<\infty
\end{equation}
for all $\tau\in [0,\bar{\tau}]$ and the spectrum of $H_\omega(\tau)$ near $E_F$ is regular, that is, with probability one, it has finite degeneracy, then the Chern number of $\{\chi(H_\omega(\tau)$$-$$E_F)\}_{\omega\in\Omega}$ is well defined and assumes the same integer value for all $\tau\in[0,\bar{\tau}]$. 
\end{corollary}

Before we start the proof, we have the following remark. The condition written in Eq.~\ref{LocLength} implies spectral localization \cite{BELLISSARD:1994xj}, that is, the spectrum near $E_F$ is pure point. In the following, when we use the wording ``localized spectrum," we refer to that part of the spectrum where Eq.~\ref{LocLength} holds true, together with the regularity condition of being finitely degenerate.\medskip

\noindent {\it Proof.} We take $Q_\omega$ to be $\phi (H_\omega$$-$$E_F)$, where $\phi(x)$ is a smooth function equal to one everywhere except for a small interval around the origin. We require that $\phi(x)$$>$0 except at $x$=0, where we require $\phi(0)$=0 and $d^N \phi/dx^N|_{x=0}$=0 for all $N$'s. Such $\phi$ can be constructed via standard procedures. Let $\phi_\pm(x)$ be the smooth functions that are identically zero for negative/positive $x$ and equal to $\phi(x)$ for positive/negative $x$. If $E_F$ is in the pure point spectrum, then $\phi(H_\omega$$-$$E_F){\cal H}$ is a closed subspace of ${\cal H}$ and since the degeneracy of the localized spectrum is finite, $\dim \mbox{Null}(\phi(H_\omega$$-$$E_F))$$<$$\infty$. In other words, $\phi(H_\omega$$-$$E_F)$ is in the Fredholm class. Moreover:
\begin{equation}
\begin{array}{c}
Q_\omega([P_\omega u P_\omega]\oplus P_\omega^\bot) Q_\omega \medskip \\
=[\phi_-(H_\omega-E_F)u\phi_-(H_\omega-E_F)]\oplus \phi_+(H_\omega - E_F).
\end{array}
\end{equation}
As opposed to $P_\omega$, the operators $\phi_\pm(H_\omega$$-$$E_F)$ are changing smoothly under the deformation and this concludes the argument. 

In the following we demonstrate that $\phi_\pm(H_\omega$$-$$E_F)$ changes continuously during the deformation. To see this, we employ a functional calculus introduced by Dynkin \cite{Dynkin:1972dn} (see also Helffer and Sjostrand \cite{Helffer:1989kx}). We consider a domain ${\mathcal D}$ in the complex plane defined by all those $z$=$u+iv$ with $|v|\leq v_0$ and $u$ in an interval taken large enough so that it contains all the spectra of $H_\omega(\tau)$. For any smooth function $\varphi(x)$, such as $\phi(x)$ and $\phi_\pm(x)$, and positive integer $N$, we can construct a function $f$:${\mathcal D}$$\rightarrow$${\bm C}$ such that:
\begin{enumerate} 
\item $f_N(z,\bar{z})$=$\varphi(z)$ when $z$ is on the real axis.
\item $|\partial_{\bar{z}}f_N(z,\bar{z})|\leq \alpha_N |\mbox{Im} \ z|^N$.\smallskip
\end{enumerate}

\noindent Such $f$ is called an almost analytic extension of $\varphi$ and one has:
\begin{equation}
\begin{array}{c}
\varphi(H_\omega(\tau)) = \frac{1}{2\pi}\int_{\mathcal D} \partial_{\bar{z}}f_N(z,\bar{z}) (z-H_\omega(\tau))^{-1}d^2 z.
\end{array}
\end{equation}
The matrix elements $\langle {\bm n},\alpha |\varphi(H_\omega(\tau))$$-$$\varphi(H_\omega(\tau')) |{\bm m},\beta\rangle$ can be computed as:
\begin{equation}\label{diff}
\begin{array}{l}
 \frac{1}{2\pi}\int_{\mathcal D} d^2 z \ \partial_{\bar{z}}f_N(z,\bar{z}) 
\sum\limits_{{\bm p},{\bm r},\alpha',\beta'}\langle {\bm n},\alpha |(z-H_\omega(\tau))^{-1}|{\bm p},\alpha' \rangle \medskip \\
\ \ \ \ \times \langle {\bm p},\alpha'|\delta H |{\bm p}+{\bm r},\beta' \rangle  \langle {\bm p}+{\bm r},\beta' |(z-H_\omega(\tau'))^{-1}|{\bm m},\beta\rangle,
\end{array}
\end{equation}
with $\delta H$=$H_\omega(\tau)$$-$$H_\omega(\tau')$ and the summation over ${\bm r}$ restricted to $|{\bm r}|$$\leq$$R$. Since the norm of an arbitrary bounded operator $B$ satisfies the relation:
\begin{equation}\label{BoundedB}
\begin{array}{c}
\|B\|^2\leq  \sup\limits_{{\bm m},\beta} \sum\limits_{{\bm n},\alpha} | \langle {\bm n},\alpha |B|{\bm m},\beta\rangle | \ \sup\limits_{{\bm n},\alpha} \sum\limits_{{\bm m},\beta} | \langle {\bm n},\alpha |B|{\bm m},\beta\rangle |,
\end{array}
\end{equation}
we need to show that sums of the matrix elements of Eq.~\ref{diff} are convergent and proportional to $|\tau$$-$$\tau'|$. We will use an estimate that follows from a standard application of the Thomas-Combes technique \cite{Combes:1973nx}, and is valid for all operators satisfying Eq.~\ref{ExpDec} \cite{Prodan:2009lo}. This estimate is:
\begin{equation}
\begin{array}{c}
|\langle {\bm n},\alpha|(u+iv-H_\omega(\tau))^{-1}|{\bm m},\beta\rangle | \leq ct. |v|^{-1} e^{-\epsilon |v||n-m|},
\end{array}
\end{equation} 
with $\epsilon$ small but finite. All constants appearing above are independent of $\omega$ and $\tau$. We then have:
\begin{equation}
\begin{array}{l}
|\langle {\bm n},\alpha |\varphi(H_\omega(\tau))-\varphi(H_\omega(\tau')) |{\bm m},\beta\rangle| 
\leq ct.|\tau-\tau'|\medskip \\ 
\ \ \ \ \times \int_{E_m}^{E_M}du \int_{-v_0}^{v_0}dv \ |v|^{N-2}
\sum\limits_{{\bm p},{\bm r}}e^{-\epsilon|v|(|{\bm n}-{\bm p}|+|{\bm p}+{\bm r}-{\bm m}|)} \medskip \\
\leq ct.|\tau-\tau'|\int_{0}^{v_0}dv \ v^{N-4} e^{-\epsilon v(|{\bm n}-{\bm m}|+1)/2} \medskip \\
\leq ct.|\tau - \tau'|(|{\bm n}-{\bm m}|+1)^{3-N}.
\end{array}
\end{equation}
Taking $N$ strictly larger than 5 will insure that the matrix elements are summable. \qed

\section{Numerical Evaluation of the Non-Commutative Chern number}\label{NumChern}

Before we turn to applications, we discuss a practical formula to evaluate the non-commutative Chern number in the presence of disorder. The formula was announced in Ref.~\cite{Prodan2010ew}. The non-commutative Chern number formula:
\begin{equation}\label{Chernxx}
\begin{array}{c}
C=2\pi i \int dP(\omega) \mbox{tr}_0\big{ \{ }P_\omega \big{[} -i[\hat{x}_1,P_\omega],-i[\hat{x}_2,P_\omega] \big{]}\big{ \} }
\end{array}
\end{equation}
makes sense only for infinite samples. Here we will look for a finite size formula that converges exponentially fast to the above exact formula. We start from the observation that, even in the clean limit where the Bloch decomposition is available, we still cannot compute $C$ exactly, and various numerical approximations must be adopted. Now, whatever formula we find for the disorder case, it must also work for the clean case. Therefore, our first goal will be to find a real-space, finite size formula for $C$ in the clean case. We start from Eq.~\ref{ChernX}.

It is clear that the integrand of Eq.~\ref{ChernX} can be evaluated only on a discrete set of points of the Brillouin torus, most often a finite $N\times N$ mesh ${\cal M}$: ${\bm k}_{\bm n}=(n_1\Delta, n_2\Delta)$, $n_{1,2}=1,\ldots,N$, where $\Delta =\frac{2\pi}{N}$. We will use the notation ${\bm \Delta}_1$=$(\Delta,0)$ and ${\bm \Delta}_2$=$(0,\Delta)$. The integral over ${\bm k}$ is then replaced by a Riemann sum:
\begin{equation}
\begin{array}{c}
\int \{\ldots\}d^2{\bm k}\rightarrow \sum_{{\bm k}_{\bm n}\in {\cal M}}\{\ldots \} \Delta ^2.
\end{array}
\end{equation}
Assuming for a moment that the integrand can be computed exactly, such action will introduce errors that vanish exponentially with $N$, because the integrand in Eq.~\ref{ChernX} is analytic of ${\bm k}$. 

We now turn our attention to the integrand. While the projector can be indeed computed exactly on the mesh considered above, its derivatives with respect to $k_{1,2}$ can be computed only approximately, using finite differences:
\begin{equation}\label{FiniteDiff}
\begin{array}{c}
\partial_{k_i}\hat{P}_{\bm k} \rightarrow \delta_{k_i}\hat{P}_{{\bm k}_{\bm n}}=\sum_{m=1}^Q c_m [\hat{P}_{{\bm k}_{\bm n}+m{\bm \Delta}_i}-\hat{P}_{{\bm k}_{\bm n}-m{\bm \Delta}_i}],
\end{array}
\end{equation}
with $Q\leq \frac{N}{2}$. We need to find the appropriate combination of the $c_m$ coefficients so that the finite difference approximation introduces only exponentially small errors. Being a smooth function defined on a torus, the projector $\hat{P}_{\bm k}$ can be decomposed as an absolutely converging sum of the form $\sum_{{\bm n}\in{\bm Z}^2} b_{\bm n} e^{i{\bm k}{\bm n}}$. Therefore we need to find a good finite difference approximation for functions of the form $e^{i{\bm k}{\bm x}}$, with ${\bm x}$ arbitrary but fixed. We have:
 \begin{equation}
 \begin{array}{c}
(\partial_{k_i}-\delta_{k_i})e^{i{\bm k}{\bm x}}=i(x_i-2\sum_{m=1}^{Q} c_m \sin m\Delta x_i)e^{i{\bm k}{\bm x}}.
\end{array}
\end{equation}
If we choose ${\bm c}$=$(c_1,c_2,\ldots,c_Q)$ to be solution of the the following linear set of equations:
\begin{equation}
\begin{array}{c}
\hat{A}{\bm c}^T=\frac{2}{\Delta}(1,0,0,\ldots,0)^T, \ \ A_{ij}=j^{2i-1},\ \ i,j\in \{1,\ldots,Q\},
\end{array}
\end{equation}
then $(\partial_{k_i}$$-$$\delta_{k_i})e^{i{\bm k}{\bm x}}$=$o(\Delta^{2Q-1})$, an exponentially small error if we take $Q$ of the order $\frac{N}{2}$.

Our next task is to translate the discrete $k$-space formula:
\begin{equation}\label{ChernDiscrete}
\begin{array}{c}
C=\frac{1}{2\pi i}\sum_{{\bm k}_{\bm n}} \mbox{tr} \{\hat{P}_{{\bm k}_{\bm n}}[\delta_{k_1}\hat{P}_{{\bm k}_{\bm n}},\delta_{k_2}\hat{P}_{{\bm k}_{\bm n}}]\}\Delta^2.
\end{array}
\end{equation}
into a dual real space formula. Note that $\hat{P}_{{\bm k}_{\bm n}}$ is nothing else but a discrete Bloch decomposition of the projector corresponding to the original real space model restricted to a $N$$\times$$N$ lattice with periodic boundary conditions. Let ${\cal H}_N$ be the Hilbert space  generated by $|{\bm n},\alpha\rangle$ with ${\bm n}$ restricted to this finite lattice. The discrete Bloch transformation is given by:
\begin{equation}
\begin{array}{c}
U:{\cal H}_N \rightarrow \bigoplus_{{\bm k}_{\bm n}\in {\cal M}} {\bf C}^K, \ 
U|{\bm x},\alpha\rangle = \bigoplus_{{\bm k}_{\bm n}\in {\cal M}} \frac{1}{N}e^{-i{\bm k}_{\bm n}{\bm x}}{\bm e}_\alpha
\end{array}
\end{equation}
and
\begin{equation}
\begin{array}{c}
U^{-1}\left ( \bigoplus_{{\bm k}_{\bm n}\in {\cal M}} {\bm \psi}({\bm k}_{\bm n}) \right )= \sum\limits_{{\bm n},\alpha} \frac{1}{N}\sum\limits_{{\bm k}_{\bm m}\in {\cal M}}e^{i{\bm k}_{\bm m}{\bm x}}\psi^\alpha({\bm k}_{\bm m}) |{\bm n},\alpha\rangle.
\end{array}
\end{equation}
Now note that Eq.~\ref{ChernDiscrete} can be equivalently written as:
\begin{equation}
\begin{array}{c}
C=-2\pi i \mbox{Tr}\big{ \{ }\bigoplus_{{\bm k}_{\bm n}\in {\cal M}} (\hat{P}_{{\bm k}_{\bm n}}[\delta_{k_1}\hat{P}_{{\bm k}_{\bm n}},\delta_{k_2}\hat{P}_{{\bm k}_{\bm n}}])\big{ \} },
\end{array}
\end{equation}
where the trace is taken over the whole Hilbert space $\bigoplus_{{\bm k}_{\bm n}\in {\cal M}} {\bf C}^K$. Since the trace is invariant under isometries, we can map this trace to the real space:
\begin{equation}
\begin{array}{c}
C=2\pi i \mbox{Tr}\big{ \{ } U^{-1} \big ( \bigoplus_{{\bm k}_{\bm n}\in {\cal M}} \hat{P}_{{\bm k}_{\bm n}}[\delta_{k_1}\hat{P}_{{\bm k}_{\bm n}},\delta_{k_2}\hat{P}_{{\bm k}_{\bm n}}]\big ) U \big{ \} }.
\end{array}
\end{equation}
The real space representation of the finite difference of Eq.~\ref{FiniteDiff} is:
\begin{equation}
\begin{array}{c}
\langle {\bm x},\alpha|U^{-1}\big ( \bigoplus_{{\bm k}_{\bm n}\in {\cal M}} \sum_{m=1}^Q c_m [\hat{P}_{{\bm k}_{\bm n}+m{\bm \Delta}_i}-\hat{P}_{{\bm k}_{\bm n}-m{\bm \Delta}_i}] \big)U|{\bm y},\beta\rangle \medskip \\
= \langle {\bm x},\alpha|\sum_{m=1}^Q c_m(e^{-im\hat{{\bm x}} {\bm \Delta}_i}Pe^{im\hat{{\bm x}}{\bm \Delta}_i}-e^{im\hat{{\bm x}}{\bm \Delta}_i}Pe^{-im\hat{{\bm x}}{\bm \Delta}_i})|{\bm y},\beta\rangle.
\end{array}
\end{equation}
which leads us to the desired real-space formula for the Chern number:
\begin{equation}\label{Chern3}
\begin{array}{c}
C=-\frac{2\pi i}{N^2}\sum_{{\bm n},\alpha}\langle {\bm n},\alpha|P \big{[} -i\lfloor \hat{x}_1,P\rfloor,-i\lfloor \hat{x}_2,P\rfloor \big{]}|{\bm n},\alpha\rangle, 
\end{array}
\end{equation}
where
\begin{equation}
\begin{array}{c}
\lfloor \hat{x}_i,P\rfloor =i \sum_{m=1}^Q c_m(e^{-im\hat{{\bm x}} {\bm \Delta}_i}Pe^{im\hat{{\bm x}}{\bm \Delta}_i}-e^{im\hat{{\bm x}}{\bm \Delta}_i}Pe^{-im\hat{{\bm x}}{\bm \Delta}_i}).
\end{array}
\end{equation} 
For the clean limit, the formula of Eq.~\ref{Chern3} is completely equivalent with the discrete $k$-space formula Eq.~\ref{ChernDiscrete}. In this case, it is known explicitly and is numerically confirmed that this Eq.~\ref{Chern3} converges exponentially fast to the exact value of the Chern number. For example, computed on a 30$\times$30 (40$\times$40) lattice for the Chern insulator model with $\eta$=0.6, the formula gives $C$=0999998 (0.999999998). In the presence of disorder, we can evaluate the Chern number via:
\begin{equation}\label{Chern4}
\begin{array}{c}
C_\omega=-\frac{2\pi i}{N^2} \sum_{{\bm n},\alpha}\langle {\bm n},\alpha|P_\omega \big{[} -i\lfloor \hat{x}_1,P_\omega \rfloor,-i\lfloor \hat{x}_2,P_\omega \rfloor \big{]}|{\bm n},\alpha\rangle. 
\end{array}
\end{equation}
When the Fermi level is in a mobility gap, we expect this formula to converge exponentially fast to the exact expression given in Eq.~\ref{Chernxx}, as $N$ is taken to infinity, a fact that we indeed observe in our numerical simulations. Note that Eq.~\ref{Chern4} is self-averaging, so the dependence on $\omega$ disappears in the thermodynamic limit.

\section{Applications}

\begin{figure}
 \center
 \includegraphics[width=5cm]{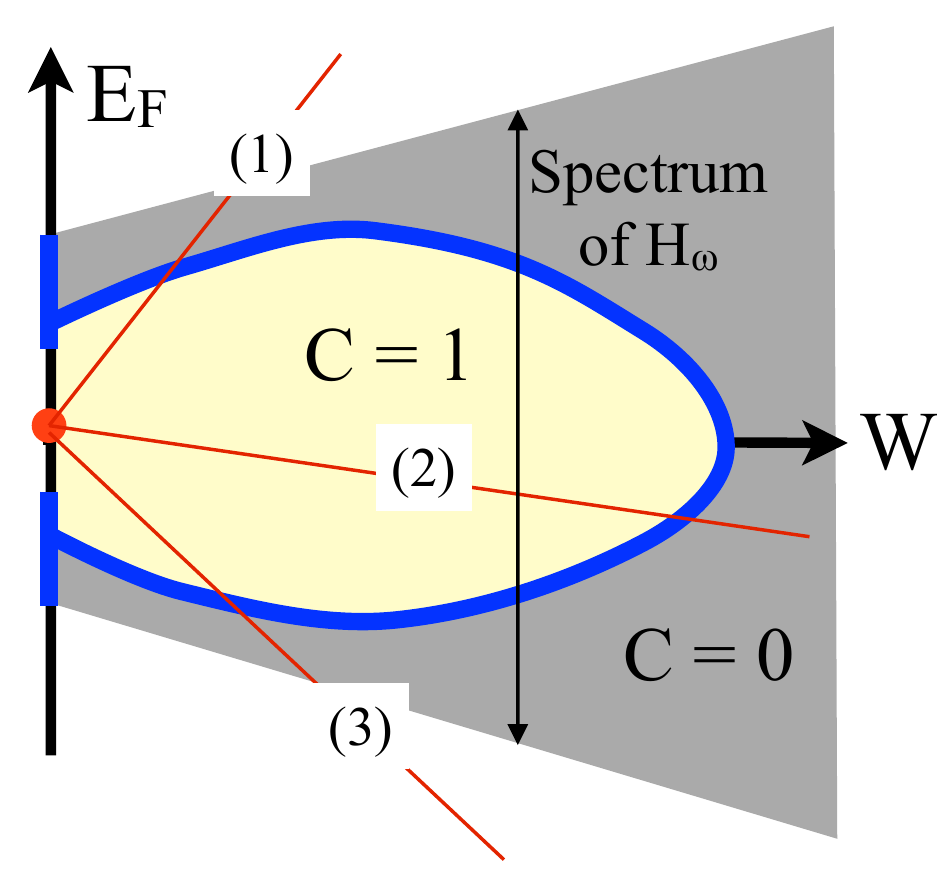}
 \caption{Deformations along any of the rays (1), (2) or (3) inherently lead to a change of the Chern number from 1 to 0. This allows one to infer the phase diagram pictured above.}
 \label{ChernPhaseDiagramExplained}
\end{figure}

\subsection{Chern Insulators}

The Non-Commutative Theory can be applied to arbitrarily complex Chern insulator models. We will, however, return to the simplest Chern insulator model discussed at the beginning. As an application, we will derive the phase diagram shown in Fig.~\ref{ChernPhaseDiagram} using only the invariant properties of the Chern number. 

We consider a smooth deformations of the model by changing one parameter that can be 1) a constant added to the Hamiltonian, which will effectively change the Fermi level $E_F$, 2) the disorder strength $W$, or 3) the hopping amplitude $\zeta_\alpha$. We denote this parameter by the generic symbol $\tau$, like we did before in the general discussion of Section \ref{Invariance}. $H_\omega$ is of course of finite range and the bound:
\begin{equation}
|\langle {\bm n},\alpha|H_\omega(\tau)-H_\omega(\tau')|{\bm m},\beta\rangle|\leq ct.|\tau - \tau'|,
\end{equation}
is uniformly satisfied for $\tau$ in an arbitrarily large but finite interval $[0,\bar{\tau}]$. In other words, we are in the conditions of the Corollary \ref{Coro}. Given all the above, we can assert with absolute confidence that the Chern number of the occupied states is an integer as long as the Fermi level resides in a localized part of the energy spectrum. The Chern number can change its value during the deformation only if the Fermi level crosses a region of delocalized spectrum. 

We now focus to what happens when we move the Fermi level or increase the disorder strength $W$. Since all the other parameters are kept fixed, we can identify the system with a point in the $(E_F,W)$ plane. Let us start from a half-filled system and a $H_0$ that is in the Chern insulator part of the clean phase diagram shown in Fig.~\ref{ChernPhaseBoundary}. We now consider deformations of the systems along different rays that start from this initial point, as shown in Fig.~\ref{ChernPhaseDiagramExplained}. Note that the energy spectrum of $H_\omega$, represented by the gray region in Fig.~\ref{ChernPhaseDiagramExplained}, becomes wider with $W$ but stays finite. Now, moving along any such ray, the Chern number will eventually become zero because one of the following reasons: the projectors $P_\omega$ become the identity (ray 1), all the bulk states localize when $W$ crosses a threshold value (ray 2) \cite{Aizenmann1993uf}, or the projectors $P_\omega$ become null (ray 3). Given these facts, a phase diagram like the one discussed in Fig.~\ref{ChernPhaseDiagram} emerges, which necessarily contains a region where $C$=1, a surrounding region where $C$=0, and a region of delocalized states separating the $C$=0 and $C$=1 regions. The numerics and scaling arguments hint that the region of delocalized states is infinitely thin.

\begin{figure}
 \center
 \includegraphics[width=12cm]{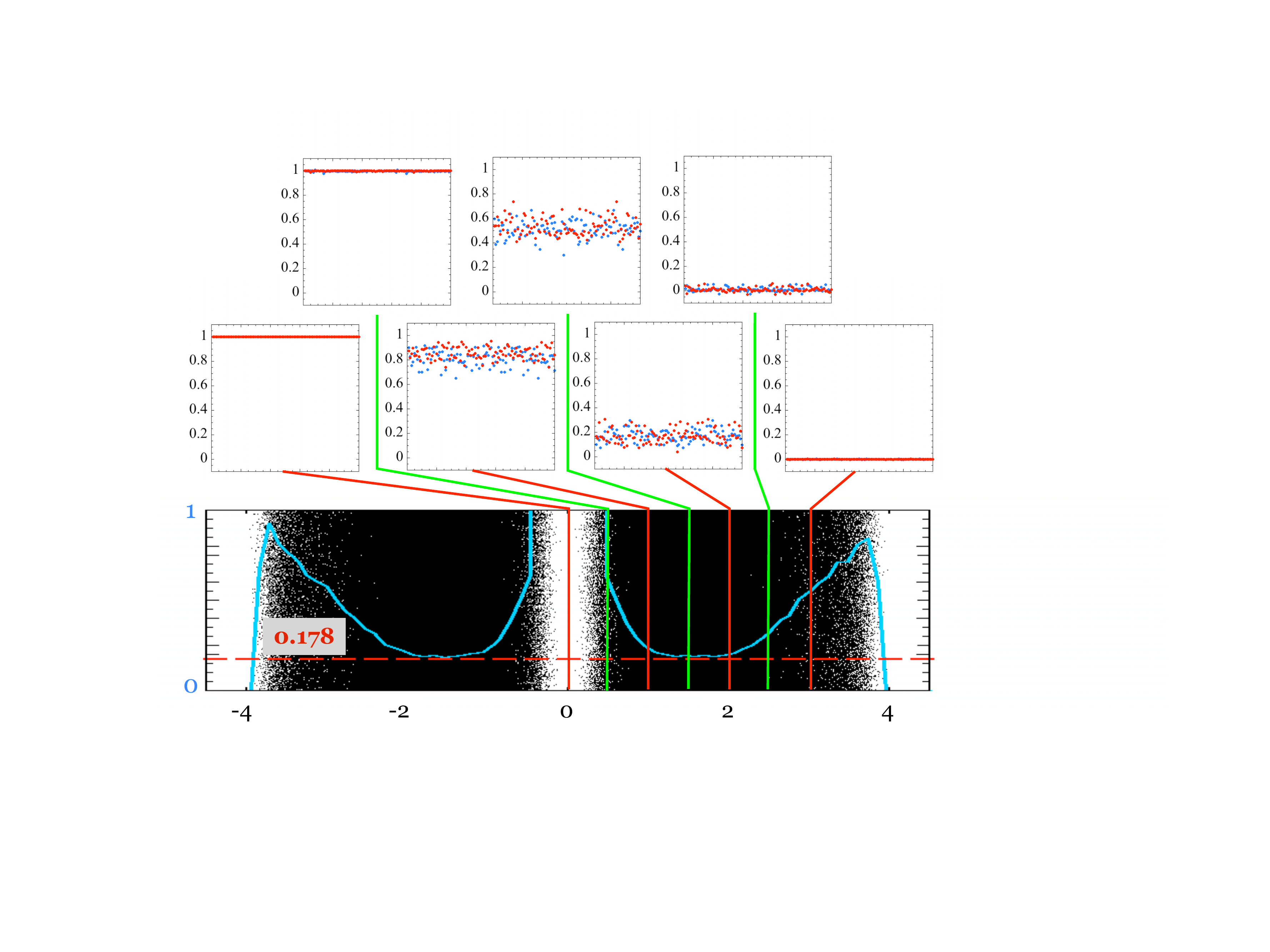}
 \caption{The lower panel shows the energy spectrum and variance of the level spacings, which already appeared in Fig.~\ref{ChernLevelStatistics}(a). The upper panels show the numerical values of $C_\omega$ for many disorder configurations, corresponding to the Fermi levels indicated by the vertical marks. The blue/red data in these panels correspond to computations completed on 30$\times$30/40$\times$40 lattices, respectively.}
 \label{NumericChern}
\end{figure}

Fig.~\ref{NumericChern} shows the numerical values of the Chern number, computed via the technique presented in Section~\ref{NumChern} for seven Fermi levels. The model parameters were given the same values as in Fig.~\ref{ChernLevelStatistics}(a) and the calculations were performed for 30$\times$30 and 40$\times$40 lattices. In absolute agreement with the general theory and with the discussion above, the calculations show that, for Fermi levels that are in between the delocalized spectral region, identified here with the region where the covariance is very closed to the value of 0.178, the Chern number has the quantized value of one or tends to this quantized value as the lattice size is increased. The Chern number decreases to zero as the Fermi level passes the region of delocalized spectrum and after that it assumes again a quantized value of zero. 

\subsection{Quantum spin-Hall Insulators}

\subsubsection{The non-commutative spin-Chern number.}

We restrict the discussion to the model of Eq.~\ref{KaneMele}. In the clean limit, we have argued that the QSH insulator can be characterized by the spin-Chern number $C_s$ defined in Eq.~\ref{SpinChern}. The definition involved the spectral projectors $P^\pm$ of the operator $P\hat{\sigma_z}P$ onto its positive/negative spectrum, more precisely the Chern number of these projector. Given that and using the non-commutative theory of the Chern number, we are now in the position to extend the definition of the spin-Chern number to the disordered case. Let $P_\omega$ be the projectors onto the states below the Fermi level and let $P_\omega^\pm$ be the spectral projectors of $P_\omega \hat{\sigma}_z P_\omega$ onto its positive/negative spectrum, respectively. Let $C_\pm$ be the corresponding Chern numbers, defined through the no-commutative formula:
\begin{equation}
C_\pm=2\pi i \int dP(\omega) \ \mbox{tr}_0\big{ \{ }P_\omega^\pm \big{[} -i[\hat{x}_1,P_\omega^\pm],-i[\hat{x}_2,P_\omega^\pm] \big{]}\big{ \} }.
\end{equation}
Then the non-commutative spin-Chern number is defined as 
\begin{equation}
\begin{array}{c}
C_s=\frac{1}{2}(C_+ - C_-).
\end{array}
\end{equation}
The next Sections will discuss the conditions in which $C_s$ is well defined and takes quantized values.
 
\subsubsection{Weak disorder regime.}

We call weak disorder the regime when the amplitude $W$ is small enough such that a gap of size $2\Delta_1$ is still present in energy spectrum of $H_\omega$, and a gap of size $2\Delta_2$ is present in the spectrum of $P_\omega\hat{\sigma_z}P_\omega$. The Fermi level is assumed to be inside the energy gap. The second condition was actually observed to hold true in all our numerical experiments, with weak and strong disorder.  

One should not mistakenly think of this weak disorder regime as the regime in which a perturbative approach is justifiable. For example, $W$=3 in Fig.~\ref{QSHLevelStatistics} belongs to our weak regime, but note that $W$=3 is larger than the width of the clean bands. Such disorder is quite large and definitely cannot be treated perturbatively.

\begin{proposition} In this weak disorder regime, the matrix elements $\langle {\bm n},\alpha,\sigma|P_\omega^\pm|{\bm m},\beta,\sigma'\rangle$ decay exponentially with the separation $|{\bm n}-{\bm m}|$. Consequently, the spin-Chern number $C_s$ is well defined and takes a quantized value. Moreover, $C_s$ cannot change its value during continuous deformations $H_\omega(\tau)$ that leaves the two gaps open at all times. 
\end{proposition}

\noindent{\it Proof.} First of all, let us establish analytically that such weak regime exists. We have seen that, in the clean limit, the gap of $P\hat{\sigma}_zP$ remains open for all $H_0$ inside the QSH part of the phase diagram (recall Fig.~\ref{Gaps}). Assume now that $H_0$ is fixed in the QSH part of the phase diagram and let us increase $W$. As long as the spectral gap of $H_\omega$ remains open, the projectors $P_\omega$ are continuous of $W$. At $W$=0, we know as a fact that $\|(P\hat{\sigma}_zP-0)^{-1}\| \approx 1$, so we can be sure that $\|(P_\omega\hat{\sigma}_z P_\omega-0)^{-1}\| < \infty$, for $W$ smaller than a critical value. This critical value depends on where exactly is $H_0$ located in the phase diagram. The conclusion here is that the origin is outside the spectrum of $P_\omega\hat{\sigma}_z P_\omega$, at least for small $W$, and that the spectrum is divided in positive and negative islands and we can safely define the projectors $P_\omega^\pm$.

 Our next task is to establish the exponential localization for $P_\omega^\pm$, which can be accomplish via a straightforward application of the Combes-Thomas technique \cite{Combes:1973nx}. We decided to present a fairly complete discussion for the readers that are not familiar with this technique. We do not seek here optimal estimates, which could be obtained with the technique developed in Ref.~\cite{Prodan:2006cr}, but rather the simplest proof possible. We will first establish the exponential localization of the full projector $P_\omega$. Consider the invertible map:
\begin{equation}
|{\bm n},\alpha,\sigma\rangle \rightarrow U_{{\bm q}}|{\bm n},\alpha,\sigma\rangle=e^{-{\bm q}\cdot{\bm n}}|{\bm n},\alpha, \sigma \rangle, \ {\bm q}\in {\bm R}^2.
\end{equation}
Due to the finite range of $H_\omega$, one will easily find that $H_\omega({\bm q})\equiv U_{{\bm q}}H_\omega U_{{\bm q}}^{-1}$ has the form ($q=|{\bm q}|$) $H_\omega({\bm q})=H_\omega + q V({\bm q})$, with $V({\bm q})$ independent of $\omega$ and uniformly bounded by some $\bar{V}$ if $q$ is restricted to a finite interval $[0,\bar{q}]$. For a $z$ located a distance $\Delta_1$ away from the spectrum of $H_\omega$, we have:
\begin{equation}
\begin{array}{l}
\|z-H_\omega({\bm q})\|=\|z-H_\omega+q V({\bm q})\|  \geq \|z-H_\omega\|-q \bar{V}=\Delta_1-q \bar{V}.
\end{array}
\end{equation}
In other words, $z-H_\omega({\bm q})$ is invertible for small $q$, and
\begin{equation}
\|(z-H_\omega({\bm q}))^{-1}\| \leq \frac{1}{\Delta_1 -q\bar{V}}.
\end{equation}
Noticing that 
\begin{equation}
(z-H_\omega({\bm q}))^{-1}=U_{\bm q}(z-H_\omega)^{-1}U_{\bm q},
\end{equation}
which can be verified by applying $z-H_\omega({\bm q})$ to both sides, we can conclude that:
\begin{equation}
e^{{\bm q}\cdot ({\bm n} - {\bm m})} |\langle {\bm n},\alpha,\sigma|(z-H_\omega)^{-1}|{\bm m},\beta,\sigma'\rangle| \leq \frac{1}{\Delta_1 -q\bar{V}}. 
\end{equation}
Taking ${\bm q}$ parallel to ${\bm n}$$-$${\bm m}$, we obtain the exponential decay of the resolvent with the separation $|{\bm n}-{\bm m}|$. The exponential decay of the projector follows from the formula
\begin{equation}
\langle {\bm n},\alpha,\sigma|P_\omega|{\bm m},\beta,\sigma'\rangle = \oint_{\cal C} \frac{dz}{2\pi i} \ \langle {\bm n},\alpha,\sigma|(z-H_\omega)^{-1}|{\bm m},\beta,\sigma'\rangle,
\end{equation}
where ${\cal C}$ is a contour surrounding the spectrum below the spectral gap. All points on this contour can be taken a $\Delta_1$ distance away from the spectrum of $H_\omega$, so the above estimates can be applied without modifications. 

\begin{figure}
 \center
 \includegraphics[width=12cm]{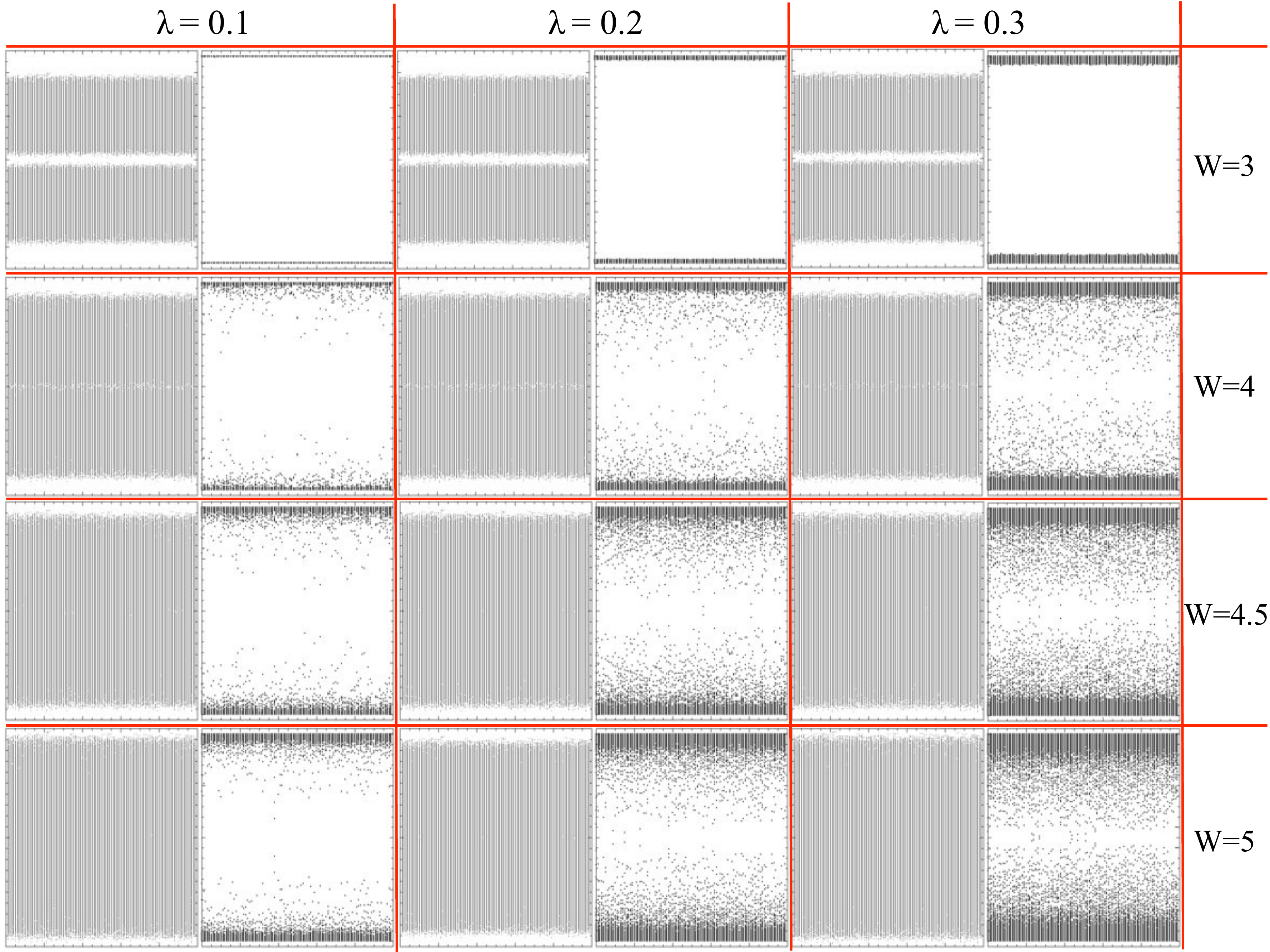}
 \caption{Each box shows the energy spectrum of $H_\omega$ (left) and next to it the spectrum of $P_\omega \hat{\sigma}_z P_\omega$ (right) for many disorder configurations at different values of W and strengths of the Rashba potential. The other parameters were fixed at $\eta$=0.6 and $t$=0. The Fermi level was fixed at the origin in these calculations.}
 \label{Separation}
\end{figure}

We now turn our attention to $P_\omega^\pm$. Since 
\begin{equation}
\langle {\bm n},\alpha,\sigma|P_\omega^\pm|{\bm m},\beta,\sigma'\rangle = \oint_{\cal C} \frac{dz}{2\pi i} \ \langle {\bm n},\alpha,\sigma|(z-P_\omega\hat{\sigma}_z P_\omega)^{-1}|{\bm m},\beta,\sigma'\rangle,
\end{equation}
we can try to repeat the arguments we used for $P_\omega$, this time working with $P_\omega\hat{\sigma}_z P_\omega$ instead of $H_\omega$. This will work because
\begin{equation}
U_{{\bm q}}P_\omega \hat{\sigma}_z P_\omega U_{{\bm q}}^{-1} =P_\omega ({\bm q}) \hat{\sigma}_z P_\omega({\bm q}),
\end{equation}
and, since $P_\omega({\bm q})$ is a smooth function of ${\bm q}$,
\begin{equation}
P_\omega ({\bm q}) \hat{\sigma}_z P_\omega({\bm q})=P_\omega \hat{\sigma}_z P_\omega + q R({\bm q}),
\end{equation}
with $R({\bm q})$ uniformly bounded by some $\bar{R}$ if $q$ is restricted to a finite interval $[0,\bar{q}]$. From here, we can repeat the arguments used for $P_\omega$, to conclude:
\begin{equation}
 |\langle {\bm n},\alpha,\sigma|P_\omega^\pm|{\bm m},\beta,\sigma'\rangle| \leq \frac{ct. e^{-q|{\bm n} - {\bm m}|}}{\Delta_2 -q\bar{R}}. 
\end{equation}
The exponential localization of $P_\omega^\pm$ can be established for any $\tau$ of a  deformation that keeps the two gaps open. Moreover, for such deformations, the projectors $P_\omega^\pm(\tau)$ change continuously with $\tau$. These two facts prevent any sudden jumps of the spin-Chern numbers.\qed
 
\subsubsection{Beyond the weak disorder regime.}

\begin{figure}
 \center
 \includegraphics[width=10cm]{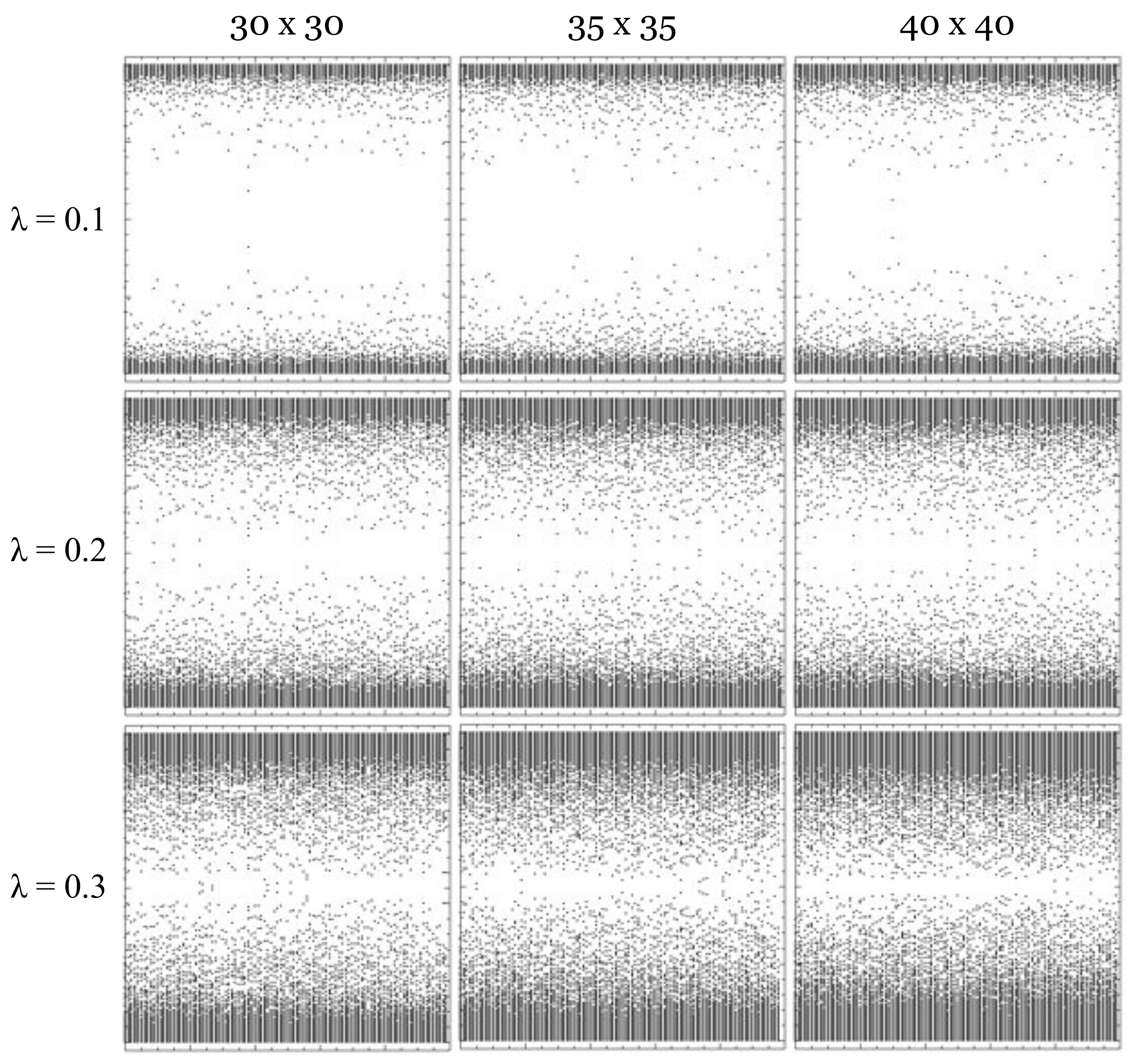}
 \caption{The spectrum of $P_\omega \hat{\sigma}_z P_\omega$ at $W$=5 and various strengths of the Rashba potential, computed on lattices of increasing size. The other parameters were fixed at $\eta$=0.6, $t$=0 and the Fermi level was pinned at the origin.}
 \label{QSHSSizeDep}
\end{figure}

In all our numerical experiments, we have observed a clear separation of the spectrum of $P_\omega \hat{\sigma}_z P_\omega$ into positive and negative parts. In fact, we consistently noticed a gap in the spectrum of $P_\omega \hat{\sigma}_z P_\omega$ near the origin, whose origins are not clear to us at this moment. We exemplify this in Fig.~\ref{Separation}, where we show the energy spectrum of $H_\omega$ and of $P_\sigma \hat{\sigma}_z P_\omega$, when $P_\omega$ is computed with $E_F$ fixed at 0. For all values of $\lambda$ and $W$ shown in this figure, we can clearly observe the spectral gap for $P_\omega \hat{\sigma}_z P_\omega$, even when the energy gap of $H_\omega$ was completely filled with localized spectrum. More importantly, the spectrum of $P_\sigma \hat{\sigma}_z P_\omega$ closer to the origin was always seen to be localized. In Fig.~\ref{QSHSSizeDep}, we show the size dependence of the spectrum of $P_\omega \hat{\sigma}_z P_\omega$ for $W$=5 and different Rashba potential strengths. A clear spectral gap near the origin remains visible in all the plots and the spectrum closer to the origin was found to be localized. Owing to the NCG theory of the Chern number, we can state the following fact.

\begin{proposition}\label{NCGspin-Chern} The spin-Chern number is well defined and takes quantized values as long as the spectrum of $P_\omega \hat{\sigma}_z P_\omega$ near the origin is localized. The invariance of $C_s$ can be also established if some additional assumptions are considered, similar to those of Refs.~\cite{Aizenman1998bf} and \cite{Richter2001jg}.
\end{proposition}

The main issue now is to establish under what conditions is the spectrum of $P_\omega \hat{\sigma}_z P_\omega$ localized near the origin. The following simple argument will show that the localization lengths of the projectors $P_\omega^\pm$ diverge whenever the localization length of the full projector $P_\omega$ diverges. This shows that the phase defined by $C_s$=1 is contained within the boundaries defined by the metallic phase established in Fig.~\ref{QSHLevelStatistics}. Here is the argument:
\begin{equation}
\begin{array}{l}
 |\langle {\bm n},\alpha,\sigma|P_\omega|{\bm 0},\alpha,\sigma\rangle|^2 = |\langle {\bm n},\alpha,\sigma|P_\omega^-+P_\omega^+|{\bm 0},\alpha,\sigma\rangle|^2 \medskip \\
\leq  2  (|\langle {\bm n},\alpha,\sigma|P_\omega^-|{\bm 0},\alpha,\sigma\rangle|^2 + |\langle {\bm n},\alpha,\sigma|P_\omega^+|{\bm 0},\alpha,\sigma\rangle|^2),
\end{array}
\end{equation}
which gives,
\begin{equation}\label{LocLengths}
\begin{array}{l}
\frac{1}{2} \Lambda(P_\omega)^2 \leq \Lambda(P_\omega^-)^2 +\Lambda(P_\omega^+)^2.
\end{array}
\end{equation}
Therefore, whenever $\Lambda(P_\omega)^2$ diverges to $+\infty$ the localization lengths of $P_\omega^{\pm}$ (which are equal due to the time-reversal symmetry) will be forced to diverge to $+\infty$. 

\begin{figure}
 \center
 \includegraphics[width=13cm]{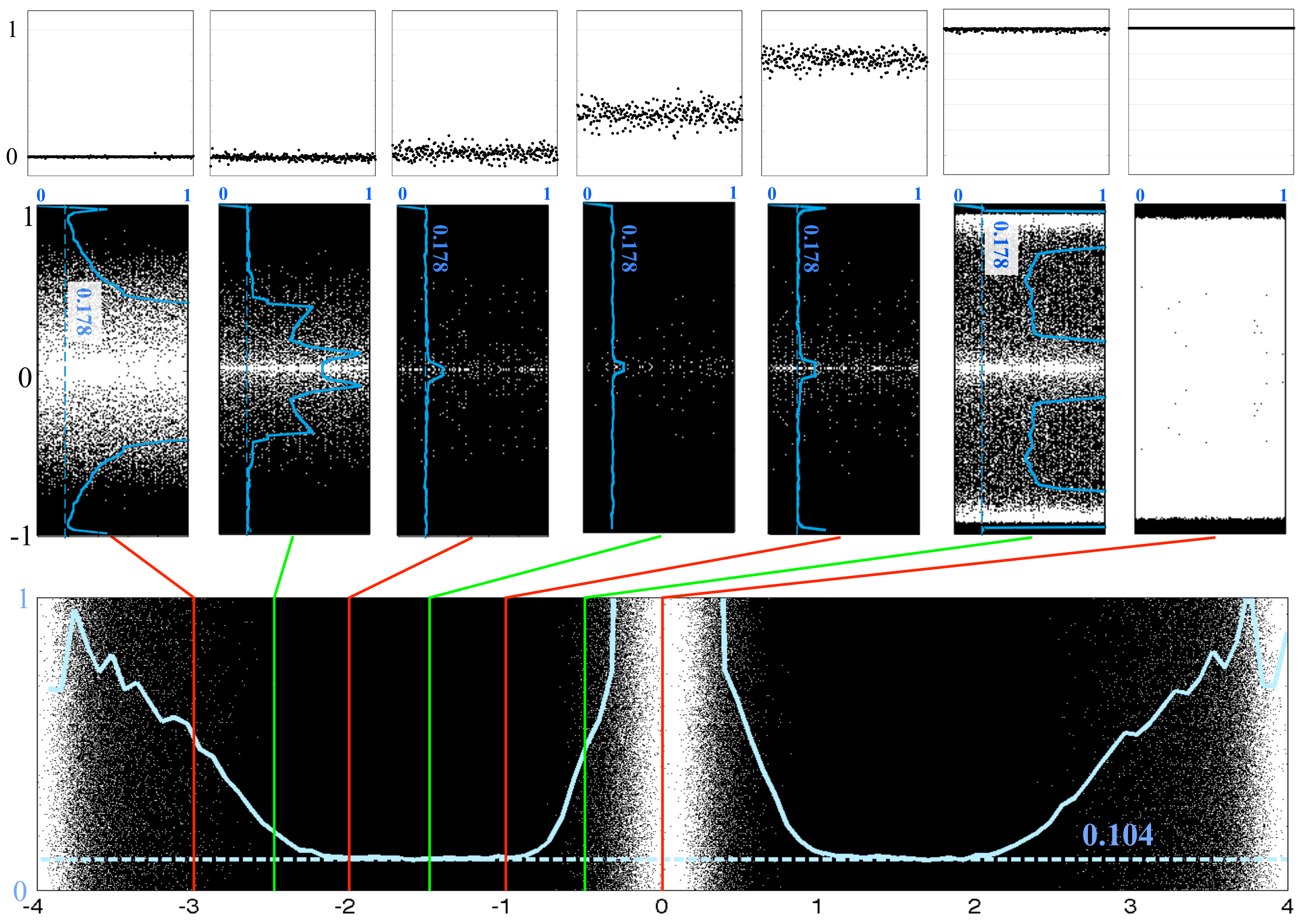}
 \caption{The lower panel shows the energy spectrum and variance of the level spacings, which already appeared in Fig.~\ref{QSHLevelStatistics}(a) (tilted horizontally here). The mid panels show the spectrum of $P_\omega \hat{\sigma}_z P_\omega$ with the Fermi level pinned at the vertical marks, and the variance of the level spacings. The upper panels show the numerical values of the spin-Chern number for many disorder configurations, corresponding to the same Fermi levels indicated by the vertical marks. The computations was completed on a 40$\times$40 lattice and for 1000 disorder configurations.}
 \label{SEig40x40W3}
\end{figure}

The bulk theory of our QSH insulators will be completed if we can establish that the phase defined by $C_s$=1 actually extends all the way to the metallic phase inferred from Fig.~\ref{QSHLevelStatistics}. For this we need to establish that the localization length of the full projector $P_\omega$ diverges whenever the localization lengths of $P_\omega^\pm$ diverge. This will not only show that the $C_s$=1 phase extends all the way to the metallic phase, but will also imply the existence of such metallic phase surrounding the QSH phase. In other words, we will be able to establish the phase diagram of the QSH model in a manner similar to what was presented for Chern insulators. This, unfortunately is a highly technical issue and falls beyond the scope of this paper. Nevertheless, the NCG statement from Proposition~\ref{NCGspin-Chern}, the inequality of Eq.~\ref{LocLengths} and the numerical technique for evaluating the spin-Chern number, have given us the ground to study the remaining issue via numerical means.    

In Fig.~\ref{SEig40x40W3}, we show what happens when we lower the Fermi level and cross the spectral region containing the extended states. The bottom panel of the figure shows the energy spectrum of $H_\omega$ at $W$=3, together with the variance of the level spacings. The mid panels show the spectrum of $P_\omega \hat{\sigma}_z P_\omega$ with $P_\omega$ computed for the marked Fermi levels. The upper panel shows the numerical value of the spin-Chern number. The calculation was performed on a 40$\times$40 lattice, for the QSH model with $t$=0, $\eta$=0.6 and $\lambda$=0.3.  We can see that, when the Fermi level is away from the spectral region containing the delocalized states, the spectrum of $P_\omega \hat{\sigma}_zP_\omega$ displays a gap near the origin, and the variance of the level spacings for this spectrum indicates that the spectrum near the origin is localized. Similarly, the spin-Chern number takes quantized values whenever the Fermi level is away from the energy region of delocalized states. This behavior was seen in all our numerical experiments, suggesting that indeed the $C_s$=1 phase extends all the way to the metallic phase. It is interesting to note that the variance in the mid panels collapses entirely onto the value 0.178 when the Fermi level is in the region of extending states, suggesting that $P_\omega \hat{\sigma}_z P_\omega$ becomes a stereotypical Gaussian random matrix when $P_\omega$ becomes delocalized.

\section{Conclusions}

The discussion we just concluded dealt with two simple but fundamental questions for the field of topological insulators: How to define quantized topological invariants in the presence of strong disorder and what are the physical properties protected by such invariants? These two questions are of absolute importance for most of the newly envisioned technological applications of these materials. In the same time, these questions are extremely difficult to deal with both analytically and numerically. Progress in this direction, even if minuscule, is extremely valuable at this stage.

We showcased two applications of the Non-Commutative Theory of the Chern invariant, which gave us a glimpse of the methods of Non-Commutative Geometry and what can be achieved by such methods. For Chern insulators, the analysis gave quite explicit and satisfactory answers:  we could spell out the most general conditions that assures the quantization and invariance of the non-commutative Chern number, we were able to conclude the existence of protected extended bulk states and even to draw the phase diagram of our simple Chern insulator model. For Quantum spin-Hall insulators, we defined the non-commutative spin-Chern number and again were able to spell out the most general conditions that assures its quantization and invariance, even if in a formal way, and to give some simple explicit physical situations, the weak disorder regime, where the quantization and invariance of the spin-Chern number can be firmly established. We have argued that the non-commutative spin-Chern number is an effective tool for mapping the topological phase of the model with strong disorder. The numerical simulations indicated that the spin-Chen number also protects the extended bulk states of the QSH insulator.  

The construction of the non-commutative spin-Chern number can be repeated for many other situations. There is a firm belief among the condensed matter scientists that every discrete symmetry has an associated topological insulator and a topological invariant that classifies its topological phases and the efforts towards identifying new classes of topological insulators are already underway \cite{Fu2010jy,Hughes2010gh,Turner2010cu}. We believe that novel topological phases can be found by looking for analogs of $\hat{\sigma}_z$, that is, $\hat{w}$ operators such that the spectrum of $P\hat{w}P$ splits in distinct spectral islands that carry non-zero Chern numbers. This is our primary reason to believe that the Non-Commutative Theory of the Chern number can become increasingly relevant for the field of topological insulators.

\ack{This work was supported by Yeshiva University and an award from Research Corporation for Science Advancement. I gratefully acknowledge extremely useful discussions with Taylor Hughes, Andrei Bernevig, Jean Bellissard, Michael Aizenman, Ai Yamakage and many more people who helped me shaped the manuscript.}

\bigskip

\bibliographystyle{iopart-num}
 

\providecommand{\newblock}{}

\end{document}